# Coulomb screening of superconductivity in magic-angle graphene

Julien Barrier[1,2,†], Liangtao Peng[3], Shuigang Xu[2], Christophe De Beule[4,5], V.I. Fal'ko[1,2,6], K. Watanabe[7], T. Tanigushi[7], A.K. Geim[1,2,*], Shaffique Adam[3, 8,*] and Alexey I. Berdyugin[8, 9, 1,*]

[1] School of Physics & Astronomy, University of Manchester, Manchester M13 9PL, United Kingdom

[2] National Graphene Institute, University of Manchester, Manchester M13 9PL, United Kingdom

[3] Department of Physics, Washington University in St. Louis, St. Louis, Missouri 63130, United States

[4] Department of Physics and Astronomy, University of Pennsylvania, Philadelphia, Pennsylvania 19104, USA

[5] Department of Physics, University of Antwerp, Groenenborgerlaan 171, 2020 Antwerp, Belgium

[6] Henry Royce Institute for Advanced Materials, The University of Manchester, Oxford Road, Manchester, M13 9PL, UK

[7] National Institute for Materials Science, 1-1 Namiki, Tsukuba 305-0044, Japan

[8] Department of Materials Science and Engineering, National University of Singapore, 9 Engineering Drive 1, Singapore 117575, Singapore

[9] Department of Physics, National University of Singapore, 2 Science Drive 3, Singapore 117551, Singapore

† present address: Max Planck Institute for Solid State Research, D-70569 Stuttgart, Germany

* correspondence should be addressed to alexey@nus.edu, geim@manchester.ac.uk, shaffique@wustl.edu

**The origin of superconductivity in magic-angle twisted bilayer graphene has been a subject of intense debate. While some experimental evidence indicated an unconventional pairing mechanism which should be sensitive to Coulomb screening, experimental attempts to tune the critical temperature by screening Coulomb interactions so far have remained unsuccessful, possibly indicating a conventional phonon-mediated pairing. Here we study a double-layer electronic system consisting of two twisted graphene bilayers in immediate proximity of each other but remaining electronically decoupled. By increasing the carrier density in one bilayer, we completely suppressed both the superconductivity and the correlated-insulator state in the adjacent magic-angle graphene. The observation of such an effect from screening offers strong support for an unconventional mechanism of Cooper pairing in magic-angle twisted bilayer graphene, shedding new light on the underlying physics governing their properties.**

Superconductivity in magic-angle twisted bilayer graphene (TBG) has attracted an intense interest [1–14] due to its presumed unconventional origin. This belief stems from several experimental observations including the combination of a relatively low carrier density with an unusually high critical temperature [5], and the striking resemblance between magic-angle TBG's phase diagram and that of Cuprate superconductors [5,6]. Also, the maximum of the critical temperature $T_C$ roughly coincides with the 'magic angle' of 1.1° between graphene layers [7,11] where the flattened energy bands maximize electron-electron interactions [15], suggesting that they may play a critical role in Cooper pairing. These observations have spurred numerous theories for unconventional mechanisms including plasmon-mediated [16–18], skyrmion-mediated [19,20], Kohn-Luttinger [21,22], and nodal pairing-enabled mediated by inter-valley coherent states [23] mechanisms. However, other studies remain supportive of the conventional phonon-mediated pairing [10–14,24–28].

Further efforts to clarify the pairing mechanism have so far been inconclusive. While certain findings (e.g., a V-shaped differential conductance in tunnelling spectra [8] and the observed interaction-driven superfluid stiffness with a small superconducting coherence length [9] support an unconventional pairing mechanism, experiments employing the proximity screening point towards a phonon-mediated superconductivity [10–12]. The latter experiments initially employed gates made of graphite to provide electrostatic screening of Coulomb interactions inside magic-angle TBG. The gates were separated by thin (6-10 nm) hexagonal boron nitride (hBN) spacers [11,12]. It was found that the screening suppressed correlated insulator (CI) states, while superconductivity persisted. However, the differing behaviors of CI and superconductivity in these experiments could arise from the variations of twist angle between different devices, with superconductivity surviving in a broader range of twist angles [11], complicating the interpretation. Moreover, it was impossible to measure the change of the superconducting critical temperature $T_c$ induced by the screening within a same device, as the screening was generally fixed by graphite's density of states and by the thickness of the dielectric spacer, while devices with different spacers inevitable had different twist angles. A more recent experiment employed Bernal bilayer graphene as a tunable screening layer that was decoupled from magic-angle TBG by thinner, 3-nm-thick hBN [10]. While screening again suppressed CI states, $T_C$ increased by 2-3%. This marginal increase was interpreted as a sign of conventional superconductivity where the screening suppressed Coulomb repulsion between electrons forming phonon-coupled Cooper pairs.

In contrast with previous experiments, the present work demonstrates a complete suppression of magic-angle superconductivity with tunable screening, with variations in $T_c$ by more than an order of magnitude, which is an attribute of unconventional pairing. As the screening efficiency of the metallic layer decays exponentially at distances as small as ~1-2 nm [29], we maximized the screening by using a double-layer electronic system consisting of two small-angle TBG layers stacked directly on top of each other. Instead of using a thin dielectric spacer, we employed a high twist angle between the two TBG layers, which is sufficient to decouple their electronic systems [30–37]. The high-twist angle configuration effectively allowed a sub-nm separation between the magic-angle TBG and the screening layer, enhancing the screening efficiency in our device. Our devices demonstrate a new experimental architecture to combine different twisted layers within the stack, while allowing the controllable interplay between their properties.

## Results

To create a double-layer electronic system, we assembled twisted tetralayer graphene using the twist angle sequence shown in Figure 1a. Briefly, the first two layers were rotated relative to each other at an angle of ~1°, the third layer was rotated at ~10° relative to the second, and the fourth layer was rotated again at ~0.5° relative to the third. This multi-twisted graphene stack was then encapsulated between thick (30 nm) hBN crystals to preserve graphene's electronic quality and placed on a silicon oxide substrate (see Supplementary Information, section 1. "Fabrication and characterization of multi-twisted graphene" for fabrication details; Fig. S1c shows the final device). Its longitudinal $\rho_{xx}$ and Hall $\rho_{xy}$ resistivities are plotted in Figs 1b,c as a function of carrier densities induced by top and bottom gates (see section 2. "Details of carrier density calculation" in Supplementary Information). The observed multiple peaks in $\rho_{xx}$ correlate with peaks and sign changes in $\rho_{xy}$. This behavior is attributed to the formation of electronic minibands inside each of the twisted bilayers forming the tetralayer stack, as expected for small twist angles.

To probe the electronic structure of our tetralayer, we measured its resistivity as a function of both gate voltages (Fig. 1d). This map revealed two distinct sets of peaks. One set (red arrows in Fig. 1d) depended primarily on the density $n_{bg}$ induced by the bottom gate and showed little sensitivity to the top gate-induced density $n_{tg}$. The other set (blue arrows) responded mostly to $n_{tg}$ with weak response to $n_{bg}$. The observed peaks in $\rho_{xx}$ are attributed to main and secondary neutrality points (NPs) within the miniband structure of the bottom and top TBGs with small twist angles. Indeed, it is expected that the bottom gate mainly affects the carrier density in the bottom TBG and does not affect the top TBG because of screening, and vice versa for the top gate. However, the gate-induced carrier densities were distributed between both TBG layers, as detailed in Supplementary Information. We note that such behavior is a simple electrostatic effect, which does not imply the coupling between the layers.

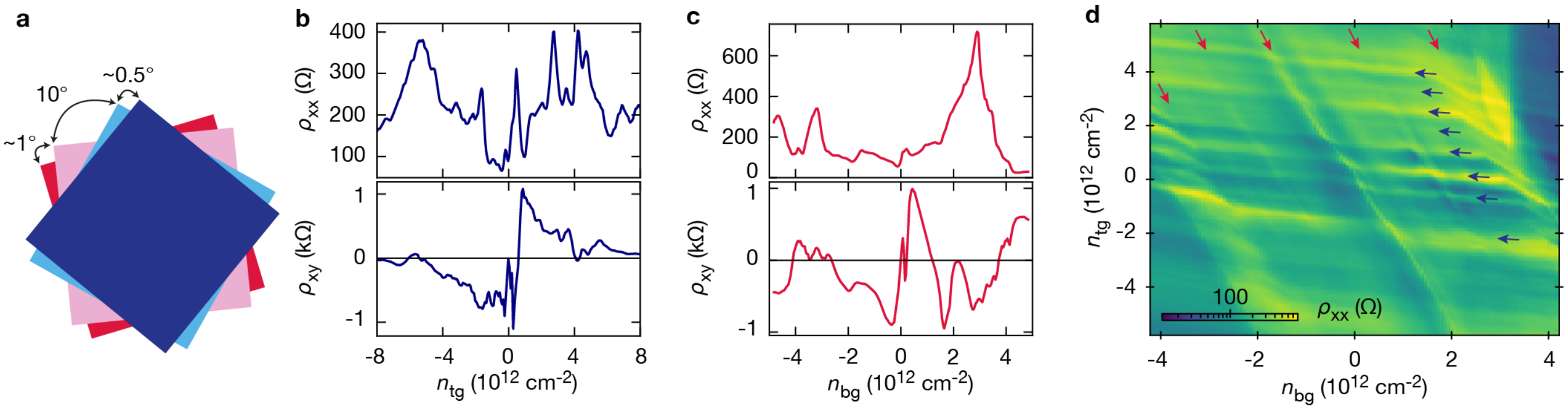

**Fig. 1| Characterization of twisted tetralayer graphene device.** (**a**) Schematics of its structure. (**b**) and (**c**) Longitudinal and Hall resistivities (top and bottom panels, respectively) as a function of top (b) and bottom (c) gate-induced carrier densities. Panel (b) was measured for $n_{bg}$=0, and panel (c) for $n_{tg}$=0. The Hall signal was anti-symmetrized using small magnetic fields of ±100 mT. (**d**) Longitudinal resistivity as a function of $n_{tg}$ and $n_{bg}$. All measurements at 2 K.

To accurately determine the carrier density inside top and bottom TBG layers ($n_{bTBG}$ and $n_{tTBG}$, respectively), we analyzed the $\rho_{xx}$ maps as detailed in Supplementary Information, section 2 "Details of carrier density calculation". The resulting resistivity map using the new density coordinates ($n_{bTBG}$ and $n_{tTBG}$) is shown in Fig. 2a. In this representation, the peaks attributed to NPs in the bottom TBG form vertical lines, and the ones attributed to the top TBG form horizontal lines. This shows that the electronic structure of our graphene tetralayer can be understood in terms of two independent electronic systems (See Supplementary Information section 3 "Decoupling between the two bilayer systems"), in agreement with the experimental design. Then, it becomes straightforward to evaluate the twist angles $\theta$ for the two TBG layers. Considering that the full filling of each miniband corresponds to 4 electrons per superlattice unit cell, the carrier density at which the secondary NPs occurred is given by $n = 4/A$ where $A = \sqrt{3}\, a^2/8 \sin^2(\theta/2)$ is the unit cell area, and $a$ is graphene's lattice constant. Applying this expression to the carrier density difference between the most pronounced vertical and horizontal resistivity peaks in Fig. 2b, we obtained twist angles $\theta_{\mathrm{bTBG}} = 1.15°$ and $\theta_{\mathrm{tTBG}} = 0.46°$ for the bottom- and top-TBG, respectively. These twist angles also agree well with the observed positions of Brown-Zak oscillations (see section 4 "Calculation of twist angles" in Supplementary Information and Fig. S2). The twist angle of the bottom TBG is within the range for magic-angle [1–14] while the top TBG's small angle allows a rather high density of states (DoS) which is important for achieving a high screening efficiency (see below).

After determining $\theta$, we analyzed the tetralayer's electronic behavior in more detail. In Fig. 2a, apart from the pronounced vertical maxima at $n_{\mathrm{bTBG}}$ around zero and $\pm 3 \times 10^{12} \mathrm{cm}^{-2}$ which correspond to zero ($\nu = 0$; main NP) and full filling ($\nu = \pm 4$; secondary NPs) of the first miniband of the bottom TBG ($\nu$ is the number of electrons per unit cell), there is also a weaker structure that emerged around $\nu = \pm 2$. We attributed those peaks to the CI states, as reported previously for magic-angle TBG [1–6,8–12]. These states were sensitive to doping of the top TBG and completely disappeared once $n_{tTBG}$ exceeded 2 $10^{12}$ cm$^{-2}$ as shown in Fig 2. The decoupling between the two TBGs was corroborated from the absence of signatures of moiré imprinting (see Supplementary Information section 5 "Absence of moiré imprinting"). In such case, one would expect to observe additional peaks at constant $n_{bTBG}$, at similar carrier densities at which peaks are observed in the top TBG layer. No such features were observed confirming that the two TBG layers were decoupled using large twist angles.

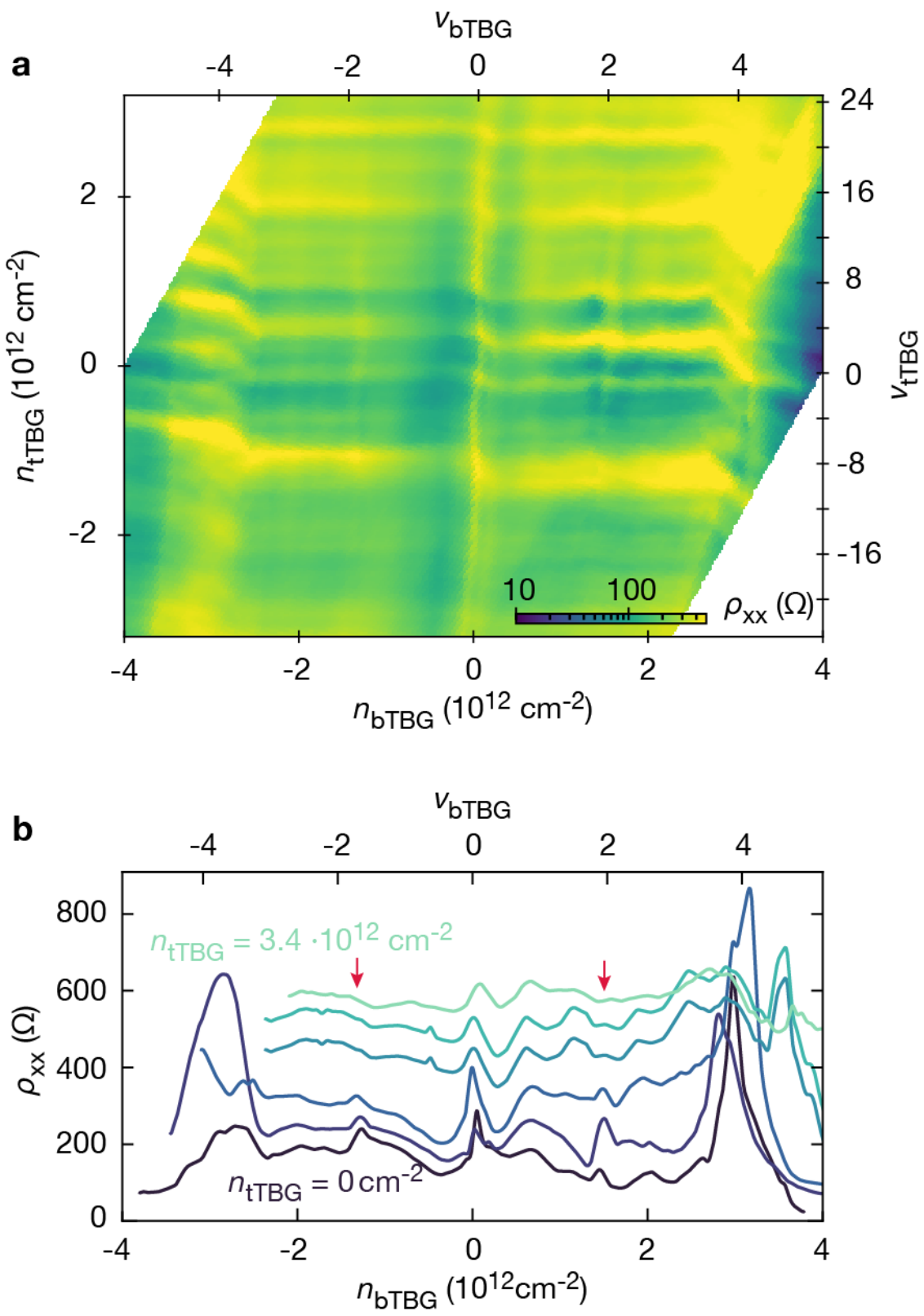

**Fig. 2| Tuning magic-angle correlated states with top-layer density.** (**a**) Longitudinal resistivity as a function of the calculated carrier densities in top and bottom TBGs. Notable curving at intersections of vertical and horizontal high resistivity states are caused by changes in screening, which are most pronounced if the Fermi level crosses a gapped state inside one of the TBG layers (see Supplementary Information section 2 "Details of carrier density calculation"). (**b**) Line cuts from panel (a), showing resistivity for fixed carrier densities in the top TBG layer. For clarity, the curves are offset vertically in steps of $0.68\times10^{12}$ cm$^{-2}$. Curves without offset are shown in Figure S3. The red arrows indicate positions of the resistivity peaks around $\nu = \pm 2$ which disappear upon increasing the carrier density in the top TBG. All measurements at 2 K. The increase in resistance observed in the top-right corner of panel (a) (see also Fig. 1d) around filling factor 4 of the bottom TBG layer can most likely be attributed to the increased resistance of the voltage probes located outside the region covered by the top gate, as highlighted in Figure S1.

To check for the presence of superconductivity, we measured the same map $\rho_{xx}(n_{\mathrm{bTBG}}, n_{\mathrm{tTBG}})$ at our lowest temperature $T$ of ~35 mK (Fig. 3a). The main difference with respect to the 2 K map of Fig. 2a is the zero resistivity regions that appeared around $\nu = \pm 2$ in the magic-angle bottom TBG (also, compare Figs 2b and 3b). The zero-resistance states quickly disappeared if $\nu$ was turned away from $\pm 2$. The states also disappeared with increasing $T$ (Fig. 3c), and upon increase of a dc current $I_{\mathrm{dc}}$ (Fig. 3d). This behavior is characteristic of the superconductivity previously reported for magic-angle graphene [1–12]. Unexpectedly, the superconductivity was also suppressed if the carrier density in the top TBG layer increased (Fig. 3a). This is further illustrated in Figs. 3d,e which show changes in $T_{\mathrm{C}}$ and the critical current as a function of $n_{\mathrm{tTBG}}$. Such suppression cannot be attributed to the simultaneously changed displacement field within the bottom TBG layer and requires an alternative explanation (see Supplementary Information section 6 "Role of the perpendicular electric field").

## Discussion

The changes in the bottom TBG induced by doping of the top TBG may originate from two distinct mechanism. First, the metallic top TBG layer may short conduction pathways at higher $|n_{\mathrm{tTBG}}|$ and reduce the correlated insulator gap. However, the sample's resistivity does not decrease at high doping levels of the top TBG layer (see Fig. 2a), as would be expected from a parallel conduction mechanism. Second, the top TBG layer may modify the electrostatic potential and tune the magnitude of Coulomb interactions within the stack. Since the large twist angle between the TBG layers decouples their electronic band structures, we attribute the changes in the properties of the bottom TBG induced by doping of the top TBG to the modified Coulomb interaction within the stack. Earlier reports [10–12] also demonstrated that screening of Coulomb interactions in magic-angle TBG reduces the CI state's gap. Based on these conclusions we also attribute the suppression of the CI states around $\nu = \pm 2$ (Figs. 2, 3a) to increased screening efficiency of the top TBG as its doping increased. On the other hand, $T_{\mathrm{C}}$ in the earlier experiments was either unaltered by screening [11,12] or only marginally (~2%) increased [10]. In stark contrast, $T_{\mathrm{C}}$ in our experiment was so strongly affected by screening that superconductivity could be destroyed even at our lowest $T$ (Fig. 3a,d,e). We believe that the critical difference between our and earlier experiments is the distance $l$ between the screening layer and the magic-angle TBG. In our case, $l$ was sub-nm, that is one order of magnitude smaller than previously. While $l$ of several nm is sufficient to suppress the CI state, the screening of superconductivity poses stronger requirements. As shown in Supplementary Information, section 7 "Screening in a double layer structure", for the half-filled miniband supporting magic-angle superconductivity, the screening efficiency decays exponentially over a characteristic distance of ~2 nm. In addition, the DoS of our top small-angle TBG is expected to be higher than that of the Bernal-stacked bilayer graphene employed previously, which also enhances the screening efficiency (Fig. S17).

For further analysis, we calculated the Coulomb interaction in the bottom TBG as $V_{\mathrm{bTBG}}(\boldsymbol{q}, \omega) = V_0/\epsilon$, where $V_0$ is the bare Coulomb interaction, and $\epsilon$ is the dielectric function of our four-layer system, which can be written as $\epsilon = \beta - V_0\, \Pi_{\mathrm{bTBG}}$. Here, $\beta$ is the screening contribution from the top layer, and $\Pi_{\mathrm{bTBG}}$ the intralayer polarizability of magic-angle TBG. The interlayer screening term $\beta$ depends on $l$ and, in the limit $l \to 0$, $\beta$ becomes proportional to the dielectric constant of the top layer, which is set by an average DoS around the Fermi energy (section 7 "Screening in a double-layer structure" in Supplementary Information). Conversely, if $l$ becomes comparable with the magic-angle superlattice period, $\beta$ approaches 1 so that the top TBG no longer contributes to screening of Coulomb interactions. To evaluate $\beta$, one needs to know the DoS of the screening (top) layer. For this purpose, we performed band structure calculations using different theoretical models that incorporate lattice relaxation, strain, and intralayer Coulomb interactions, all of which are important in marginally twisted bilayers. The results are shown in Fig. S19. We find that the magnitude of DoS and its behavior at high carrier densities are highly sensitive to the modelling assumptions, which render predictions unreliable.

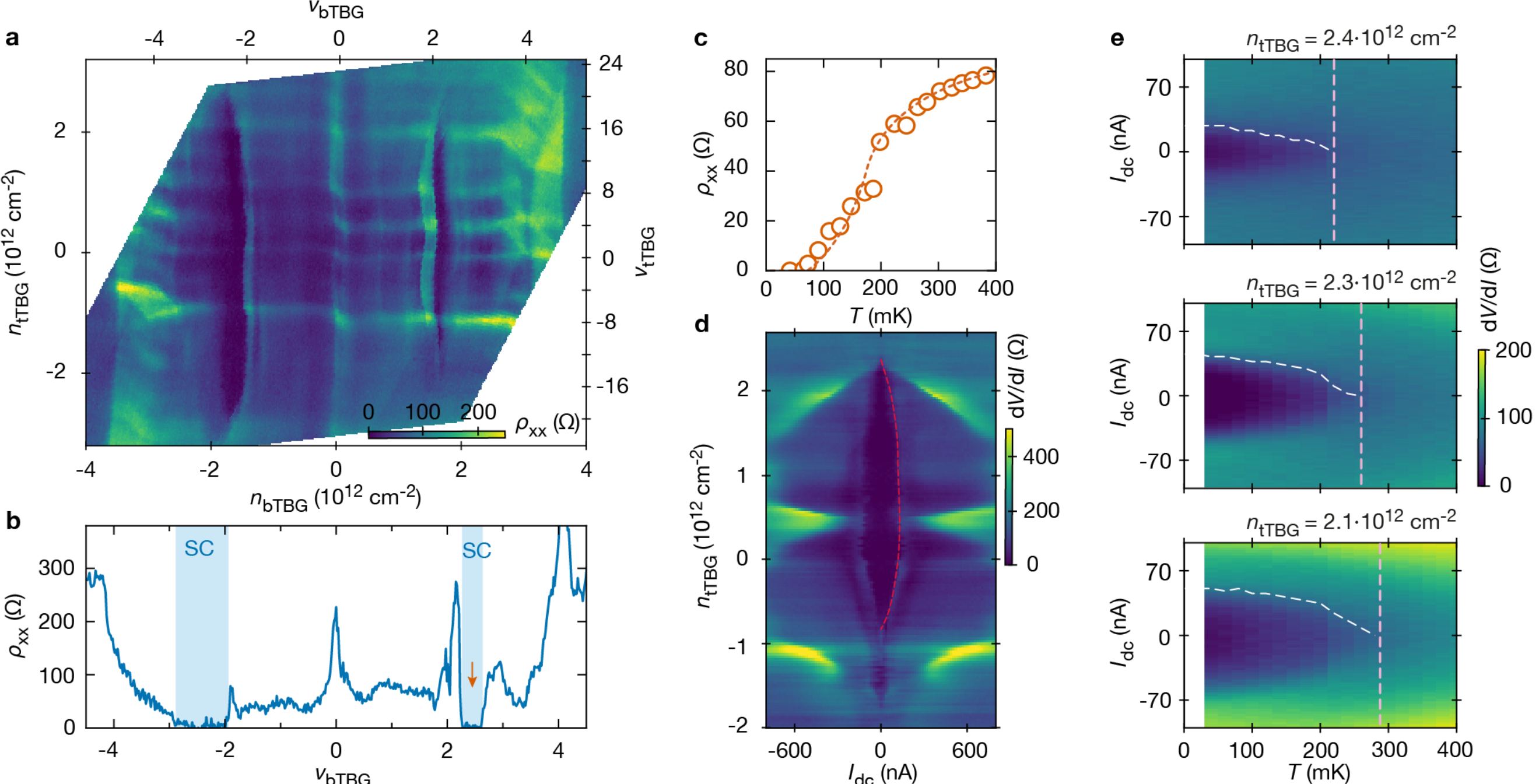

**Fig. 3| Screening of magic-angle superconductivity.** (**a**) Longitudinal resistivity as a function of carrier density in the top and bottom TBGs at 35 mK. Same color scale as in Fig. 1E. (**b**) $\rho_{xx}$ as a function of $n_{\mathrm{bg}}$ measured at zero $n_{\mathrm{tg}}$. (**c**) Temperature dependence of $\rho_{xx}$ at the density indicated by the arrow in panel (b). Dashed curve, guide to the eye. At zero top-gate voltage, increased contact resistivity leads to a slightly reduced $T_c$ compared with panel (e) at higher carrier density. All other measurements were performed under applied top-gate voltage (Fig. S10). (**d**) Differential resistance as a function of $I_{\mathrm{dc}}$ and $n_{\mathrm{tTBG}}$ for the same $n_{\mathrm{bTBG}}$ as in panel (c). Red dashed line is a guide to the eye highlighting superconducting critical current. The asymmetry critical currents range for positive and negative $n_{\mathrm{tTBG}}$ possible originate from the asymmetry in the band structure of the top TBG layer. The additional arc-like pattern which appears at high current can be attributed to the out-of-equilibrium criticalities in one of the TBG layers [38]. A magnified view at lower bias currents is presented in Fig. S20 (**e**) Differential resistance as a function of $I_{\mathrm{dc}}$ and $T$ for different $n_{\mathrm{tTBG}}$. The arc-like dashed curves mark the critical current defined as the current at which $\rho_{xx}$ reaches half the normal-state resistivity. The vertical lines mark $T_{\mathrm{C}}$.

Nevertheless, a model may capture the qualitative trends observed in our experiment by satisfying two essential criteria: (1) an increasing DoS with the carrier density and (2) the DoS's order of magnitude should be consistent with experimental observations. To avoid over-modelling, we chose to simplify the problem and approximate the DoS of the top TBG layer $\mathcal{N}_{\mathrm{tTBG}}$ as a scaled single-layer graphene band structure, characterized by a scaling factor $\alpha$. We model it as $\mathcal{N}_{\mathrm{tTBG}} = \alpha \times \mathcal{N}_{\mathrm{Gr}}$ where $\mathcal{N}_{\mathrm{Gr}}$ is the DoS of monolayer graphene. The proportionality coefficient $\alpha$ can be estimated by modelling the suppression of the CI state in a way that the CI gap becomes smaller than 2 K if the carrier density in the top TBG layer exceeds $2{\times}10^{12}$ cm$^{-2}$, as per our experiment (Fig. 2). Further details are discussed in Supplementary Information section 8 "Density of states approximation". This simple estimation gives $\alpha \sim 10$, which is just a factor of two away from the top TBG DoS obtained using sophisticated models accounting for the lattice relaxation, and intra-layer Coulomb interactions at the characteristic carrier density $2{\times}10^{12}$ cm$^{-2}$ (see Fig. S18). Given the complexity of the problem, we believe this level of agreement is reasonable and supports the validity of our simplified approach.

Using the modelled $\mathcal{N}_{\mathrm{tTBG}}$, we could estimate how different mechanisms of superconductivity would be affected by the proximity screening. In conventional superconductors, where electron-electron repulsion is stabilized by phonon-induced attraction, the screening should generally suppress the repulsion and, hence, increase $T_C$ [10], which contradicts our observations. In principle, the screening might also affect the deformation potential and reduce the electron-phonon coupling. This scenario is considered in Supplementary Information, section 9 "Phonon-mediated superconductivity" and shown to be unrealistic. Accordingly, the screening behavior observed in our experiment rules out the conventional (BCS-like) mechanisms of Cooper pairing and lends support to an electronic pairing mechanism. As an example of the latter mechanisms, we considered electron-plasmon pairing [16–18] for which screening is expected to strongly impact superconductivity [17,18]. Our numerical modelling shows that, indeed, screening by the top TBG should greatly reduce $T_C$ (section 10 "Plasmon-mediated superconductivity" in Supplementary Information). The comparison between critical temperatures observed in experiment and modelling results are shown in Figure S10. Furthermore, we find that in this mechanism the critical temperature is sensitive to the average DoS of the screening layer over a broad energy window, rather than to its precise value at the Fermi level, see Supplementary Information section 11 "Models incorporating relaxation". This explains the absence of pronounced modulation of critical temperature and the critical current near the CNPs of the top TBG layer, where the DoS exhibits a local minimum. Finally, we numerically calculated the dependence $T_C(l)$ for this particular pairing mechanism and found that the screening effect should be most pronounced at sub-nm distances and rapidly decrease with increasing $l$ above ~2 nm, in agreement with the exponential dependence derived in Supplementary Information, section 7 "Screening in a double layer structure".

Although our modelling using a plasmon-induced pairing yields reasonable agreement with the experimental observations, a complete theoretical treatment would require a fully self-consistent GW calculation of the entire four-layer system, which is technically unfeasible at this point. Additionally, we point out that, while plasmon-mediated superconductivity seems a plausible explanation for magic-angle superconductivity, we cannot rule out other unconventional mechanisms such as, e.g., skyrmion-mediated [19,20], Kohn-Luttinger [21,22], and nodal pairing-enabled mediated by inter-valley coherent states [23]. The latter mechanisms are similarly susceptible to proximity screening. Finally, we considered the possibility that the observed destruction of superconductivity originates from the reversed proximity effect [39] between the magic-angle superconductor and the normal-metal top TBG. This scenario would require a highly transparent superconductor-normal metal interface, which is not the case with our system (Supplementary Information section 12 "Estimations of the interface transparency for the proximity effect").

## Conclusions

Our experiments revealed a striking sensitivity of the magic-angle superconductivity to proximity screening, which rules out conventional phonon pairing and tightens constraints on possible unconventional mechanisms of superconductivity in magic-angle TBG. Additionally, our approach of combining small and large twist angles within the same heterostructure shows a way to design novel electronic systems where Coulomb interactions in one layer are screened by controlling the carrier density in the other. Looking ahead, in future work it would be interesting to combine superconducting magic-angle TBG or other flat-band systems with monolayer graphene, exploiting the more controllable and well-characterised density of states of the screening layer. Finally, we note that while conceptually similar to previously studied double-layer systems separated by thin dielectric spacers [40–42], the high-twist-angle approach provides ultimately small separation between electronic systems, which is essential for efficient screening, and could also be used to explore inter- and intra- layer interaction effects in moiré quantum materials [1,2,43].

During the preparation of this manuscript, we became aware of another work sharing the same conclusion on the role of screening in magic-angle TBG [44].

## Supplementary Information

### 1. Fabrication and characterization of multi-twisted graphene

The twisted graphene tetralayer device was prepared using the ‘cut and stack’ method [1,2]. First, we exfoliated hexagonal boron nitride (hBN) and graphene on top of $SiO_2$ substrates to find large and clean flakes. Next, using an AFM tip attached to the micromanipulator stage we cut the target graphene flake into four pieces, as shown in Figure S1a. After that we used polycarbonate (PC)/polydimethylsiloxane (PDMS) stamps to pick up the hBN layer and used that layer to sequentially pick up graphene pieces while twisting them relative to each other on a given set of angles (≈0.5°, 10°, 1°). Finally, we realized the stack on top of the clean hBN flake. The final heterostructure is shown in Figure S1b.

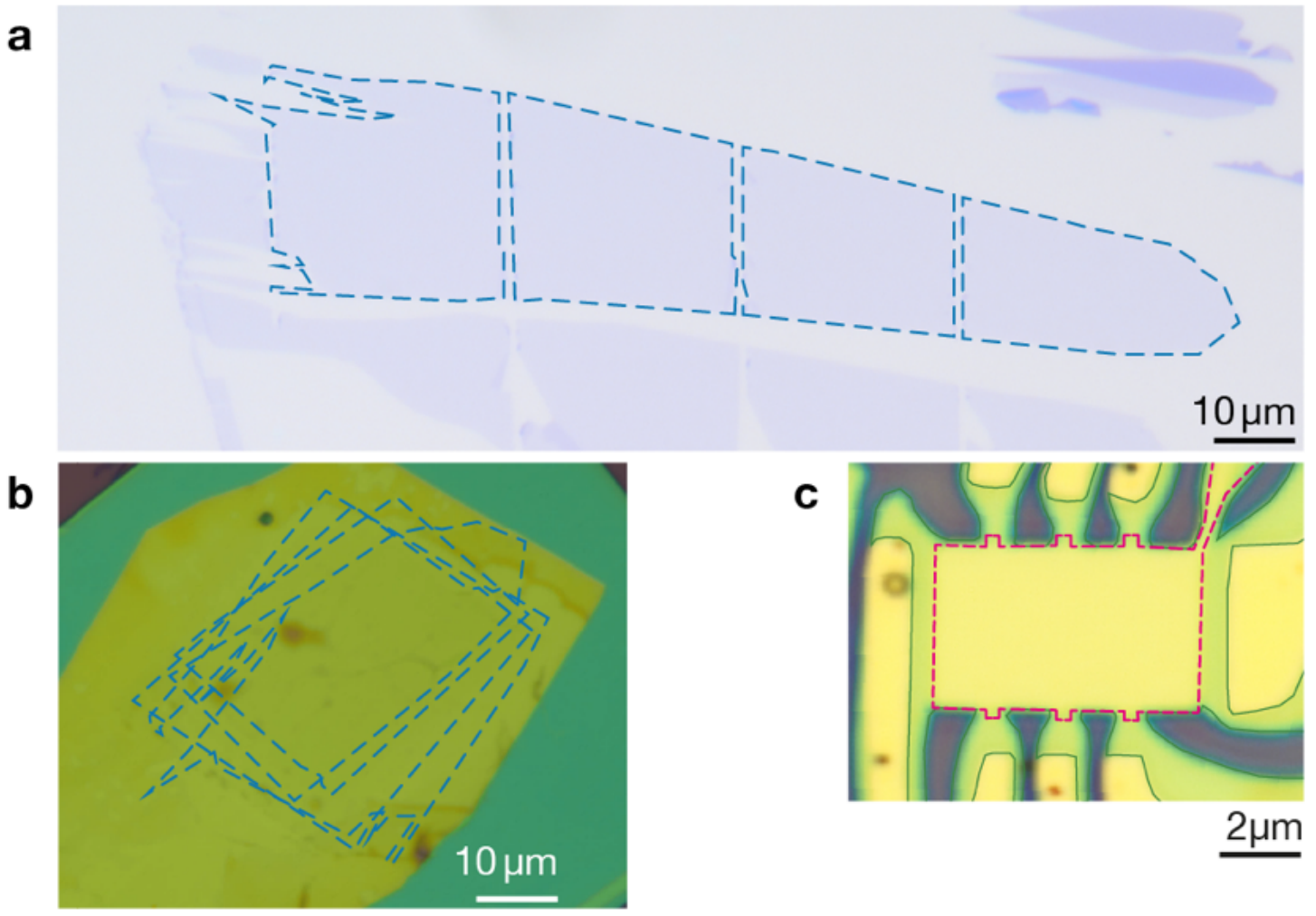

**Figure S1 | Optical micrographs during fabrication.** (**a**) Single layer graphene flake cut in 4 pieces using an AFM tip. Dashed lines highlight the boundaries of the used graphene pieces. (**b**) Final van der Waals heterostructure used for the fabrication of the device. Dashed lines highlight the graphene pieces. (**c**) Optical micrograph of the final device. The red dashed (green solid) contour highlights the extent of the metallic top gate (etched graphene, resp.).

Next, using standard nanofabrication techniques (electron-beam lithography, reactive ion etching with $O_2/CHF_3$ plasma), we shaped the tetralayer stack into a Hall bar geometry. We then used electron-beam lithography followed by metal evaporation to embed the electrodes and top gate at the desired locations. As electrodes and top gate, we used 3 nm of Cr followed by 60 nm of Au, deposited by RF sputtering. The final device is shown in Figure S1c.

Electrical measurements were carried out in a dilution refrigerator setup with a mixing chamber temperature $T_{mc}$ = 7 mK. Each line was equipped with custom low-pass RC filters located at the mixing chamber stage and at room temperature. Using commercial Coulomb Blockade Thermometers, we estimated the electronic temperature $T_e$ of our setup. It remained equal to $T_{mc}$ above 50 mK, but significantly deviated at lower temperatures, reaching $T_e$ = 35 mK at $T_{mc}$ = 7 mK.

The device’s resistivity was measured using standard low-frequency (<150 Hz) lock-in techniques with excitation currents between 1 and 5 nA. Differential resistance (*dV/dI*) characteristics (e.g. Fig. 3d, 3e of the main text) were measured with lock-ins after mixing AC and DC input currents with a homemade precision V-I converter.

Among the multiple devices fabricated for this study, two contained a layer with a twist angle close to the magic angle. Of these, only one exhibited superconductivity. Another device, with twist angles of approximately 0.95°, 10°, and 0.3°, did not show superconductivity.

This outcome is not unusual. Similar observations have been reported in other studies of magic-angle twisted bilayer graphene, including devices in which magic-angle TBG was stacked with monolayer graphene or with other small-angle TBGs [3, 4]. Importantly, the absence of superconductivity in a nominally magic-angle device is not by itself conclusive evidence of a screening effect. It may equally reflect device-quality limitations unrelated to screening.

To our knowledge, no study has yet demonstrated a superconducting magic-angle device paired with a deliberately mismatched large-angle screening layer. Future experiments using a more controllable screening layer, such as monolayer graphene or high angle TBGs, whose well-characterised band structure would simplify interpretation, may provide a cleaner platform for disentangling screening effects from the additional complexity introduced by the 0.5° twisted graphene layer studied here.

**2. Details of carrier density calculation**

As discussed further, we were able to find the carrier density inside each of the TBGs layers as a function of top and bottom gate voltages by using fan diagram measurements as a function of top and bottom gates, and double gate measurements in a zero magnetic field.

First, we considered that the carrier density induced by each of the gates was distributed between both TBG layers, so the total top and bottom gate induced carrier densities ($n_{\mathrm{tg}}(V_{\mathrm{tg}})$ and $n_{\mathrm{bg}}(V_{\mathrm{bg}})$ respectively) could be written as:

$$n_{\mathrm{bg}}(V_{\mathrm{bg}}) = n_{\mathrm{bg}}^{\mathrm{bTBG}}(V_{\mathrm{bg}}) + n_{\mathrm{bg}}^{\mathrm{tTBG}}(V_{\mathrm{bg}}), \tag{1}$$

$$n_{\mathrm{tg}}(V_{\mathrm{tg}}) = n_{\mathrm{tg}}^{\mathrm{bTBG}}(V_{\mathrm{tg}}) + n_{\mathrm{tg}}^{\mathrm{tTBG}}(V_{\mathrm{tg}}), \tag{2}$$

where $n_{\mathrm{bg}}^{\mathrm{bTBG}}(V_{\mathrm{bg}})$ corresponds to the carrier density induced by the bottom gate in the bottom TBG, $n_{\mathrm{bg}}^{\mathrm{tTBG}}(V_{\mathrm{bg}})$ is the carrier density induced by the bottom gate in the top TBG, $n_{\mathrm{tg}}^{\mathrm{tTBG}}(V_{\mathrm{tg}})$ is the carrier density induced by the top gate in the top TBG, and $n_{\mathrm{tg}}^{\mathrm{bTBG}}(V_{\mathrm{tg}})$ is the carrier density induced by the top gate in the bottom TBG layer.

The diagonal coefficients $n_{\mathrm{bg}}^{\mathrm{bTBG}}(V_{\mathrm{bg}})$, and $n_{\mathrm{tg}}^{\mathrm{tTBG}}(V_{\mathrm{tg}})$ in eq. (1) and (2) could be found by analyzing quantum oscillations in magnetic field-dependent measurements as shown in Figure S2. To obtain these maps we either fixed $V_{\mathrm{bg}} = 0$ and measured the longitudinal resistance as a function of the top gate voltage and magnetic field (Figure S2a), or fixed $V_{\mathrm{tg}} = 0$ and measured the longitudinal resistance as a function of the bottom gate voltage and magnetic field (Figure S2b).

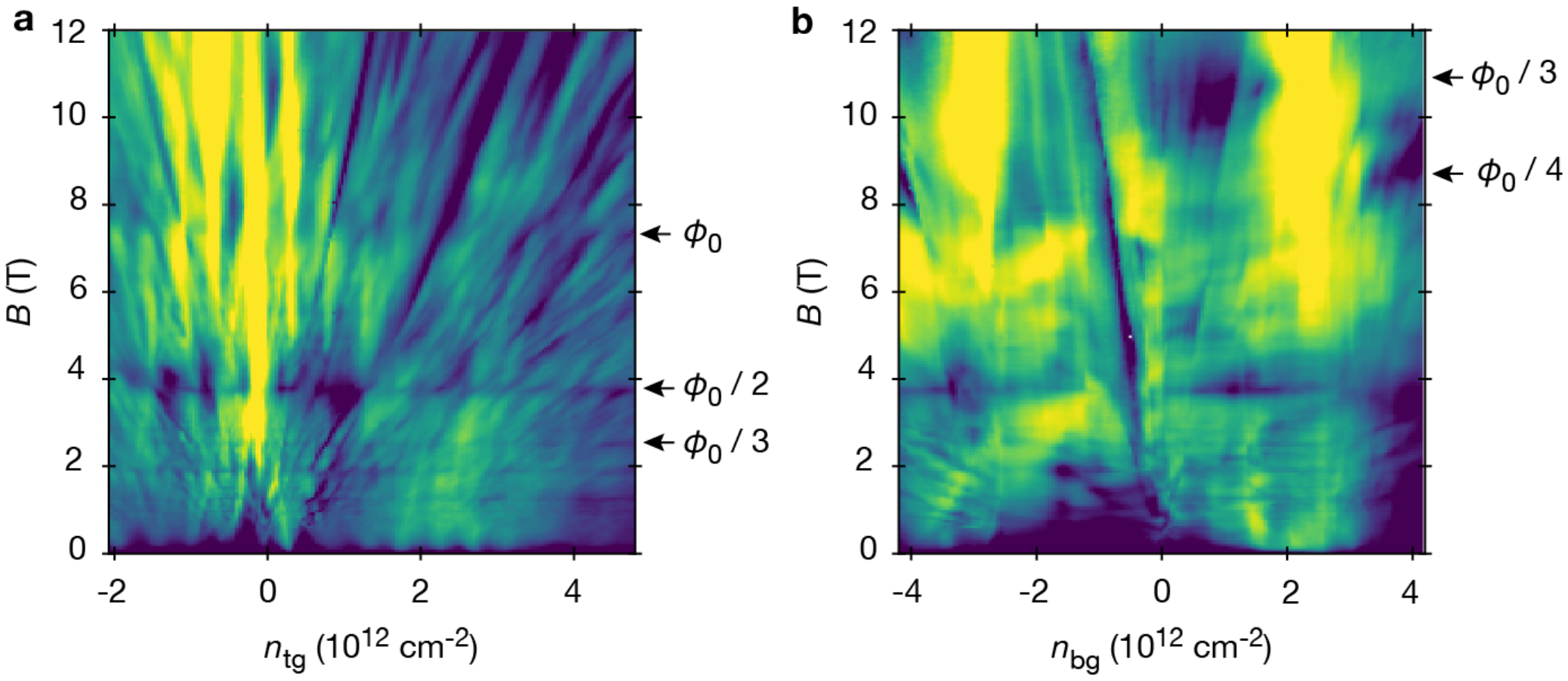

**Figure S2 |** $R_{\mathrm{xx}}(V_g, B)$ measured at 2K for either $V_{bg}$=0 (a) or $V_{tg}$=0 (b). The arrows on the right indicate low resistivity lines, corresponding to Brown-Zak oscillations. Dark tilted lines correspond to $R_{\mathrm{xx}} = 0$ for a series of Landau levels, for which their $V_g, B$ dependence allowed us to refine the induced densities. Color scale: indigo (160Ω) to yellow (2kΩ) in log scale.

In Figure S2, under applied magnetic field we could distinguish multiple fans corresponding to the multiple NPs in top and bottom TBG layers in the Hofstadter's butterfly picture. Apart from these fans, we observed carrier density-independent Brown-Zak oscillations, which (as discussed in the next section) we used to find the twist angles in our device. We note that each of the gates mainly changed the carrier density in the closest TBG layer and only slightly affected the carrier density in the second one. Thus, using the known dielectric thickness of our gates we got a first approximation for the carrier density induced in the closest TBG layer, and found the approximate distance between the NPs. Using this distance in terms of carrier densities we were able to find the correspondence between BZ oscillations periodicities and top and bottom TBG layers.

Next, using filling factors found in previous works (0.5° [5] and 1.1° [6]) we fitted the filling factors for the main Landau levels shown in Figure S2 as dark tilted lines (fans) as a function of the gate-induced carrier densities and magnetic field. It is important to note that the fan from a given layer is sensitive only to its own carrier density and is practically unaffected by the carrier density induced by the same gate in the second TBG layer. As a result, from such fits we extract $n_{\mathrm{bg}}^{\mathrm{bTBG}}(V_{\mathrm{bg}})$, and $n_{\mathrm{tg}}^{\mathrm{tTBG}}(V_{\mathrm{tg}})$.

Next, we used the double gate resistivity map shown in Figure S3 to find the non-diagonal terms $n_{\mathrm{tg}}^{\mathrm{bTBG}}(V_{\mathrm{tg}})$ and $n_{\mathrm{bg}}^{\mathrm{tTBG}}(V_{\mathrm{bg}})$. To do this, we noted that the peaks on the double gate map in Figure S3 correspond to the NPs in the top and bottom TBGs. Those peaks have linear trajectories in the $(V_{\mathrm{bg}}, V_{\mathrm{tg}})$ space, as shown in Figure S3, and the peak with the higher slope in the $(V_{\mathrm{bg}}, V_{\mathrm{tg}})$ space (shown in red) is attributed to $n^{\mathrm{bTBG}} = \mathrm{const}$, the one with the lower slope (shown in navy) is attributed to $n^{\mathrm{tTBG}} = const$. Here, $n^{\mathrm{bTBG}} = 0$ (main neutrality point of the bottom TBG layer) follows the trajectory $V_{\mathrm{bg}} = -3.35\, V_{\mathrm{tg}} + 0.3\mathrm{V}$.

Following these lines, the total carrier density in the bottom TBG as a function of top and bottom gate voltages can therefore be written as $n^{\mathrm{bTBG}}(V_{\mathrm{bg}}, V_{\mathrm{tg}}) = n_{\mathrm{bg}}^{\mathrm{bTBG}}(\mathrm{V_{bg}}) + n_{\mathrm{tg}}^{\mathrm{bTBG}}(V_{\mathrm{tg}}) = \frac{1}{e}\left[C_{\mathrm{bg}}^{\mathrm{bTBG}} V_{\mathrm{bg}} + C_{\mathrm{tg}}^{\mathrm{bTBG}} V_{\mathrm{tg}}\right]$, where $C_{\mathrm{bg}}^{\mathrm{bTBG}}$ is the effective capacitance between the bottom gate and the bottom TBG, and $C_{\mathrm{tg}}^{\mathrm{bTBG}}$ is the effective capacitance between the top gate and bottom TBG, and $e$ is the electron charge. Next, using the equation for the condition $n^{\mathrm{bTBG}} = 0$ we obtained:

$$n^{\mathrm{bTBG}}(V_{\mathrm{bg}}, V_{\mathrm{tg}}) = \frac{1}{e}\left[C_{\mathrm{bg}}^{\mathrm{bTBG}}(-3.35V_{\mathrm{tg}} + 0.3) + C_{\mathrm{tg}}^{\mathrm{bTBG}} V_{\mathrm{tg}}\right] = 0. \quad (3)$$

As discussed above, we extracted $C_{\mathrm{bg}}^{\mathrm{bTBG}}$ from the magnetic field-dependent measurements; thus, using equation 3 we found $C_{\mathrm{tg}}^{\mathrm{bTBG}} = 3.35 C_{\mathrm{bg}}^{\mathrm{bTBG}}$ and thus we found $n_{\mathrm{tg}}^{\mathrm{bTBG}}(V_{\mathrm{tg}})$. Similarly, using the condition for $n^{\mathrm{tTBG}}(V_{\mathrm{bg}}, V_{\mathrm{tg}}) = 0$ we found $C_{\mathrm{bg}}^{\mathrm{tTBG}}$ and $n_{\mathrm{bg}}^{\mathrm{tTBG}}(V_{\mathrm{tg}})$.

As a result, we found all four coefficients in eq. 1 and 2, allowing us to calculate the total gate induced carrier densities as well as carrier densities within each of the TBG layer used to plot the data in Fig.2 and 3 of the main text.

It is important to highlight that, to obtain equation 3 we assumed that the coefficients between the gate voltage and the carrier density in each TBG layer are constant as a function of top and bottom gate voltages. This assumption is reasonable as long as NPs follow linear trajectories (see, Figure S3). We note, however, that the resistivity peaks clearly deviated from linear trajectories at the crossing of NPs from top and bottom TBGs. Examples of such crossings are highlighted in Figure S3. Such deviation originated from the change of the screening when the Fermi level crosses the gap in one of the TBG layers, for example, a large single particle gap at the full filling of the first energy band of the magic angle TBG. However, such changes did not produce a big correction for the total carrier density, which is evident from the fact that the twist angles found from the positions of the NPs agreed well with the twist angles found from the BZ oscillations, which are independent on carrier density calibrations.

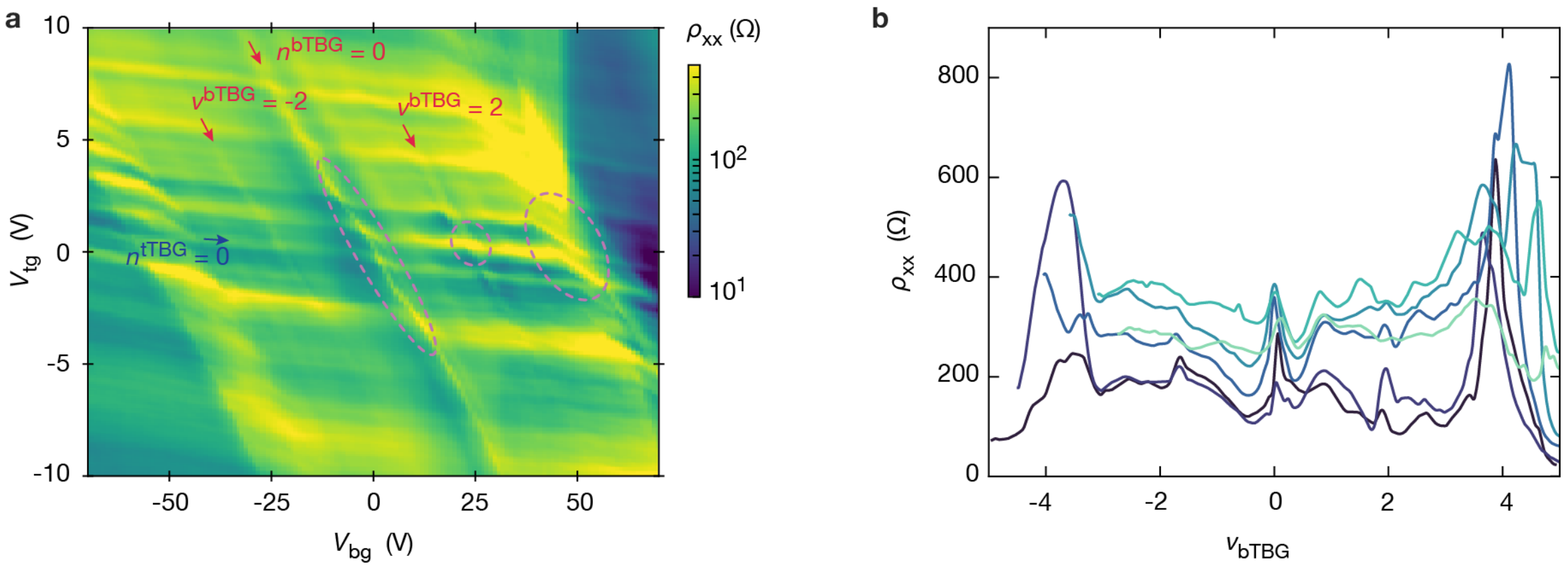

**Figure S3| (a)** $\rho_{\mathrm{xx}}(V_{\mathrm{bg}}, V_{\mathrm{tg}})$ measured at T = 2 K. The blue line indicates $n_{\mathrm{tTBG}} = 0$, and the red line $n_{\mathrm{bTBG}} = 0$. The purple dashed ellipse shows an example of a region where the carrier density distribution between the layers is affected by the large single particle gap in the bottom TBG layer. (**b**) Line cuts showing resistivity for fixed carrier densities in the top TBG layer in steps of $0.68\times10^{12}$ $\mathrm{cm}^{-2}$. Same data as in fig. 2b, without the offset.

### 3. Decoupling between the two bilayer systems

It is known that two graphene layers, stacked on top of one another with a large (>10°) between their crystallographic axis behave as two independent electronically decoupled layers. Specifically, the moiré potential couples only the states of energies $\pm\hbar v_F \Delta K = \pm\hbar v_F K \times 2\sin\theta/2$ [7] where $K = 4\pi/(3a_0)$.

Consequently, two electronic systems are decoupled in momentum space, eliminating momentum-conserving interlayer tunnelling.

Here we show that such decoupling also generalizes to a system of 2 + 2 layers, where two bilayers are stacked together at large angles. In section 6 “role of the perpendicular electric field”, we have established (see Figure S6), that the flat bands are insensitive to the perpendicular electric field.

To further substantiate this observation, we calculated the band structure of the full twisted tetralayer system using a tight-binding model (see Figure S4a), which we can compare to the superimposed band structure of two TBG layers (see Figure S4b), for a set of representative commensurate twist angles ($\theta_{12} \approx 1.25°$, $\theta_{23} \approx 43.75°$, and $\theta_{34} \approx 6.01°$). The choice of commensurate twist angles is detailed in supplementary section 13. Close to the center of the Brillouin zone, the low-energy bands of the tetralayer system are found to be identical to those of the individual TBG layers, demonstrating that our device enables us to probe the properties of each TBG independently. This can be understood from the large momentum separation between the Dirac cones for large twist angles, which reduces the effective coupling between the layers.

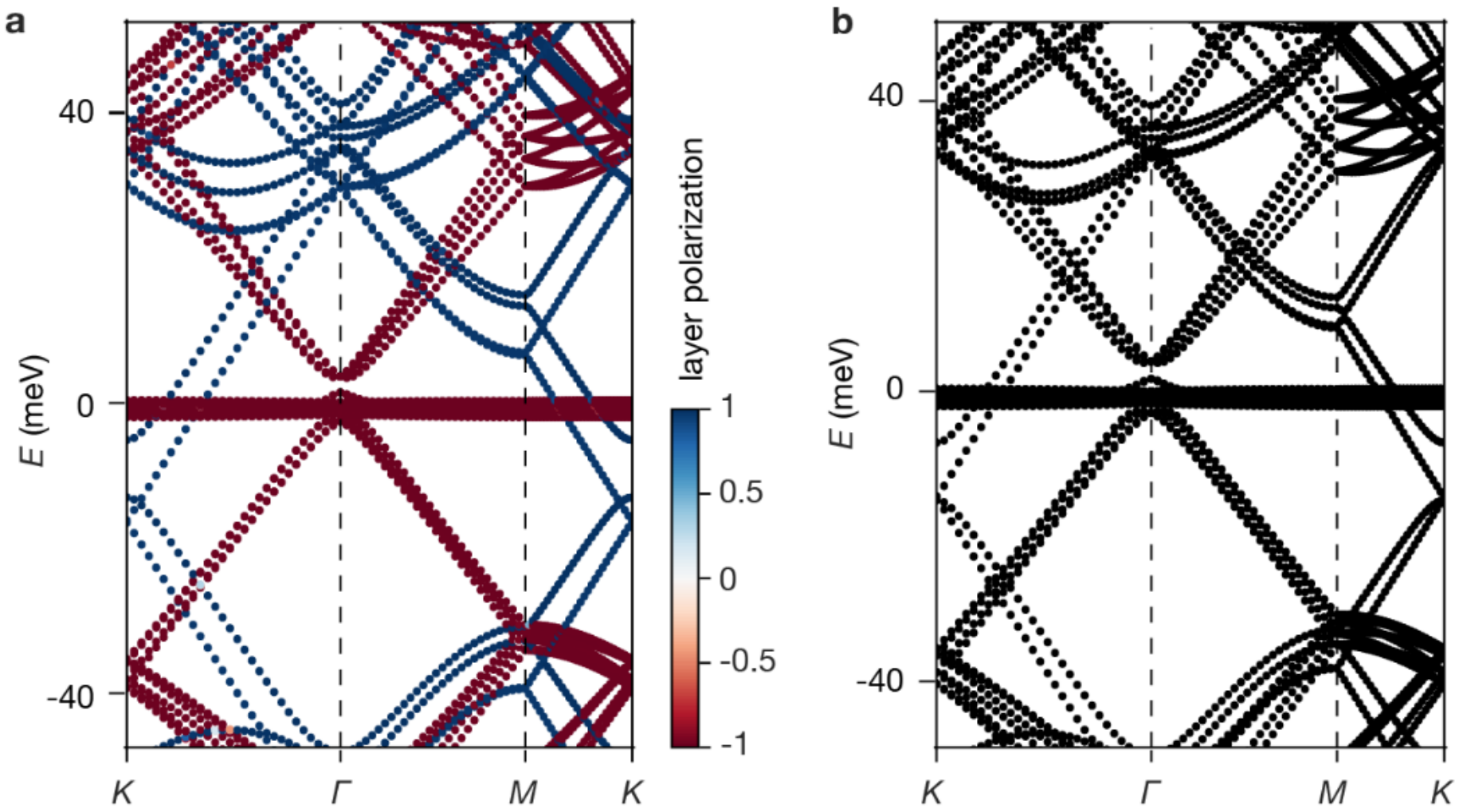

**Figure S4 |** Comparison of the band structures of the tetralayer system and the middle-layer–decoupled twisted tetralayer. a) Tight-binding band structure of the full coupled tetralayer system. The color left panel corresponds to the polarization in the bottom (red) and top (blue) twisted bilayer subsystem. b) Tight-binding band structure of the two individual twisted bilayer subsystems. The low-energy bands of the coupled case (a) are almost the same as those of the decoupled system (b), demonstrating that our device enables us to probe the properties of each TBG independently. For simplicity, we consider the rigid system without lattice relaxation.

## 4. Calculation of twist angles

We used two independent approaches to find twist angles in our device, both of which gave us similar results ensuring the reliability of our analyses.

First, the twist angle $\theta$ in a given TBG layer could be found using the distance in terms of doping between main and secondary NPs in the miniband spectrum:

$$\theta \approx a\sqrt{\frac{\sqrt{3}}{8}n_0}, \quad (4)$$

where $a$ is graphene’s lattice constant, and $n_0$ is the carrier density distance between neutrality points.

In the case of the top TBG, $n_0 = 0.5 \pm 0.13 \times 10^{12}$ cm$^{-2}$, therefore $\theta_{\mathrm{tTBG}} = 0.46 \pm 0.06°$. For the bottom TBG, $n_0 = 3.05 \pm 0.26 \times 10^{12}$ cm$^{-2}$, therefore $\theta_{\mathrm{bTBG}} = 1.15 \pm 0.05°$.

Second, we used Brown-Zak oscillations which appeared as a result of the recovery of electrons Bloch wavefunctions under an applied magnetic field [8,9]. Their fundamental frequency was directly connected to the superlattice unit cell area $A$ as:

$$A = \frac{\Phi_0}{B_0}, \tag{5}$$

where $\Phi_0$ is the flux quantum and $B_0$ is the fundamental frequency of oscillations.

Using (5), the twist angle between two graphene layers could be found as:

$$\theta \approx 2a \sqrt{\frac{\sqrt{3}}{8} \frac{B_0}{\Phi_0}}. \tag{6}$$

Figure S2a,b show fan diagrams as a function of either top or bottom gate voltages while keeping the second gate at zero volts. In both maps we observed a clear set of horizontal lines, which corresponded to the Brown-Zak oscillations either in top or in bottom layer. We found the fundamental frequency for the top TBG layer $B_{0,\mathrm{tTBG}} = 7.35 \pm 0.2$ T and for the bottom $B_{0,\mathrm{bTBG}} = 32 \pm 1$ T which corresponded to twist angles of 0.55±0.01° and 1.15±0.02° for the top and bottom TBGs, respectively.

**5. Absence of moiré imprinting**

The decoupling between the two TBGs was corroborated from the absence of signatures for moiré imprinting [10,11]. In such case, one would expect to observe additional peaks at constant $n_{\mathrm{bTBG}}$, at similar carrier densities at which peaks are observed in the top TBG layer. No such features were observed confirming that the two TBG layers were decoupled using large twist angles. To highlight the absence of such signatures, we differentiated the measured resistivity with both the top and back gates (see, Figure S5). Sign changes in d*R*/d*n* are seen for each correlated insulator on the bottom TBG, and for each secondary neutrality point on the top TBG. Apart from the crossing points and the boundaries of the superconducting dome, the features in the derivative maps are strictly linear as a function of either top-layer or bottom-layer carrier density, further illustrating the absence of moiré imprinting.

In addition, the derivative maps reveal a presence of more fragile vertical features at fillings $\nu = 1$ and $\nu = 3$ of the bTBG layer (see, Figure S5). These correspond to additional correlated states with smaller excitation gaps, which are typically observed in magic-angle devices.

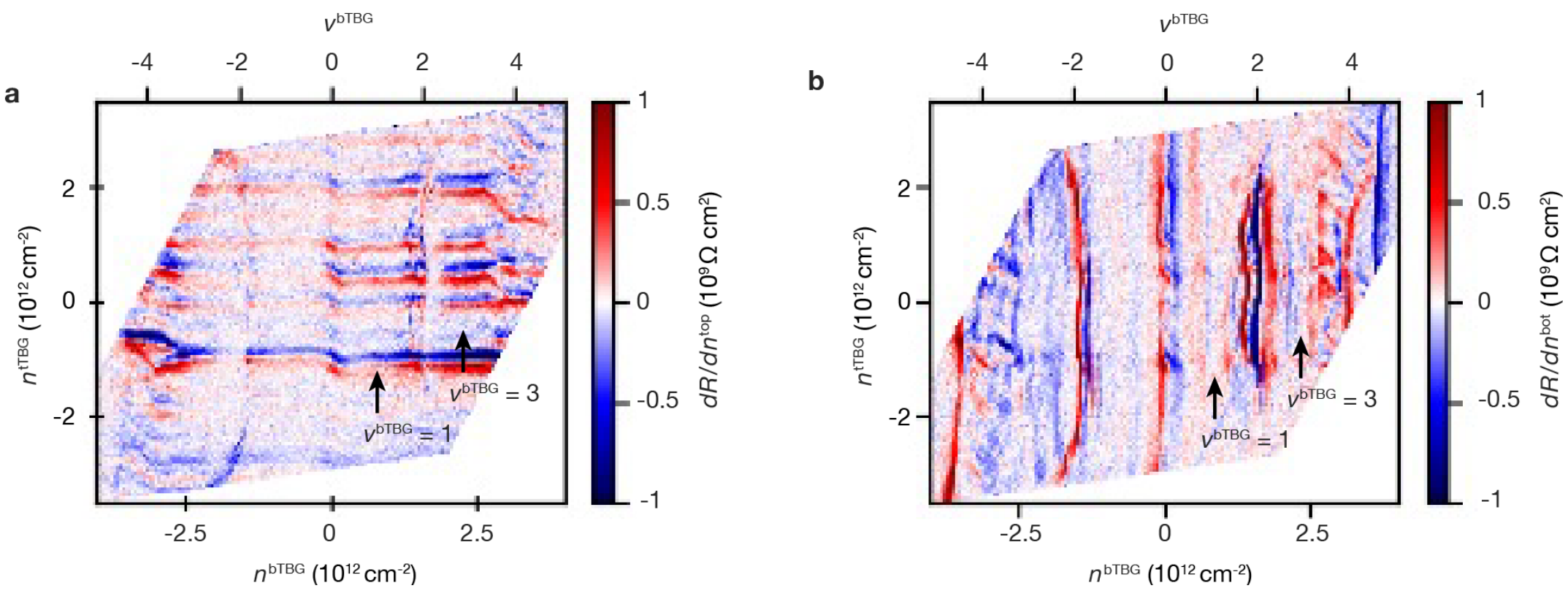

**Figure S5** | Resistance derivatives with top-gate-induced density (left) and back-gate-induced density (right), as a function of bottom and top tBG carrier densities. Data are same as in Fig. 3a.

## 6. Role of the perpendicular electric field

Apart from tuning the carrier densities of the top and bottom TBG layers and applied gate voltages also introduce a perpendicular electric field (displacement field) across the magic-angle TBG.

In order to understand the role of the displacement field on the magic-angle TBG, we replotted the data from figure 3a as a function of carrier density within the bottom TBG $n_{bTBG}$ and displacement field applied through that layer $D_{bTBG}$ (see Figure S6a) using gate calibrations discussed in a previous section. For positive carrier densities, superconductivity is suppressed within a relatively narrow displacement field range, of approximately $D$≈-0.3 to 0.1V/nm. In contrast, at negative carrier densities, superconductivity is suppressed around $D$≈+0.4V/nm for positive displacement fields, and $D$≈-0.05V/nm for negative displacement fields. While the suppression is nearly symmetric with respect to the screening-layer doping $n_{tTBG}$ (see, Fig 3a), the pronounced asymmetry in displacement field would not be expected if the suppression were driven by the displacement field itself. This points towards a DOS-influence of the superconducting phase.

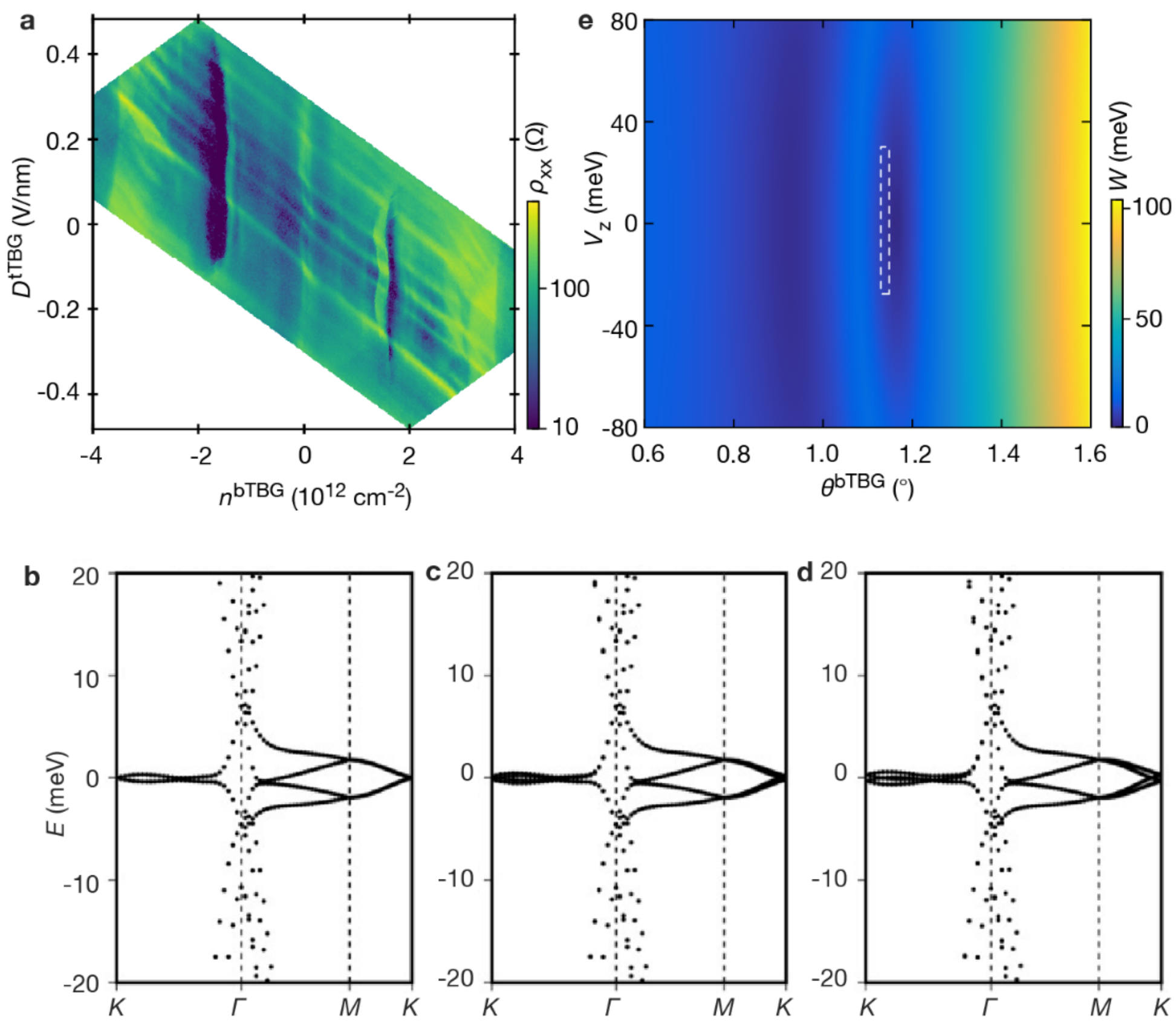

**Figure S6 |** Influence of the perpendicular electric field on the magic-angle TBG flat bands. a) Longitudinal resistivity as a function of carrier density and displacement field at 35mK (same data as in figure 3A). b-d) Low-energy band structures of MA-TBG (1.05°) for different displacement fields, $D$ = 0, 0.2 and 0.4 V/nm, respectively. e) Bandwidth $W$ at the Γ point of the mini Brillouin zone of the low-energy moiré bands near charge neutrality of twisted bilayer graphene, computed with the relaxed continuum model (including the relaxed moiré tunnelling and shift fields). Dashed square: conditions of our experiment for the bottom TBG layer.

In addition, independent experimental studies have shown that superconductivity in magic-angle TBG is insensitive to displacement fields over a range of at least ±0.6V/nm [12]. To further support this claim, we performed a theoretical analysis on the influence of the displacement field on the band structure of magic-angle TBG. To this end, we calculated the low-energy bands near the magic angle for 3 representative displacement fields: $D$ = 0, 0.2 and 0.4V/nm, corresponding to interlayer energies of 0, 10 and 20meV, estimated taking into account the electrostatic screening with an hBN dielectric constant of $2.5\cdot\varepsilon_0$ [13] (see, Figure S6b-d). For the range of displacement field in our experiment, the bandwidth variations are smaller than the bandwidth of the flat band by more than an order of magnitude, which can be explained by the strong hybridization of the magic-angle TBG layers. For this reason, we consider the influence of the displacement field negligible in our study. We also show the bandwidth of the flat bands near charge neutrality at the Γ point of the mini Brillouin zone, as a function of twist angle and interlayer bias (see Figure S6e), using the hopping function from ref [14] and relaxation parameters from ref [15]. We highlight the conditions of our experiment in which the bandwidth is essentially insensitive to the interlayer potential.

In fact, the weak dependence of the bandwidth on the applied displacement field is not specific to the particular twist angle used in this study, but is expected over a broad range of twist angles around ~1°.As shown in Figure S6b-d, the bandwidth sensitivity to the displacement field with variation of twist angle remains negligible within the experimental $D$ range, confirming the generality of this behavior.

**7. Screening in a double-layer structure**

To calculate the screening in our device, our model considers two TBG layers separated by a distance $l$ with a spacer of homogeneous dielectric constant (see details in section 12 "Estimation of screening length scale"). The top TBG layer acts as a screening layer on the bottom TBG layer. When the top TBG layer is doped, the induced charge density changes the dielectric environment of the stack, thereby influencing the Coulomb interaction in the bottom layer. Following ref. [16], the screened Coulomb interaction for the bottom TBG layer could be found as:

$$V(\boldsymbol{q}, i\omega) = \frac{V_{\text{eff}}(\boldsymbol{q}, i\omega)}{1 - V_{\text{eff}}(\boldsymbol{q}, i\omega)\Pi_2(\boldsymbol{q}, i\omega)}, \tag{7}$$

where $\Pi_2(\boldsymbol{q}, i\omega)$ is the dynamic polarizability of bottom layer and $V_{\text{eff}}(\boldsymbol{q}, i\omega)$ is the effective bare Coulomb interaction accounting for screening from the top TBG layer, given by

$$V_{\text{eff}}(\boldsymbol{q}, i\omega) = V_{22}(\boldsymbol{q})\left[1 - \frac{V_{12}(\boldsymbol{q})V_{21}(\boldsymbol{q})}{V_{11}(\boldsymbol{q})V_{22}(\boldsymbol{q})}\left(1 - \frac{1}{\epsilon_1(\boldsymbol{q}, i\omega)}\right)\right], \tag{8}$$

where $\epsilon_1(\boldsymbol{q}, i\omega) = 1 - V_{11}(\boldsymbol{q})\Pi_1(\boldsymbol{q}, i\omega)$, and $V_{ij}(\boldsymbol{q})$ are the intra- and inter-layer Coulomb interaction ($i$ and $j$ denoting the considered layer, 1 for top, 2 for bottom), which, for a given momentum $\boldsymbol{q}$ can be written as [16]:

$$V_{11}(\boldsymbol{q}) = V_{22}(\boldsymbol{q}) = \frac{2\pi e^2}{\kappa q}; \quad V_{12}(\boldsymbol{q}) = V_{21}(\boldsymbol{q}) = \frac{2\pi e^2}{\kappa q} e^{-lq}. \tag{9}$$

Here, we used $\kappa = 10$ to account for both the background dielectric environment and the remote bands screening, and $l = 0.35$nm represent the separation between the top and bottom TBG layers. In the static limit ($i\omega \to 0$), our screening model is the same as the one of ref. [17].

The dynamic polarization function for both layers was calculated from the random phase approximation (RPA),

$$\Pi(\boldsymbol{q}, i\omega) = \sum_{\boldsymbol{k}} \sum_{\gamma,\gamma'} \frac{\left(f_{\boldsymbol{k}+\boldsymbol{q}}^{\gamma} - f_{\boldsymbol{k}}^{\gamma'}\right) F_{\boldsymbol{k},\boldsymbol{k}+\boldsymbol{q}}^{\gamma\gamma'}}{E_{\boldsymbol{k}+\boldsymbol{q}}^{\gamma} - E_{\boldsymbol{k}}^{\gamma'} - i\omega}, \tag{10}$$

where the integration over $\boldsymbol{k}$ is restricted in moiré Brillouin zone and the composite indices $\gamma, \gamma'$ run over electron bands, valleys and spins. Here $f_{\boldsymbol{k}}^{\gamma}$ is the Fermi-Dirac distribution for a state with energy $E_{\boldsymbol{k}}^{\gamma}$, and $F_{\boldsymbol{k},\boldsymbol{k}+\boldsymbol{q}}^{\gamma\gamma'} = \left|\int_{\Omega} d^2\boldsymbol{r}\, \psi_{\gamma,\boldsymbol{k}+\boldsymbol{q}}^{\dagger}(\mathbf{r}) e^{i\boldsymbol{q}\cdot\boldsymbol{r}}\, \psi_{\gamma',\boldsymbol{k}}(\mathbf{r})\right|^2$ is the form factor associated with different Bloch states. The DoS obtained in the previous section was applied when calculating the polarization function of the top TBG layer.

From eqs. 8 and 9, it is possible to rewrite equation 7 as:

$$V(\boldsymbol{q}, i\omega) \quad = \frac{V_0(\boldsymbol{q})\left[1 - e^{-2lq}\left(1 - \frac{1}{\epsilon_1(\boldsymbol{q}, i\omega)}\right)\right]}{1 - V_0(\boldsymbol{q})\left[1 - e^{-2lq}\left(1 - \frac{1}{\epsilon_1(\boldsymbol{q}, i\omega)}\right)\right]\Pi_2(\boldsymbol{q}, i\omega)} = \frac{V_0(q)}{\Gamma(l, \boldsymbol{q}, i\omega) - V_0(q)\Pi_2(\boldsymbol{q}, i\omega)}, \tag{11}$$

where we introduced a screening factor $\Gamma$ which quantifies the strength of screening provided by the top layer, given by:

$$\Gamma(l, \boldsymbol{q}, i\omega) = \frac{1}{1 - e^{-2lq}\left(1 - \frac{1}{\epsilon_1(\boldsymbol{q}, i\omega)}\right)}. \tag{12}$$

A higher screening factor signifies a stronger screening effect from the top layer. In the limit of a large interlayer separation ($lq \gg 1$) we get:

$$\lim_{lq \gg 1} \Gamma(l, \boldsymbol{q}, i\omega) = 1, \tag{13}$$

suggesting that the contribution from the top layer becomes negligible. In contrast, for small interlayer distances ($lq \ll 1$), screening factor expands as follows:

$$\Gamma(l, \boldsymbol{q}, i\omega) \approx \epsilon_1(\boldsymbol{q}, i\omega) - 2ql\epsilon_1(\boldsymbol{q}, i\omega)[\epsilon_1(\boldsymbol{q}, i\omega) - 1] + O(l^2). \tag{14}$$

And, in the limit of vanishing interlayer distance ($l \approx 0$), we get:

$$\Gamma(l, \boldsymbol{q}, i\omega) \approx \epsilon_1(\boldsymbol{q}, i\omega) = 1 - V_0(\boldsymbol{q})\Pi_1(\boldsymbol{q}, i\omega). \tag{15}$$

Which, in the static limit with small momentum $\boldsymbol{q}$ becomes simply proportional to the DoS of the screening layer which sets an upper limit on the screening factor.

$$\Gamma(l, \boldsymbol{q}, 0) \approx 1 + V_0(\boldsymbol{q})\mathcal{N}(E_\mathrm{F}) \sim \mathcal{N}(E_\mathrm{F}), \tag{16}$$

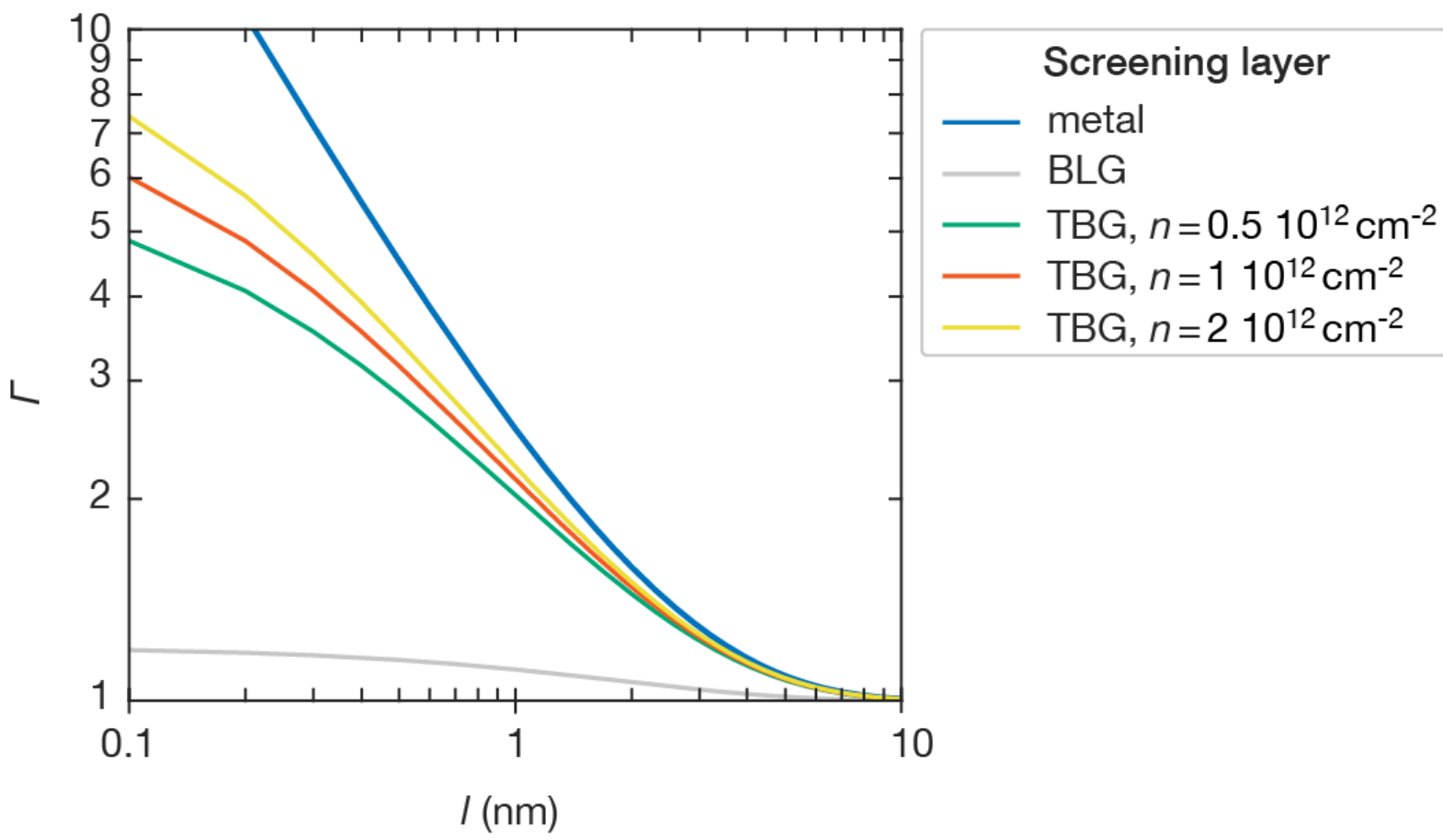

**Figure S7|** Screening factor $\Gamma$ as a function of interlayer distance calculated for different DoS of the screening layer, and for the DoS of Bernal graphene bilayer, assuming an effective mass $m^* = 0.033m$, and a carrier density $n$ =

2x10$^{12}$ cm$^{-2}$. The calculation was done in a static limit, for the characteristic wave vector of the half of the mini-Brillouin zone of the magic angle TBG.

Figure S7 shows the behavior of the screening factor as a function of the interlayer separation distance $l$, for various screening materials. When the interlayer distance is large, the influence of the top layer on the bottom layer becomes negligible. Consequently, all curves become indistinguishable for high $l$, consistent with the behavior described in eq. 13. For small interlayer distances, the impact from the induced screening becomes more pronounced. To allow for easier comparison with the previous experiments, we also plotted the screening factor calculated for Bernal graphene bilayer as a screening layer. The screening efficiency of such layer is expected to be much lower than the screening from the TBG layer used in our experiment, even at sub nanometer distances.

**8. Density of state approximation**

The observed sensitivity of the critical temperature in magic-angle twisted bilayer graphene (TBG) to the doping of the top layer can be attributed to electrostatic screening effects, which are determined by the density of states (DoS) of the top TBG band structure. Unfortunately, it is challenging to get reliable calculations of the band structure of a TBG layer with twist angles much smaller than 1 degree due to lattice reconstruction effects and strains, which vary in every device. To overcome this problem, we assumed that the DoS of the top TBG layer always increases with doping. This approximation is motivated by our experimental results in Fig. 2 (main text) which show that correlated insulators are suppressed at increased carrier density of the top layer, which indicates an increasing screening. Moreover, Fig. 3d (main text) shows that the superconducting critical current is also continuously suppressed as a function of top TBG doping. To capture this behavior, we calculated the DoS of the top TBG layer to be proportional to the DoS of single-layer graphene, providing a useful comparison to quantify the DoS of our system:

$$\mathcal{N}_{\mathrm{tTBG}} = \alpha \times \mathcal{N}_{\mathrm{Gr}}, \quad (17)$$

where $\mathcal{N}_{\mathrm{tTBG}}$ is the density of the top layer graphene, $\mathcal{N}_{\mathrm{Gr}}$ is the DoS of monolayer graphene, and $\alpha$ is a numerical constant. To quantify $\alpha$ we used the fact, that in our experiment the correlated insulating gap $\Delta_{\mathrm{CI}}$ around positive half filling of the magic angle TBG became lower than 2K (completely smeared by temperature) when the top TBG carrier density exceeded n = $2\cdot10^{12}$ cm$^{-2}$. This suppression provides a clear reference point to estimate the magnitude of screening in our device.

Our calculations of $\Delta_{\mathrm{CI}}$ were based on the Bistritzer–MacDonald model [18], incorporating Hartree-Fock interactions in a plane-wave basis following the method developed in ref. [19]**.** In this study, we adopted the parameters $\theta = 1.1°$, $\hbar v_F = 5250$ meV $\cdot$ Å and $a = 2.46$ Å is monolayer graphene's lattice constant. For interlayer hopping, we chose $w_{\mathrm{AA}} = 79.7$ meV, and $w_{\mathrm{AB}} = 97.5$ meV as in ref. [20] to account for the relaxation effect**.**

The Hartree-Fock interactions can be introduce by defining the density matrix $\rho(\boldsymbol{k})$ in the plane wave basis as:

$$\rho_{\gamma,\boldsymbol{G};\gamma',\boldsymbol{G}'}(\boldsymbol{k}) = \sum_{n} z^{n*}_{\gamma',\boldsymbol{k},\boldsymbol{G}'}\, z^{n}_{\gamma,\boldsymbol{k},\boldsymbol{G}}, \quad (18)$$

where $\boldsymbol{k}$ is the Bloch vector constrained within the moiré Brillouin zone (mBZ), $\boldsymbol{G}$ represents Fourier component of plane wave expansion, $\gamma$ and $\gamma'$are dummy indices running over spin and valley degrees of freedom, $n$ represents band indices, run over all filled bands, and $z^{n}_{\gamma,\mathbf{k},\mathbf{G}}$ is the numerical eigenvector in the plane wave expansion,

$$|\psi_{n,\boldsymbol{k}}\rangle = \sum_{\gamma,\boldsymbol{G}} z^{n}_{\gamma,\boldsymbol{k},\boldsymbol{G}} \,|\psi_{\gamma,\boldsymbol{k}+\boldsymbol{G}}\rangle. \tag{19}$$

We therefore obtained Hartree self-energy in the plane wave basis, $\Sigma^{\mathrm{H}}_{\gamma,\boldsymbol{G};\gamma',\boldsymbol{G}'}$ as:

$$\Sigma^{\mathrm{H}}_{\gamma,\boldsymbol{G};\gamma',\boldsymbol{G}'}(\boldsymbol{k}) = \frac{1}{\Omega}\sum_{\gamma'',\boldsymbol{k}',\boldsymbol{G}''} V\,(\boldsymbol{G}-\boldsymbol{G}')\rho_{\gamma'',\boldsymbol{G}''+\boldsymbol{G}-\boldsymbol{G}';\gamma'',\boldsymbol{G}''}(\boldsymbol{k}')\delta_{\gamma\gamma'}. \tag{20}$$

Here, $\Omega$ is the total area of sample, $V$ is the bare Coulomb interaction including the screening from the top layer calculated from eq. 7 and $\delta_{\gamma\gamma'}$ is the Kronecker delta function.

Likewise, the Fock self-energy $\Sigma^{\mathrm{F}}_{\gamma,\boldsymbol{G};\gamma',\boldsymbol{G}'}$ is given by

$$\Sigma^{\mathrm{F}}_{\gamma,\boldsymbol{G};\gamma',\boldsymbol{G}'}(\boldsymbol{k}) = -\frac{1}{\Omega}\sum_{\boldsymbol{k}',\boldsymbol{G}''} V\,(\boldsymbol{k}'+\boldsymbol{G}''-\boldsymbol{k})\rho_{\gamma,\boldsymbol{G}''+\boldsymbol{G};\gamma',\boldsymbol{G}''+\boldsymbol{G}'}(\boldsymbol{k}'). \tag{21}$$

Following refs [19,21], we substituted the density matrix by $\tilde{\rho}(\boldsymbol{k}) = \rho(\boldsymbol{k}) - \rho_0(\boldsymbol{k})$, where $\rho_0(\boldsymbol{k})$ represent the density matrix of decoupled graphene bilayer filled up to charge neutral point, to avoid the double counting problem. Starting from a random initialization for $\Sigma^{\mathrm{H}}$ and $\Sigma^{\mathrm{F}}$, we proceeded by diagonalizing the full Hamiltonian $H = H_0 + \Sigma^{\mathrm{H}} + \Sigma^{\mathrm{F}}$ to obtain new eigenvalues and eigenvectors, where $H_0$ is the Hamiltonian from Bistritzer-MacDonald model. New values for $\Sigma^{\mathrm{H}}$ and $\Sigma^{\mathrm{F}}$ were computed iteratively to find the self-consistent solution. The correlated insulating gap $\Delta_{\mathrm{CI}}$ was then calculated from the self-consistent band structures. For the CI, we consider Kramers intervalley-coherent state which has been shown is the ground state at half filling [21,22]. We believe any Coulomb interaction driven CI ground state will have similar behavior with screening.

Figure S8 shows the correlated insulator gap at $\nu = +2$ calculated for different proportionality factors $\alpha$. We found $\alpha$ as the minimum value for which $\Delta_{\mathrm{CI}} < 0.2$ meV for the top TBG filling factor 20, which yielded $\alpha \approx 12$. Importantly, $\alpha \approx 12$ means that the DoS necessary to suppress the superconductivity in magic-angle TBG is inaccessible using bilayer graphene, for which DoS is approximately 10 times smaller than the DoS of the top TBG layer at the carrier density $2\text{x}10^{12}$ cm$^{-2}$ obtained from our modelling. It explains the difference between our findings and the results of the previous work with proximity bilayer graphene.

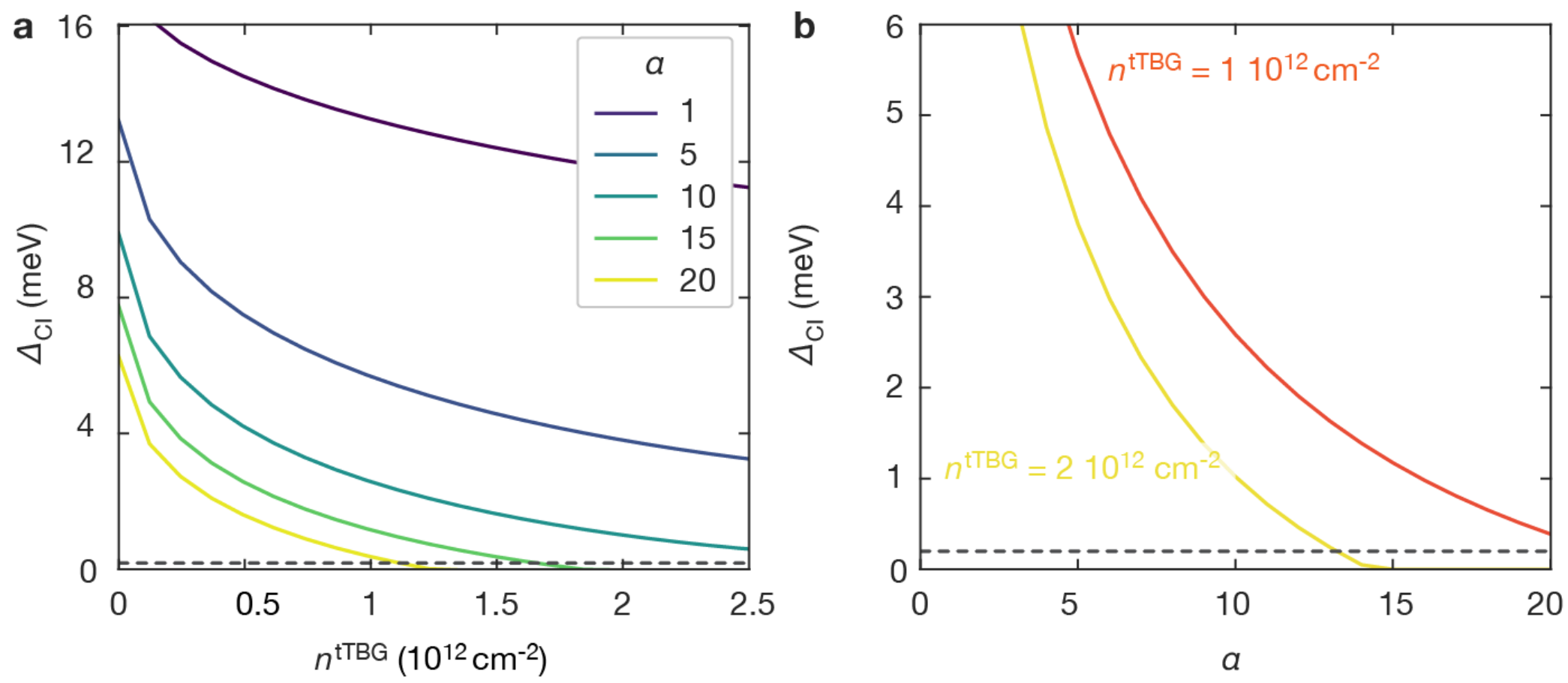

**Figure S8|** Self-consistent correlated insulating gap calculated from density of state approximation. (**A**) Correlated insulating gap as function of top TBG carrier density for different values of $\alpha$. (**B**) Correlated insulating gap as a

function of $\alpha$ for different carrier densities, from which we estimate $\alpha \approx 10$. Both panels: grey dashed line indicates $\Delta_{\mathrm{CI}} = 0.2$ meV.

### 9. Phonon-mediated superconductivity

After we calculated how Coulomb interactions change with the top TBG doping, it became possible to estimate the effect of screening on different superconducting mechanisms. First, we explored the possibility of phonon-mediated superconductivity. While it is still debated whether the screening can change the electron-phonon coupling, we considered both scenarios. On the one hand, if screening does not affect the electron-phonon coupling, then the critical temperature will not change with increased screening. This prediction contradicts our experiment. On the other hand, it is possible to consider that screening changes the deformation potential, which in turns reduces the electron-phonon coupling. In TBG, the acoustic phonon modes can couple with either (i) the deformation potential $D_{\mathrm{A}}$, which arises from longitudinal acoustic phonons, or (ii) the gauge field $\beta_{\mathrm{A}}$, originating from both longitudinal and transverse phonons. However, the gauge field term is not expected to couple to the charge and, therefore, is expected to be immune to screening [23]. In contrast, the deformation potential induces charge separation, and it is possible to consider that it exhibits sensitivity towards screening. Therefore, as detailed below, we considered only superconductivity mediated by screened deformation modes.

We first estimated the critical temperature $T_{\mathrm{c}}$ within the BCS theory as [24,25]:

$$T_{\mathrm{c}} \approx 1.14\, \omega_{\mathrm{D}} e^{-\frac{1}{\lambda}}, \tag{22}$$

where $\omega_{\mathrm{D}}$ represents the Debye frequency and $\lambda$ is a dimensionless pairing strength. In the spectral representation, these two quantities can be recast as:

$$\omega_{\mathrm{D}} = \frac{2}{\lambda} \int d\omega \alpha^2 F(\omega)\,; \quad \lambda = 2 \int d\omega \frac{\alpha^2 F(\omega)}{\omega}, \tag{23}$$

where we introduced the phonon spectral function:

$$\alpha^2 F(\omega) = \frac{1}{\mathcal{N}(E_{\mathrm{F}})} \sum_{m,n,\boldsymbol{k},\boldsymbol{q}} \left|g_{\boldsymbol{k},\boldsymbol{q}}^{mn}\right|^2 \delta(E_{\mathrm{F}} - E_{\boldsymbol{k}}^{m}) \delta\left(E_{\mathrm{F}} - E_{\boldsymbol{k}+\boldsymbol{q}}^{n}\right) \delta\left(\omega - \omega_{\boldsymbol{q}}\right). \tag{24}$$

Here $\mathcal{N}(E_{\mathrm{F}})$ is the density of states at Fermi level, $m, n$ denote band indices, $E_{\boldsymbol{k}}^{m}$ is the energy dispersion of bottom TBG layer $E_{\mathrm{F}}$ is the Fermi energy, $\omega_{\boldsymbol{q}} = c_{\mathrm{s}} q$ is the phonon dispersion with $c_{\mathrm{s}}$ being the sound speed, and $g$ is the electron-phonon coupling strength, given by

$$\left|g_{\boldsymbol{k},\boldsymbol{q}}^{mn}\right|^2 = \frac{\hbar D_{\mathrm{A}}^2 q}{2A\rho c_{\mathrm{s}}} \left|\langle \psi_{\boldsymbol{k}}^{m} | \psi_{\boldsymbol{k}+\boldsymbol{q}}^{n} \rangle\right|^2, \tag{25}$$

where $D_{\mathrm{A}}$ is the deformation potential, $\rho$ is the mass density, and $A$ is area of unit cell. To get the numerical results we used values found in refs. [26,27]: $D = 25$ eV, $c_s = 12000\ \frac{\mathrm{m}}{\mathrm{s}}$, $\rho = 7.6 \times 10^{-8}\ \frac{\mathrm{g}}{\mathrm{cm}^2}$.

The role of screening to the deformation phonon modes is to renormalize both the electron-phonon coupling strength (eq. 25) and phonon propagator considering the retarded effects [28–30]. The BCS limit could be retrieved by substituting the renormalized sound velocity $\tilde{c}_{\mathrm{s}} = \frac{c_{\mathrm{s}}}{\sqrt{\epsilon(\boldsymbol{q})}}$ and renormalized deformation potential $\widetilde{D}_{\mathrm{A}} = \frac{D_{\mathrm{A}}}{\epsilon(\boldsymbol{q})}$ into Eq. 24 [31]. After that it became straightforward to obtain the critical temperature using Eqs. 22 and 23.

Figure S9 shows the phonon spectra function for different carrier densities of the top TBG layer, and the inset shows the critical temperature which we found following the above mentioned BCS formalism. One can see that the deformation phonon modes are strongly screened even when the top layer is undoped, which indicates that the strong self-screening near TBG can suppress the BCS critical temperature even without external screening. As the carrier density of the top TBG increases, phonon modes get screened further.

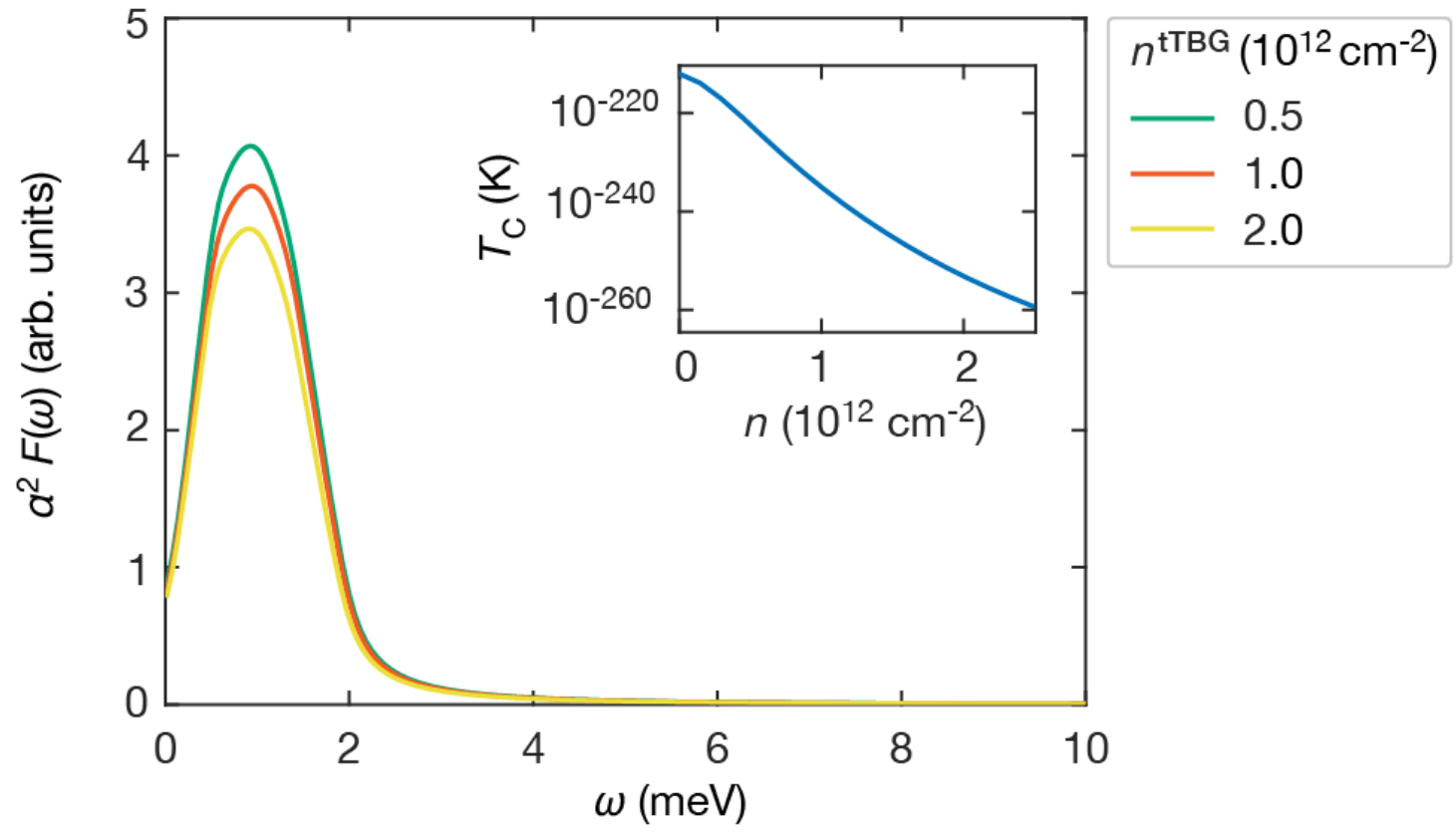

**Figure S9 |** Phonon spectral function for different fillings in the top TBG layer. Inset: resulting critical temperature modelled from the BCS theory.

We also found the corresponding transition temperature within BCS theory is unphysically small: $T_{\mathrm{c}}$ would be exponentially suppressed by screening as $T_{\mathrm{c}} \sim \omega_{\mathrm{D}} \exp\left(-\epsilon^2/\lambda\right)$ ($\omega_{\mathrm{D}} \sim 1$ meV, $\lambda \sim 1$, and $\epsilon \sim 100$ near the magic angle). Therefore, we concluded that the deformation potential is too weak to support superconductivity, and the gauge phonon modes could not explain the experimental data.

### 10. Plasmon-mediated superconductivity

As the conventional phonon-mediated pairing could not explain our data, we turned to the possibility of unconventional superconductivity. We picked plasmon-mediated superconductivity as one of the possibilities. This mechanism draws analogies with the phonon-mediated pairing, for which the attractive interactions between electrons can be described by considering a single electron exciting phonons in a crystal. Since the phonons are much slower than electrons, a second electron would be attracted to the first electron through a time-dependent coupling to phonons. The plasmon-mediated mechanism can be understood in similar terms: as an electron travels through the medium, it excites density fluctuations (plasmons). In this regime, a second electron passing by will feel a retarded effective attraction interaction due to the screening from density fluctuations. This attractive interaction serves as the binding force that enables the formation of Cooper pairs. The electron-electron interaction, mediated by plasmon, is also affected by the external screening. However, unlike the phonon mechanism, the role of screening to the plasmon-mediated superconductivity is expected to have non-monotonic effects on the critical temperature (see, e.g. ref [16]).

It was shown that superconductivity mediated by a purely electronic mechanism can be studied by introducing a momentum-averaged frequency-dependent Coulomb interaction [16,30,32]:

$$\lambda(i\omega_n) = \mathcal{N}(E_\mathrm{F})\langle\langle V(i\omega_n)\rangle\rangle, \tag{26}$$

Here, $\lambda(i\omega_n)$ is the pairing strength of Cooper pairs, and $\omega_n = (2n+1)\pi T$ with integer $n$ denoting the Fermion Matsubara frequency. $\mathcal{N}(E_\mathrm{F})$ represents the density of states (DoS) at the Fermi level, and $\langle\langle V(i\omega_n)\rangle\rangle$ is the momentum-averaged Coulomb interaction:

$$\langle\langle V(i\omega_n)\rangle\rangle = \frac{\sum_{\boldsymbol{k},\boldsymbol{p}} \Theta\,(\boldsymbol{k}_\mathrm{c} - \boldsymbol{k})\Theta(\boldsymbol{k}_\mathrm{c} - \boldsymbol{p})V(\boldsymbol{k} - \boldsymbol{p}, i\omega_n)}{\sum_{\boldsymbol{k},\boldsymbol{p}} \Theta\,(\boldsymbol{k}_\mathrm{c} - \boldsymbol{k})\Theta(\boldsymbol{k}_\mathrm{c} - \boldsymbol{p})}, \tag{27}$$

where $\boldsymbol{k}_\mathrm{c} = 2\boldsymbol{k}_\mathrm{F}$ represents the momentum cutoff, $\Theta$ is the Heaviside function, and $V(\boldsymbol{k}, i\omega_n)$ denotes the dynamically screened Coulomb interaction. Following the procedures proposed by Grabowski and Sham [32], the linearized isotropic Eliashberg gap equation can be formulated as:

$$\Delta_n = -2\,\tilde{T}_c \sum_{m=-\infty}^{\infty} \frac{1}{Z_m\tilde{\omega}_m} \arctan\frac{1}{Z_m\tilde{\omega}_m}\lambda_{n,m}\Delta_m, \tag{28}$$

with the mass renormalization function considering self-energy corrections:

$$Z_m = 1 - \frac{\tilde{T}_c}{\tilde{\omega}_n}\int_{-1}^{1} d\tilde{E} \sum_m \frac{\lambda_{n,m}\tilde{\omega}_m}{\tilde{\omega}_m^2 + \tilde{E}^2}, \tag{29}$$

where $\tilde{T}_\mathrm{c} = \frac{k_\mathrm{B}T_\mathrm{c}}{E_\mathrm{F}}$ represents the superconducting critical temperature, and $\tilde{\omega}_n = (2n+1)\pi\tilde{T}_\mathrm{c}$ denotes the dimensionless Matsubara frequency. Both quantities are scaled by the Fermi energy to be dimensionless. $\Delta_m$ is the order parameter, and $\lambda_{n,m}$ is shorthand for $\lambda(i\omega_n - i\omega_m)$. Equation 29 has the form of an eigenvalue equation $\bar{\Delta} = \hat{C}(\tilde{T})\bar{\Delta}$, where the critical temperature $\tilde{T}_\mathrm{c}$ can be determined by identifying the largest eigenvalue of the matrix $\hat{C}$ that equals 1 [33,34], which allows a solution for the gap equation.

Figure S10 shows the numerical results for the superconducting critical temperature $T_\mathrm{c}$ in the plasmon-mediated mechanism within the density of state approximation. We found that increasing the screening by changing the doping of the top TBG layer dramatically reduced the superconducting critical temperature, which agreed well with our observations, both from a qualitative consideration, and from absolute values. Additionally, we calculated $T_\mathrm{c}$ as a function of the interlayer distance $l$ and found that a subnanometer distances are necessary to observe a pronounce tuning of $T_\mathrm{c}$ from screening, even with the high DoS of our top TBG. Overall, our numerical results suggest that the plasmon-enabled pairing mechanism is a plausible explanation for superconductivity in magic-angle TBG.

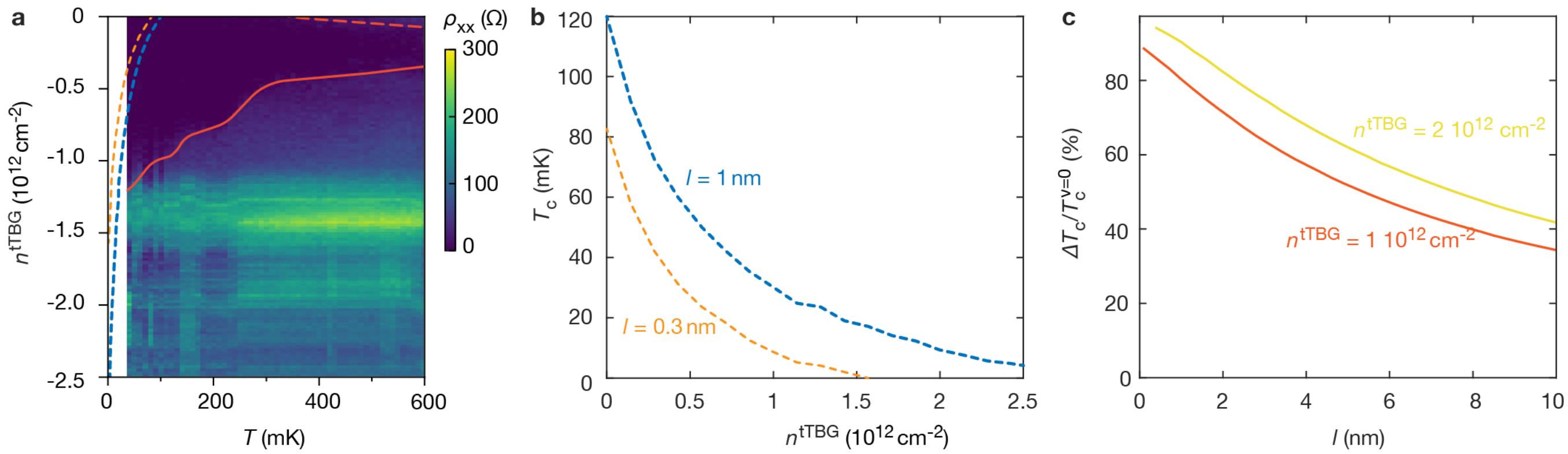

**Figure S10**| (**a**) d$V$/d$I$($T$, $n^{\mathrm{tTBG}}$) measured at $n^{\mathrm{bTBG}}$ = 2. Solid red line serves as a guide for the eye for the superconducting critical temperature $T_c$, with dashed red line indicating an apparent reduction in the measured $T_c$ close to $n^{\mathrm{tTBG}}$ = 0 due to increased contact resistance. Blue and orange dashed lines: theory for the plasmon-mediated mechanism with the density of state approximation for discussed in the main text and an interlayer distance of 0.3 nm, for interlayer distances of 1 and 0.3nm, respectively (as shown in the panel b). (**b**) Critical temperature for the plasmon-mediated mechanism within the density of state approximation, for two different interlayer distances, as a function of top TBG doping. (**c**) Relative changes in the critical temperature for different fillings of the top TBG layer, normalized by the critical temperature at $\nu = 0$, as a function of the interlayer distance

Finally, we note that the screening of superconductivity depends not solely on the DoS at the Fermi energy but on an average DoS over the characteristic energy scale (e.g. the plasmon energy). This broader energy range (~10 meV) includes several minibands of the top TBG layer. So, it is expected that the screening should not be strongly modulated with the shift of the Fermi level in the top TBG layer, but instead it is smoothly changed, which agrees with our observations.

## 11. Models incorporating relaxation

In this section, we present details on the theoretical calculation of the moiré bands of twisted bilayer graphene (tBG) including the effect of lattice relaxation, as well as the Hartree potential at finite doping. These effects are important for the modelling of the electronic properties of our device. Especially important is the moiré reconstruction for the marginally twisted bilayer with twist angle $\theta \approx 0.5°$. This atomic reconstruction of the moiré changes the interlayer tunnelling, which modifies the moiré band structure. Furthermore, the electronic density is strongly localized in the moiré cell for the twist angles we consider in this work which drives a large Hartree effect at finite doping due to Coulomb interactions. Here, in order to lower the electrostatic energy, the moiré bands are renormalized to reduce the density fluctuations. Therefore, the energetically flattened moiré bands in twisted bilayer graphene (and other moiré systems) are strongly dependent on doping of these bands. We first consider the elastic properties of the twisted bilayer graphene (13.1). In particular, we use the continuum theory of elasticity to model the reconstruction of the moiré pattern. This long-wavelength theory is justified from the slow variation of the moiré with respect to the atomic lattice for the twist angles we consider. Next, we benchmark our calculations of the relaxed moiré band structures with the literature using a tight-binding model in the Slater-Koster approximation (13.2). However, modelling the Hartree effect using tight-binding calculations becomes increasingly costly as the twist angle is reduced. For this reason, we use the electronic continuum model for that purpose, which is described in detail in title 13.3. The details of the Hartree calculation are presented in title 13.4. We then demonstrate that the band structure of marginally twisted bilayer graphene depends sensitively on the choice of model assumptions e.g. inclusion of relaxation (13.5), Hartree renormalization of bands, and strain. Finally, we discuss how for conventional superconductivity (i.e. BCS) the critical temperature depends on the density of states at the Fermi level (13.6), but for the more complete Eliashberg theory it could in addition depend on the density of states averaged over the frequency scale of the bosonic mode responsible for pairing.

**11.1. Reconstruction of the moiré lattice from continuum elasticity**

Reconstruction in twisted van der Waals materials can be understood from the fact that a moiré pattern can locally be viewed as a spatial modulation of the layer stacking. Certain layer stacking configurations for untwisted bilayers have a lower energy than others, and therefore the moiré relaxes to expand energetically favorable stacking configurations and shrink regions with unfavorable layer stacking. This comes at the cost of elastic energy in the two layers required to move the atoms around. In twisted bilayer graphene, the lowest stacking configuration is Bernal stacking (AB/BA) while orthorhombic stacking (AA) is the least favorable configuration [35]. As the twist angle decreases, the moiré period increases, and the gain in interlayer adhesion energy by forming favorable stacking domains wins over the elastic energy cost. For twisted bilayer graphene, relaxation becomes appreciable for twist angles close to and below the magic angle $\theta \approx 1^\circ$ [36]. Because there are two degenerate low-energy stacking configurations for tBG, the moiré reconstructs towards a triangular domain wall network with expanded AB/BA domains connected by partial shear dislocations that meet at AA stacking centers.

Here, we model the relaxed configuration of tBG using continuum elasticity theory. The central object one has to determine is the interlayer stacking configuration

$$\boldsymbol{\phi}(\mathbf{r}) = M\mathbf{r} + \mathbf{u}(\mathbf{r}), \tag{30}$$

in the moiré cell $\mathbf{r} \in s_1\mathbf{L}_1 + s_2\mathbf{L}_2$ where $s_{1,2} \in [0,1)$. Here $M$ is the displacement gradient that defines the rigid (i.e. unrelaxed) configuration and $\mathbf{u}$ is the displacement field due to the reconstruction of the moiré pattern. More specifically, $\mathbf{u} = \mathbf{u}_1 - \mathbf{u}_2$ with $\mathbf{u}_{1,2}$ the acoustic atomic displacement field of the individual layers where the subscript gives the layer index. In general, the displacement gradient contains contributions of both twist, lattice mismatch, and external heterostrain. It can be written as

$$M = \left(1 + \frac{\mathcal{E}}{2}\right) R\left(\frac{\theta}{2}\right) - \left(1 - \frac{\mathcal{E}}{2}\right) R\left(-\frac{\theta}{2}\right) \approx \mathcal{E} + \theta \begin{pmatrix} 0 & -1 \\ 1 & 0 \end{pmatrix}, \tag{31}$$

where $\mathcal{E}$ is a symmetric matrix that encodes lattice mismatch and heterostrain, and $R(\theta)$ is the rotation matrix for a counterclockwise rotation by angle $\theta$. In the last step we assumed that the twist angle is small. The moiré lattice with lattice vectors $\mathbf{L}$ and reciprocal lattice vectors $\mathbf{g}$ is then defined by $\boldsymbol{\phi}(\mathbf{r} + \mathbf{L}) = \boldsymbol{\phi}(\mathbf{r}) + \mathbf{a} \equiv \boldsymbol{\phi}(\mathbf{r})$ which yields

$$\mathbf{g} = M^\top \mathbf{b}, \qquad \mathbf{L} = M^{-1}\mathbf{a}, \tag{32}$$

where $\mathbf{a}$ ($\mathbf{b}$) is a monolayer (reciprocal) lattice vector with $\mathbf{a} \cdot \mathbf{b} \in 2\pi\mathbb{Z}$ and $\mathbf{L} \cdot \mathbf{g} \in 2\pi\mathbb{Z}$. For example, for a twist moiré we have $\mathbf{g} = (a/L)\mathbf{b} \times \hat{z}$ with $L = a/[2\sin(\theta/2)] \approx a/\theta$ the moiré lattice constant, $a \approx 2.46$ Å the monolayer lattice constant, and where $\hat{z}$ is the unit vector perpendicular to the twisted layers. In the following, we assume that the displacement field has the periodicity of the original moiré with

$$\mathbf{u}(\mathbf{r}) = \sum_{\mathbf{g}} \mathbf{u}_{\mathbf{g}}\, e^{i\mathbf{g}\cdot\mathbf{r}} = \sum_{\mathbf{b}} \mathbf{u}_{\mathbf{b}}\, e^{i\mathbf{b}\cdot M\mathbf{r}}, \tag{33}$$

where $\mathbf{u}_\mathbf{g}$ are the Fourier components which can be labeled by either the monolayer or moiré reciprocal vectors. It will be more convenient to use the monolayer vectors because they are always defined on a hexagonal grid even in the presence of heterostrain.

The ground state configuration is obtained by minimizing the total energy, which is given by the sum of elastic and adhesive contributions:

$$\frac{\delta}{\delta \boldsymbol{\phi}(\mathbf{r})}\left(F_{\text{elas}} + F_{\text{adh}}\right) = 0. \tag{34}$$

The elastic and adhesion energies $F_{\text{elas}}$ and $F_{\text{adh}}$ are explicitly given by [37,38]

$$\begin{aligned} F_{\text{elas}}[\boldsymbol{\phi}(\mathbf{r})] &= \int d^2\mathbf{r}\left[\frac{\lambda}{4}(\partial_i \phi_i)^2 + \frac{\mu}{8}\left(\partial_i \phi_j + \partial_j \phi_i\right)^2\right], \\ F_{\text{adh}}[\boldsymbol{\phi}(\mathbf{r})] &= \int d^2\mathbf{r}\, V[\boldsymbol{\phi}(\mathbf{r})], \end{aligned} \tag{35}$$

where $\lambda$ and $\mu$ are the in-plane Lamé parameters, or equivalently with $\lambda + \mu$ the bulk modulus and $\mu$ the shear modulus, which describe the linear elastic response of graphene. The adhesion potential $V$ can be written as

$$V(\boldsymbol{\phi}) = \sum_{\mathbf{b}} e^{i\mathbf{b}\cdot\boldsymbol{\phi}}\, V_{\mathbf{b}}, \tag{36}$$

and gives the energy of each layer stacking, encoded by $\boldsymbol{\phi}$. It has to be a periodic function of $\boldsymbol{\phi}$ with periodicity given by the monolayer lattice. This is because shifting one layer over a monolayer lattice constant with respect to the other returns the same bilayer. Using

$$\partial_j \phi_i = M_{ij} + \partial_j u_i, \tag{37}$$

we obtain

$$\begin{aligned} F_{\text{elas}} &= \int d^2\mathbf{r}\left[\frac{\lambda}{4}(M_{ii} + \partial_i u_i)^2 + \frac{\mu}{8}\left(M_{ij} + M_{ji} + \partial_i u_j + \partial_j u_i\right)^2\right] \\ &= \int d^2\mathbf{r}\left[\frac{\lambda}{4}(\partial_i u_i)^2 + \frac{\mu}{8}\left(\partial_i u_j + \partial_j u_i\right)^2\right], \end{aligned} \tag{38}$$

where we discarded constants and boundary terms. After performing a Fourier transform on the displacement fields, the elastic energy density becomes

$$\begin{aligned} f_{\text{elas}} &= -\sum_{\mathbf{g},\mathbf{g}'} \frac{1}{A_{\text{m}}}\int_{\text{moiré cell}} d^2\,\mathbf{r}\left[\frac{\lambda}{4}(\mathbf{g}\cdot\mathbf{u}_\mathbf{g})(\mathbf{g}'\cdot\mathbf{u}_{\mathbf{g}'}) + \frac{\mu}{8}\left(g_i u_{\mathbf{g},j} + g_j u_{\mathbf{g},i}\right)\left(g_i' u_{\mathbf{g}',j} + g_j' u_{\mathbf{g}',i}\right)\right] e^{i(\mathbf{g}+\mathbf{g}')\cdot\mathbf{r}} \\ &= \frac{1}{4}\sum_{\mathbf{g}}\left[\lambda\left|\mathbf{g}\cdot\mathbf{u}_\mathbf{g}\right|^2 + \frac{\mu}{2}\left|g_i u_{\mathbf{g},j} + g_j u_{\mathbf{g},i}\right|^2\right] \\ &= \frac{1}{4}\sum_{\mathbf{g}}\left[(\lambda + 2\mu)\left|\mathbf{g}\cdot\mathbf{u}_\mathbf{g}\right|^2 + \mu\left|(\hat{z}\times\mathbf{g})\cdot\mathbf{u}_\mathbf{g}\right|^2\right]. \end{aligned} \tag{39}$$

Using a Helmholtz decomposition, which is always possible for a vector field on a torus,

$$\mathbf{u}_{\mathbf{g}} = \frac{a}{L}\frac{u_{\mathbf{g}}^{\parallel}\mathbf{g} + u_{\mathbf{g}}^{\perp}\hat{z}\times\mathbf{g}}{ig^2}, \tag{40}$$

where we defined the longitudinal ($u_{\mathbf{g}}^{\parallel}$) and transverse ($u_{\mathbf{g}}^{\perp}$) parts of the Fourier component $\mathbf{u}_{\mathbf{g}}$ of the displacement field, respectively. The prefactor $L/a$ is chosen for convenience. And since in the presence of heterostrain $L_1 = |\mathbf{L}_1| \neq |\mathbf{L}_2| = L_2$ we define the average moiré lattice constant $L = (L_1 + L_2)/2$. In terms of these coefficients, the elastic energy density can be written as

$$f_{\text{elas}} = \frac{a^2}{4L^2}\sum_{\mathbf{g}}\left[(\lambda + 2\mu)\left|u_{\mathbf{g}}^{\parallel}\right|^2 + \mu\left|u_{\mathbf{g}}^{\perp}\right|^2\right], \tag{41}$$

where we defined $L = (L_1 + L_2)/2$ for the general case with $L_1 = |\mathbf{L}_1| \neq |\mathbf{L}_2| = L_2$. We also have

$$\begin{aligned} f_{\text{adh}} &= \frac{1}{A_{\text{m}}}\int_{\text{moiré cell}} d^2\,\mathbf{r}\,V[\boldsymbol{\phi}(\mathbf{r})] \\ &= \sum_{\mathbf{b}} V_{\mathbf{b}}\int_0^1 d\,s_1\int_0^1 d\,s_2 \exp(i\mathbf{b}\cdot[s_1\mathbf{a}_1 + s_2\mathbf{a}_2 + \mathbf{U}(s_1\mathbf{a}_1 + s_2\mathbf{a}_2)]), \end{aligned} \tag{42}$$

where we used

$$\begin{aligned} \mathbf{u}(s_1\mathbf{L}_1 + s_2\mathbf{L}_2) &= \sum_{\mathbf{b}} \mathbf{u}_{\mathbf{b}} \exp\big(i\mathbf{b}\cdot(s_1\mathbf{a}_1 + s_2\mathbf{a}_2)\big) \\ &\equiv \mathbf{U}(s_1\mathbf{a}_1 + s_2\mathbf{a}_2). \end{aligned} \tag{43}$$

Next, we minimize the total elastic and adhesion energy with respect to the $u_{\mathbf{g}}^{\parallel}$ and $u_{\mathbf{g}}^{\perp}$ components. This yields the ground state configuration $\boldsymbol{\phi}_0(\mathbf{r})$. Note that the problem of minimizing the energy functional $F[\boldsymbol{\phi}(\mathbf{r})]$ with respect to the configuration $\boldsymbol{\phi}(\mathbf{r})$ has become an ordinary minimization problem with respect to $u_{\mathbf{g}}^{\parallel}$ and $u_{\mathbf{g}}^{\perp}$ by going to reciprocal space. Therefore, we calculate the partial derivatives

$$\begin{aligned} \frac{\partial f}{\partial u_{-\mathbf{g}}^{\parallel}} &= \frac{a^2}{2L^2}(\lambda + 2\mu)u_{\mathbf{g}}^{\parallel} + \frac{1}{A_{\text{m}}}\int_{\text{moiré cell}} d^2\,\mathbf{r}\,\frac{\partial\mathbf{u}}{\partial u_{-\mathbf{g}}^{\parallel}}\cdot\frac{\partial V}{\partial\boldsymbol{\phi}}, \\ \frac{\partial f}{\partial u_{-\mathbf{g}}^{\perp}} &= \frac{a^2}{2L^2}\mu u_{\mathbf{g}}^{\perp} + \frac{1}{A_{\text{m}}}\int_{\text{moiré cell}} d^2\,\mathbf{r}\,\frac{\partial\mathbf{u}}{\partial u_{-\mathbf{g}}^{\perp}}\cdot\frac{\partial V}{\partial\boldsymbol{\phi}}, \end{aligned} \tag{44}$$

with

$$\begin{aligned} \frac{\partial V}{\partial\boldsymbol{\phi}} &= i\sum_{\mathbf{b}}\mathbf{b}\,e^{i\mathbf{b}\cdot\boldsymbol{\phi}}V_{\mathbf{b}}, \\ \frac{\partial\mathbf{u}}{\partial u_{-\mathbf{g}}^{\parallel}} &= -\frac{a}{L}\frac{\mathbf{g}}{ig^2}e^{-i\mathbf{g}\cdot\mathbf{r}}, \\ \frac{\partial\mathbf{u}}{\partial u_{-\mathbf{g}}^{\perp}} &= -\frac{a}{L}\frac{\hat{z}\times\mathbf{g}}{ig^2}e^{-i\mathbf{g}\cdot\mathbf{r}}. \end{aligned} \tag{45}$$

We thus obtain

$$
\begin{aligned}
\frac{\partial f}{\partial u^{\parallel}_{-\mathbf{g}}} &= \frac{a^2}{2L^2}(\lambda+2\mu)u^{\parallel}_{\mathbf{g}} - \frac{a\mathbf{g}}{iLg^2}\cdot\left(\frac{\partial V}{\partial \boldsymbol{\phi}}\right)_{\mathbf{g}}, \\
\frac{\partial f}{\partial u^{\perp}_{-\mathbf{g}}} &= \frac{a^2}{2L^2}\mu u^{\perp}_{\mathbf{g}} - \frac{a\hat{z}\times\mathbf{g}}{iLg^2}\cdot\left(\frac{\partial V}{\partial \boldsymbol{\phi}}\right)_{\mathbf{g}}.
\end{aligned} \tag{46}
$$

Setting these two equations to zero, we find that the energy is minimized for

$$
\begin{aligned}
\frac{a}{L}u^{\parallel}_{\mathbf{g}} &= \frac{2\mathbf{g}}{(\lambda+2\mu)ig^2}\cdot\left(\frac{\partial V}{\partial \boldsymbol{\phi}}\right)_{\mathbf{g}}, \\
\frac{a}{L}u^{\perp}_{\mathbf{g}} &= \frac{2\hat{z}\times\mathbf{g}}{\mu i g^2}\cdot\left(\frac{\partial V}{\partial \boldsymbol{\phi}}\right)_{\mathbf{g}},
\end{aligned} \tag{47}
$$

with $\mathbf{g} = M^{\top}\mathbf{b}$ and explicitly,

$$
\begin{aligned}
\left(\frac{\partial V}{\partial \boldsymbol{\phi}}\right)_{\mathbf{g}} &= i\sum_{\mathbf{b}'}\mathbf{b}' V_{\mathbf{b}'}\frac{1}{A_{\mathrm{m}}}\int_{\text{moiré cell}} d^2\,\mathbf{r}\, e^{i\mathbf{b}'\cdot\boldsymbol{\phi}(\mathbf{r}) - i\mathbf{b}\cdot M\mathbf{r}} \\
&= i\sum_{\mathbf{b}'}\mathbf{b}' V_{\mathbf{b}'}\int_0^1 d\,s_1\int_0^1 d\,s_2\exp\Big(i(\mathbf{b}'-\mathbf{b})\cdot(s_1\mathbf{a}_1+s_2\mathbf{a}_2) + i\mathbf{b}'\cdot\mathbf{u}(s_1\mathbf{L}_1+s_2\mathbf{L}_2)\Big) \\
&= i\sum_{\mathbf{b}'}\mathbf{b}' V_{\mathbf{b}'}\int_0^1 d\,s_1\int_0^1 d\,s_2\exp\left(i(\mathbf{b}'-\mathbf{b})\cdot(s_1\mathbf{a}_1+s_2\mathbf{a}_2) + i\mathbf{b}'\cdot\sum_{\mathbf{b}''}\mathbf{u}_{\mathbf{b}''}\, e^{i\mathbf{b}''\cdot(s_1\mathbf{a}_1+s_2\mathbf{a}_2)}\right).
\end{aligned} \tag{48}
$$

The equations (47) form a pair of self-consistent equations that need to be solved numerically. We implemented this in JULIA and the code is available upon reasonable request. Moreover, to simplify this calculation, we use the symmetries of the moiré to constrain $\mathbf{u}(\mathbf{r})$ [39]. For a general rigid moiré in the presence of heterostrain, and an even adhesion potential $V(\boldsymbol{\phi}) = V(-\boldsymbol{\phi})$, which is the case for twisted bilayer graphene, the displacement field only has the following symmetry:

$$
\mathbf{u}(-\mathbf{r}) = -\mathbf{u}(\mathbf{r}) = 2\sum_{\boldsymbol{b}}{}' u(\boldsymbol{b})\sin(\mathbf{b}\cdot M\,\mathbf{r}), \tag{49}
$$

where the primed sum only runs over half of the reciprocal vectors, excluding zero. In the following, we use the parameters of Ref [15]. for the elastic constants and the adhesion potential. For bilayer graphene, the adhesion potential has $C_{6v}$ symmetry and by limiting to the first three reciprocal stars in the Fourier series, we have

$$
\begin{aligned}
V(\boldsymbol{\phi}) &= V_0 + 2V_1\sum_{n=1}^{3}\cos\left(\mathcal{C}_{3z}^{n}\mathbf{b}_1\cdot\boldsymbol{\phi}\right) \\
&\quad + 2V_2\sum_{n=1}^{3}\cos\left[\mathcal{C}_{3z}^{n}(\mathbf{b}_1-\mathbf{b}_2)\cdot\boldsymbol{\phi}\right] \\
&\quad + 2V_3\sum_{n=1}^{3}\cos\left(2\mathcal{C}_{3z}^{n}\mathbf{b}_1\cdot\boldsymbol{\phi}\right),
\end{aligned} \tag{50}
$$

where $\mathbf{b}_1 = 4\pi\hat{y}/\sqrt{3}$ and $\mathbf{b}_2 = \mathcal{C}_{3z}\mathbf{b}_1$ with $\mathcal{C}_{3z}$ a counterclockwise rotation about the $z$ axis by 120 degrees. In the following, we set $V_0 = 0$ as it amounts to a global energy shift independent on the stacking configuration. The elastic coefficients and the Fourier components of the adhesion potential are listed in the top row of Table S1, together with the Poisson ratio $\sigma = 1/(1+2\mu/\lambda)$ used for a finite uniaxial heterostrain.

**Table S1|** Two-dimensional Lamé parameters $\lambda$ and $\mu$, and Fourier coefficients of the adhesion potential $V_n$ ($n = 1,2,3$) in units of meV per monolayer unit cell. We also give the Poisson ratio $\sigma = 1/(1+2\mu/\lambda)$ and the dimensionless parameters $c_n = V_n/\mu$. Parameters taken from two different references as indicated.

| | $\lambda$ | $\mu$ | $\sigma$ | $V_1$ | $V_2$ | $V_3$ | $c_1$ | $c_2$ | $c_3$ |
|---|---|---|---|---|---|---|---|---|---|
| Ref. [15] | 22166 | 47352 | 0.190 | 2.032 | −0.187 | −0.0475 | $4.29 \times 10^{-5}$ | $-3.95 \times 10^{-6}$ | $-1.00 \times 10^{-6}$ |
| Ref. [40] | 17026 | 50135 | 0.145 | 8.4 | 0 | 0 | $1.675 \times 10^{-4}$ | 0 | 0 |

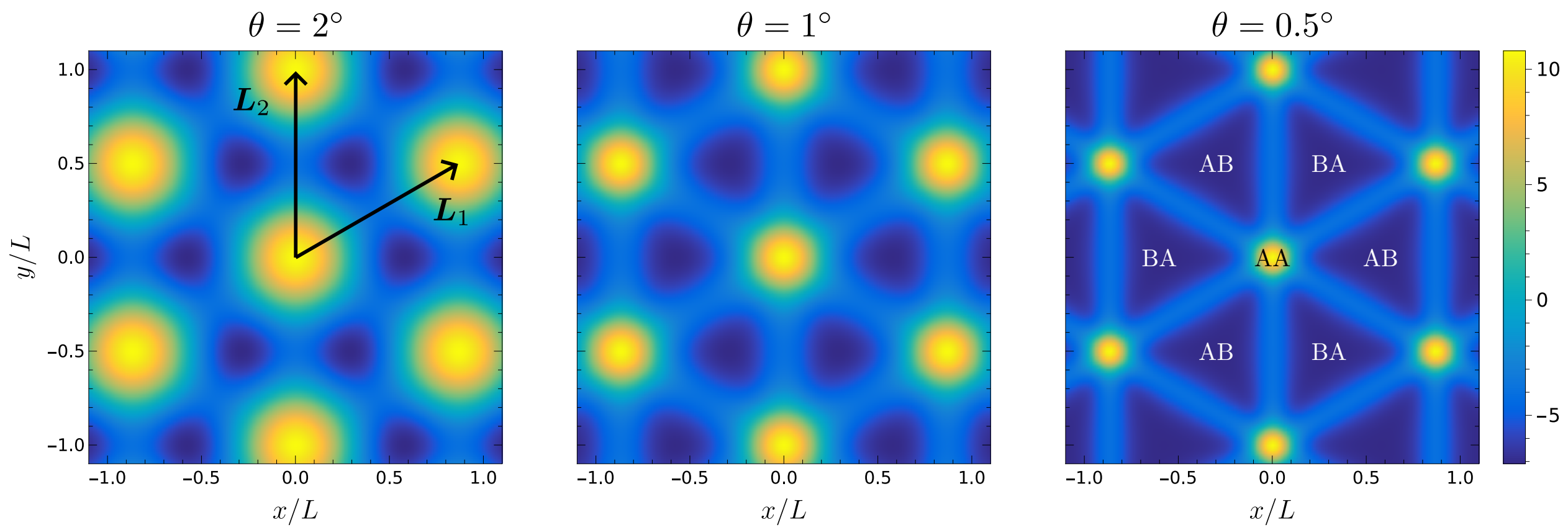

**Figure S11|** Adhesion potential $V[\boldsymbol{\phi}(\mathbf{r})]$ in units of meV per monolayer unit cell, for the ground state stacking configuration of twisted bilayer graphene as a function of position in units of the moiré lattice constant. Shown for three representative twist angles as indicated in each panel, using the parameters listed in the top row of Table 1. Low-energy regions correspond to AB/BA stacking and high-energy regions to AA stacking as indicated in the rightmost panel.

We illustrate the relaxed configurations by plotting the adhesion potential of the ground state stacking configuration in Figure S11 the absence of external heterostrain. We also show the local twist angle $\theta(\mathbf{r}) = (1/2)\nabla \times \boldsymbol{\phi}(\mathbf{r})$ of the ground state in Figure S12. We consider three cases: weak ($\theta = 2°$, intermediate ($\theta = 1°$), and strong ($\theta = 0.5°$) relaxation. In the numerical calculation, we need to restrict the number of Fourier components of the displacement field. Here we always take 16 reciprocal shells giving a total of 136 reciprocal stars, where we define a reciprocal star as the six reciprocal vectors with the same length related by 60° rotations.

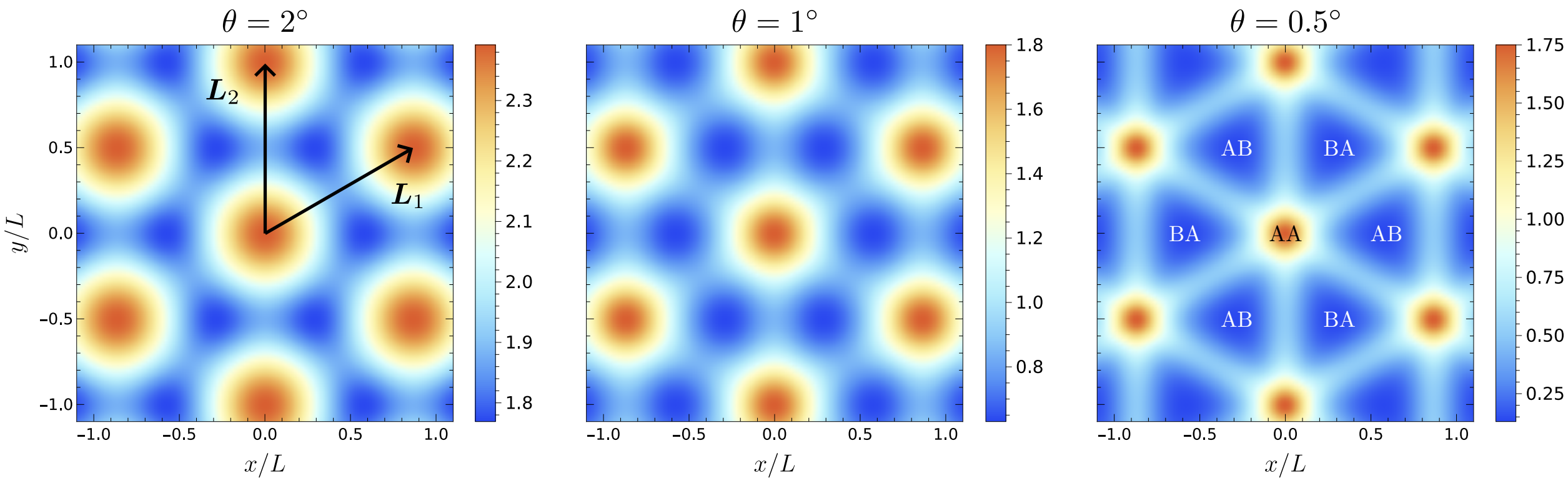

**Figure S12**| Local twist angle $(1/2)\nabla \times \boldsymbol{\phi}(\mathbf{r})$ in degrees, for the ground state stacking configuration of twisted bilayer graphene as a function of position in units of the moiré lattice constant. Shown for three representative twist angles as indicated in each panel, using the parameters listed in the top row of Table 1. The local twist angle increases in AB/BA regions which locally untwist and expand, while it increases in AA regions which twist even more to reduce the high-energy stacking region.

In Figure S13 we illustrate the reconstructed moiré in the presence of uniaxial heterostrain [41]:

$$\mathcal{E}_{\text{uniaxial}} = R(-\phi) \begin{pmatrix} \epsilon & 0 \\ 0 & -\sigma\epsilon \end{pmatrix} R(\phi), \tag{51}$$

where $\phi$ and $\epsilon$ give the direction and magnitude of the uniaxial strain, respectively. We always take $\epsilon = 0.03$ and consider uniaxial heterostrain along both the zigzag ($\phi = 0$) and armchair direction ($\phi = \pi/2$) defined with respect to the untwisted bilayer ($\theta = 0$).

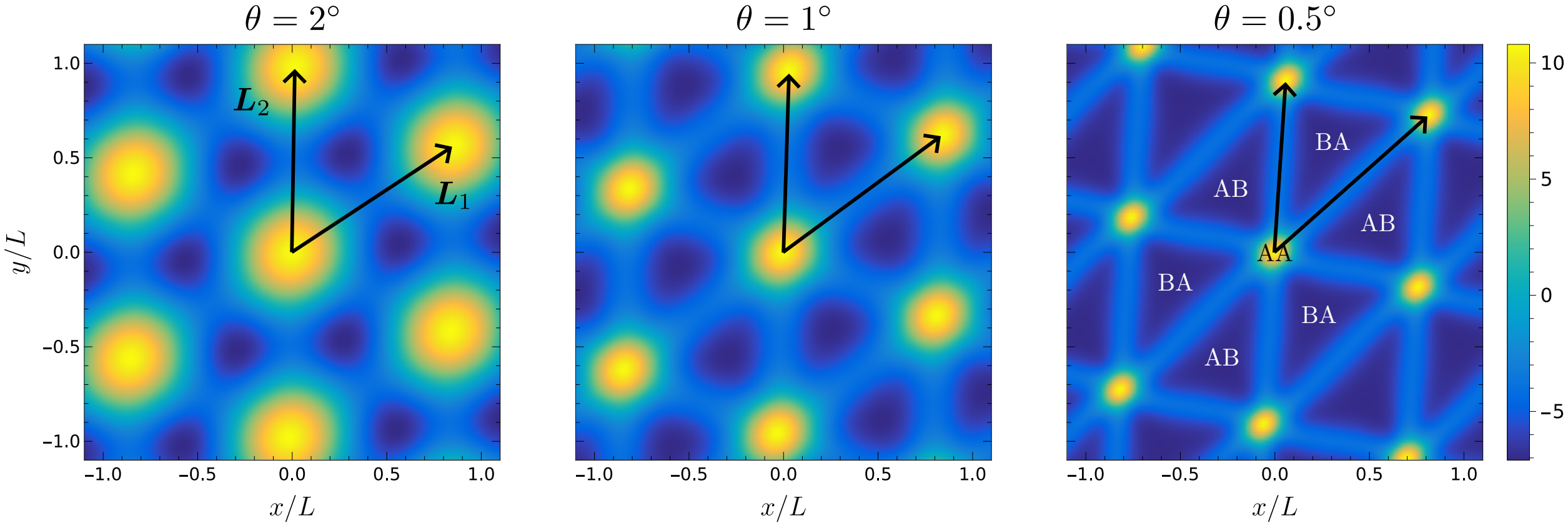

**Figure S13**| Adhesion potential $V[\boldsymbol{\phi}(\mathbf{r})]$ in units of meV per monolayer unit cell, for the ground state stacking configuration of twisted bilayer graphene in the presence of uniaxial heterostrain along the zigzag direction ($\phi = 0$) with $\epsilon = 0.003$. Shown for three representative twist angles as indicated in each panel, using the parameters listed in the top row Table S1. Low-energy regions correspond to AB/BA stacking and high-energy regions to AA stacking as indicated in the rightmost panel

**11.2 Tight-binding model in the Slater-Koster approximation**

In this section, we benchmark the relaxed moiré bands with results from the literature [37]. To this end, we take the relaxed atomic position in each layer

$$\mathbf{r}_{1,2} = \mathbf{r}_{1,2}^0 \pm \frac{1}{2}\left[\mathbf{u}\left(\mathbf{r}_{1,2}^0\right) + h\left(\mathbf{r}_{1,2}^0\right)\hat{z}\right], \tag{52}$$

where $\mathbf{r}_{1/2}^0 = R(\pm\theta/2)\mathbf{r}$ are the atomic positions of the rigid moiré for the two layers, respectively, $\mathbf{u}(\mathbf{r})$ is the in-plane displacement field, and $h(\mathbf{r})$ is the out-of-plane corrugation. Here $\mathbf{r} \in n_1\mathbf{a}_1 + n_2\mathbf{a}_2 + \boldsymbol{\delta}_\sigma$ are the atomic positions of untwisted graphene with $n_1$ and $n_2$ integers and $\boldsymbol{\delta}_\sigma$ the position of the two sublattices in the unit cell. In our choice of coordinates, we place the zigzag direction along the $x$ axis and $\boldsymbol{\delta}_\mathrm{A} - \boldsymbol{\delta}_\mathrm{B} = a_0\hat{y}$ with $a_0 = a/\sqrt{3} \approx 1.42$ Å the nearest-neighbour distance. The corrugation can be obtained from the local-stacking approximation:

$$h(\mathbf{r}) = h[\boldsymbol{\phi}(\mathbf{r})] = h_0 + 2h_1 \sum_{n=1}^{2} \cos\left[\mathbf{b}_n \cdot \boldsymbol{\phi}(\mathbf{r})\right], \tag{53}$$

with

$$h_0 = \frac{h_\mathrm{AA} + 2h_\mathrm{AB}}{3}, \qquad h_1 = \frac{h_\mathrm{AA} - h_\mathrm{AB}}{9}, \tag{54}$$

where $h_\mathrm{AA} \approx 3.6$ Å and $h_\mathrm{AB} \approx 3.35$ Å is the interlayer distance for AA $[\boldsymbol{\phi} = (0,0)]$ and AB/BA $[\boldsymbol{\phi} = (0, \pm a_0)]$ stacking, respectively [42]. The tight-binding model can be written as

$$H = \sum_{n,m} t\left(\mathbf{r}_n - \mathbf{r}_m\right) c_n^\dagger c_m, \tag{55}$$

which describes hopping of electrons in the carbon $p_z$ orbitals from site $m$ to $n$ with amplitude $t(\mathbf{r}_n - \mathbf{r}_m)$. Here we work in the two-centre approximation, i.e. we assume that the hopping amplitude only depends on the separation between the two atoms and not on their environment. For the hopping amplitude, we use the Slater-Koster form,

$$t(\mathbf{r}) = V_{pp\pi}(r)\,\frac{x^2 + y^2}{r^2} + V_{pp\sigma}(r)\,\frac{z^2}{r^2}, \tag{56}$$

where $\mathbf{r} = (x, y, z)$ is a bond vector with $r = |\mathbf{r}|$. Following Ref. [14], we take a simple exponential decay

$$\begin{aligned} V_{pp\pi}(r) &= V_{pp\pi}^0 \exp\left[-\frac{r - a_0}{r_0}\right], \\ V_{pp\sigma}(r) &= V_{pp\sigma}^0 \exp\left[-\frac{r - h_\mathrm{AB}}{r_0}\right], \end{aligned} \tag{57}$$

with $V_{pp\pi}^0 = -2.7$ eV, $V_{pp\sigma}^0 = 0.48$ eV, $a_0 = 1.42$ Å is the nearest-neighbour distance, $h_\mathrm{AB} = 3.35$ Å is the interlayer distance of Bernal bilayer graphene, and $r_0 = 0.184a$ [14,37].

We can benchmark our approach by reproducing the results of Ref. [40]. All the parameters that were used for this benchmark are given in the bottom row of Table S1 which are corroborated by Table I of Ref. [41]. Using the method described in Section 2 we then reproduced Table II of Ref. [40]. However, we believe that the adhesion potential in this work was accidentally overestimated by a factor four. Nevertheless, we have used the relaxed configurations obtained with these parameters to calculate the electronic band structure with the tight-binding model. Note that we only include in-plane displacements. As in Ref. [40] we do not consider out-of-plane corrugations. This may be justified by the encapsulation of the device with hexagonal boron nitride, which suppresses corrugations. The results are shown in Figure S14 where we show both the rigid (unrelaxed) and relaxed moiré bands close to charge neutrality for several twist angles. In the tight-binding method we are limited to commensurate structures. In particular, we choose commensurate structures for which the moiré lattice coincides with the microscopic lattice periodicity. In this case, the commensurate twist angles are given by

$$\theta_m = \arccos\left(\frac{3m^2 + 3m + \frac{1}{2}}{3m^2 + 3m + 1}\right), \tag{58}$$

for $m = 1,2,3, \ldots$. We conclude that we correctly reproduced the results of Ref. [40]. In the following, we will use the tight-binding model of Refs. [14,37] to construct the moiré potentials of the electronic continuum model in the presence of lattice relaxation. However, for these calculations the relaxed moiré lattice is computed using the more accurate parameters of Ref. [15] for the elastic constants and adhesion potential shown in the top row of Table S1.

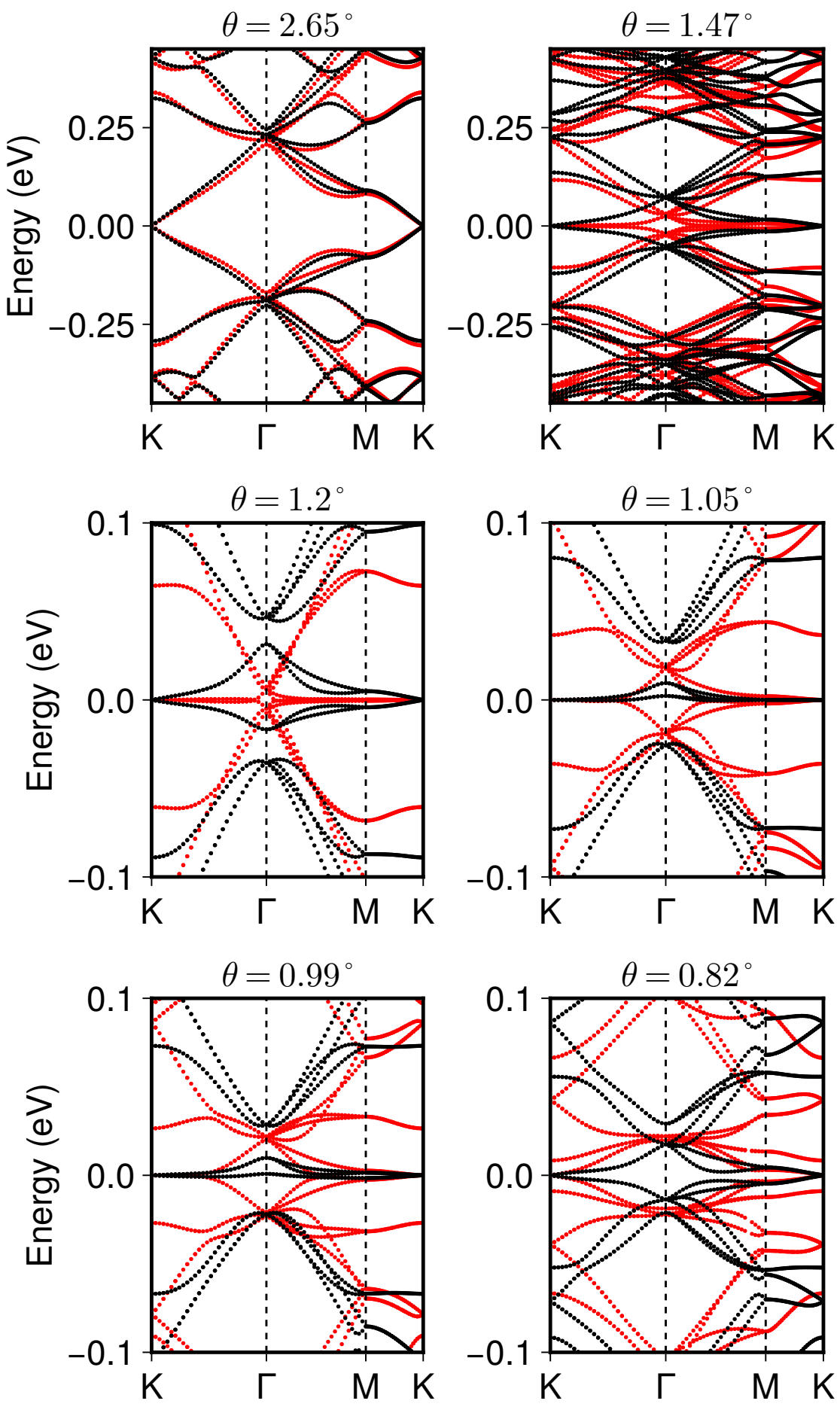

**Figure S14**| Reproduction of Fig. 3 of Ref. [40] using the tight-binding model with atomic displacement field $\mathbf{u}(\mathbf{r})$ computed from continuum elasticity using the parameters in the bottom row of Table S1 for a series of commensurate twist angles with $m = 12,22,27,31,33,40$ (see Eq. 58). Black dots represent eigenvalues for the relaxed tBG and red dots for the unrelaxed tBG.

**<u>11.3 Electronic continuum model in the presence of lattice relaxation and heterostrain</u>**

In this section, we construct the electronic continuum model for twisted bilayer graphene taking into account reconstruction of the moiré pattern. The inputs for this theory are a hopping function, which is taken from Ref. [14], and the in-plane and out-of-plane atomic displacement fields $\mathbf{u}(\mathbf{r})$ and $h(\mathbf{r})$, respectively. The latter are calculated using continuum elasticity as described in (13.1). Here we loosely follow the methodology of Refs. [43,44].

We consider twisted bilayer graphene where the top and bottom layer (denoted by index 1 and 2) is rotated counterclockwise by an angle $\pm\theta/2$, respectively. We choose coordinates where the zigzag direction lies along the $x$ axis for the untwisted system. The low-energy (noninteracting) electronic Hamiltonian can be decomposed as

$$H = H_{\text{intra}} + H_{\text{inter}}, \tag{59}$$

where $H_{\text{intra}}$ is the intralayer Hamiltonian and $H_{\text{inter}}$ is the interlayer Hamiltonian. In the following we consider the intralayer and interlayer terms separately and consider the lowest-order corrections due to lattice relaxation and external heterostrain.

**a. Intralayer Hamiltonian**

To first order in the gradients of the fermion fields and atomic displacement, the intralayer Hamiltonian can be written as

$$H_{\text{intra}} = \hbar v_{\text{F}} \sum_{l=1,2} \int d^2\mathbf{r}\, \psi_l^\dagger(\mathbf{r}) \left\{ R\left(\mp\frac{\theta}{2}\right)\left[-i\nabla + \tau_z \frac{e}{\hbar}\mathbf{A}_l(\mathbf{r})\right] \cdot \left(\tau_z\sigma_x, \sigma_y\right) \right\} \psi_l(\mathbf{r}), \tag{60}$$

where $\hbar = h/(2\pi)$ is the reduced Planck constant, $v_{\text{F}}$ is the Fermi velocity, $\sigma_{x,y}$ are Pauli matrices corresponding to sublattice, $\tau_z = \pm 1$ is the valley index, and $\psi_l$ are two-component fermion operators defined as $\psi_l(\mathbf{r}) = [\psi_{lA}(\mathbf{r}), \psi_{lB}(\mathbf{r})]^\top$ for layer $l = 1,2$. In the nearest-neighbour approximation, we have $\hbar v_F/a = \sqrt{3}t_0/2 \approx 2338.3$ meV. However, taking into account longer-range hopping processes in the tight-binding model for (13.2) we have $\hbar v_F/a \approx 2125.6$ meV.

Here we also introduced the pseudo gauge field due to strain, which is given by [45]:

$$\mathbf{A}_l(\mathbf{r}) = \frac{\hbar\gamma}{e} R(3\phi_l) \begin{pmatrix} u_{l,yy} - u_{l,xx} \\ u_{l,xy} + u_{l,yx} \end{pmatrix}, \tag{61}$$

where $e$ is the electron charge and $\gamma = \beta/(2a_0)$ in the nearest-neighbour and two-centre approximation with $\beta = (a_0/t_0)dt/dr|_0$ the electron Grüneisen parameter with $t_0 = -2.7$ eV the nearest-neighbour hopping amplitude in the absence of strain. From the hopping function in eq. 57 we find $\beta = a_0/r_0 \approx 3.14$. However, taking into account all hopping processes from the tight-binding model, we find $\hbar v_F\gamma \approx 6728.7$ meV. We further introduced $u_{l,ij} = \partial_i u_{l,j}$ where we neglect the quadratic contributions to the strain tensor from out-of-plane corrugations, and $\phi_l$ is the angle between the $x$ axis and the zigzag direction of layer $l$. In our choice of coordinates, the zigzag direction lies along the $x$ axis for $\theta = 0$ and we have $\phi_{1/2} = \pm\theta/2$.

Using that $\mathbf{u}_{1,2} = \pm\mathbf{u}/2$, we have that $u_{l,ij} = (-1)^{l+1}\,\partial_i u_j/2$ with

$$u_{ij} = \partial_i u_j = \sum_{\mathbf{g}} i\, g_i u_{\mathbf{g}j} e^{i\mathbf{g}\cdot\mathbf{r}}, \tag{62}$$

where we plugged in the Fourier series for $\mathbf{u}(\mathbf{r})s$. Therefore,

$$\mathbf{A}_l(\mathbf{r}) = (-1)^{l+1}\mathbf{A}(\mathbf{r}), \qquad \mathbf{A}(\mathbf{r}) = \frac{\hbar\gamma}{2e} R(3\phi_l) \sum_{\mathbf{g}} i \begin{pmatrix} g_y u_{\mathbf{g}y} - g_x u_{\mathbf{g}x} \\ g_x u_{\mathbf{g}y} + g_y u_{\mathbf{g}x} \end{pmatrix} e^{i\mathbf{g}\cdot\mathbf{r}}. \tag{63}$$

In the Helmholtz decomposition (eq. 40) this becomes

$$\mathbf{A}(\mathbf{r}) = \frac{\hbar\gamma a}{2eL} R(3\phi_l) \sum_{\mathbf{g}} \frac{1}{g^2}\left[ u_{\mathbf{g}}^{\parallel} \begin{pmatrix} g_y^2 - g_x^2 \\ 2g_x g_y \end{pmatrix} + u_{\mathbf{g}}^{\perp} \begin{pmatrix} 2g_x g_y \\ g_x^2 - g_y^2 \end{pmatrix} \right] e^{i\mathbf{g}\cdot\mathbf{r}}. \tag{64}$$

We note that in the presence of uniaxial heterostrain, the pseudo gauge fields also contain a constant contribution from the rigid displacement gradient [46].

Finally, we want to point out that technically $\mathbf{A}_l(\mathbf{r})$ is not a gauge field. Instead, we think it is more appropriate to refer to these as *shift fields* as they locally shift the momentum in each layer. This is because $\mathbf{A}_l(\mathbf{r})$ has the opposite sign for layer $l = 1,2$. Hence, a gauge transformation:

$$\psi_l(\mathbf{r}) \mapsto e^{i\Lambda(\mathbf{r})}\psi_l(\mathbf{r}), \tag{65}$$

results in

$$\mathbf{A}_{1,2} \mapsto \mathbf{A}_{1,2} \pm \nabla\Lambda. \tag{66}$$

Consequently, the divergence cannot be gauged away for both $\mathbf{A}_1$ and $\mathbf{A}_2$ simultaneously, and therefore it has a physical meaning. Hence, the shift fields are characterized by both a pseudomagnetic field and their divergence.

**b. Interlayer Hamiltonian**

The interlayer moiré coupling in twisted bilayer graphene is given by

$$H_{\text{inter}} = \sum_{\sigma\sigma'} \int d^2\mathbf{r}\ \psi_{\sigma 1}^{\dagger}(\mathbf{r}) T_{\sigma\sigma'}(\mathbf{r}) \psi_{\sigma' 2}(\mathbf{r}) + \text{h.c.}, \tag{67}$$

with $T_{\sigma\sigma'}(\mathbf{r})$ the moiré tunnelling. Note that we do not include gradient corrections, i.e. momentum-dependent tunnelling which is the moiré version of the skew-hopping processes ($\gamma_3$ and $\gamma_4$) of untwisted bilayer graphene. Here $\sigma$ and $\sigma'$ give the two sublattices on layer 1 and layer 2, respectively.

In the gauge we have chosen for the intralayer Hamiltonian, we have

$$T_{\sigma\sigma'}(\mathbf{r}) = e^{i\mathbf{q}_1\cdot\mathbf{r}} \sum_{\mathbf{g}} T_{\sigma\sigma'}(\mathbf{g}) e^{i\mathbf{g}\cdot\mathbf{r}}, \tag{68}$$

with $\mathbf{q}_1 = \mathbf{k}_2 - \mathbf{k}_1 = -4\pi\hat{y}/(3L)$. Here $\mathbf{k}_{1,2} = R(\pm\theta)\mathbf{K}$ are the rotated $\mathbf{K}$ points of layer 1 and 2 with $\mathbf{K} = 4\pi\hat{x}/(3a)$. In order to simplify the discussion, we only consider twist, and we also restrict to valley $K$. The inclusion of heterostrain is straightforward using the same formalism, and the moiré tunnelling of the other valley can be obtained from time-reversal symmetry by complex conjugation. We now go to momentum space by taking a Fourier transform,

$$\psi_{\sigma l}(\mathbf{r}) = \frac{1}{\sqrt{A_{\mathrm{m}}}}\sum_{\mathbf{k}}\sum_{\mathbf{g}} e^{i(\mathbf{k}-\mathbf{k}_l+\mathbf{g})\cdot\mathbf{r}}\, c_{\sigma l,\mathbf{k}+\mathbf{g}}, \tag{69}$$

where $A_{\mathrm{m}} = |\mathbf{L}_1 \times \mathbf{L}_2|$ is the area of the moiré unit cell. The interlayer Hamiltonian becomes

$$\begin{aligned} H_{\mathrm{inter}} &= \sum \frac{1}{A_{\mathrm{m}}}\int d^2\mathbf{r}\, c^{\dagger}_{\sigma 1,\mathbf{k}+\mathbf{g}} e^{-i(\mathbf{k}-\mathbf{k}_1+\mathbf{g})\cdot\mathbf{r}} T_{\sigma\sigma'}(\mathbf{r}) e^{i(\mathbf{k}'-\mathbf{k}_2+\mathbf{g}')\cdot\mathbf{r}} c_{\sigma 2,\mathbf{k}'+\mathbf{g}'} + \mathrm{h.c.} \\ &= \sum_{\sigma\sigma'}\sum_{\mathbf{k}}\sum_{\mathbf{g},\mathbf{g}'} c^{\dagger}_{\sigma 1,\mathbf{k}+\mathbf{g}}\, T_{\sigma\sigma'}(\mathbf{g}-\mathbf{g}') c_{\sigma 2,\mathbf{k}+\mathbf{g}'} + \mathrm{h.c.}. \end{aligned} \tag{70}$$

For $\mathbf{u}_{1,2} = \pm\mathbf{u}/2$ and in the two-centre approximation, we use the formalism of Ref. [43,44] (for example, see Eq. 28 of Ref. [44]) and obtain

$$T_{\sigma\sigma'}(\mathbf{r}) = \sum_{\mathbf{b}} e^{i\mathbf{b}\cdot\boldsymbol{\delta}_{\sigma\sigma'}}\, e^{i(\mathbf{b}+\mathbf{K})\cdot\boldsymbol{\phi}(\mathbf{r})}\check{t}_{\perp}[\mathbf{b}+\mathbf{K}, h(\mathbf{r})], \tag{71}$$

with $\boldsymbol{\delta}_{\sigma\sigma'} = \boldsymbol{\delta}_{\sigma} - \boldsymbol{\delta}_{\sigma'}$ where $\boldsymbol{\delta}_{\sigma}$ is the sublattice position in the (untwisted) graphene unit cell and $h(\mathbf{r})$ is the out-of-plane displacement field. We further defined the two-dimensional Fourier transform (with respect to the in-plane coordinate $x$ and $y$) of the interlayer hopping function $t_{\perp}(\mathbf{r})$. It is given by

$$\begin{aligned} \check{t}_{\perp}(\mathbf{k},h) &= \frac{1}{A_{\mathrm{g}}}\int d^2\mathbf{r}\, e^{-i\mathbf{k}\cdot\mathbf{r}} t_{\perp}(\mathbf{r}+h\hat{z}) \\ &= \frac{2\pi}{A_{\mathrm{g}}}\int_0^{\infty} d\,r\, r J_0(kr) t_{\perp}(\mathbf{r}+h\hat{z}), \end{aligned} \tag{72}$$

with $A_{\mathrm{g}} = \sqrt{3}a^2/2$ the unit cell area of monolayer graphene and $J_0(kr)$ a Bessel function of the first kind. The second line holds only in the two-centre approximation (we placed $x$ along $\mathbf{k}$ to simplify the integral, such that $\mathbf{k}\cdot\mathbf{r} = kr\cos\phi$ in standard polar coordinates). In this case, the Fourier components are real and only depend on $k = |\mathbf{k}|$. In eq. 71 we also introduced the local stacking configuration (in the absence of external heterostrain)

$$\boldsymbol{\phi}(\mathbf{r}) = M\mathbf{r} + \mathbf{u}(\mathbf{r}), \tag{73}$$

where $M$ is the rigid displacement gradient defined in eq. 31. In the absence of external heterostrain, we have $M\mathbf{r} = (a/L)\hat{z}\times\mathbf{r}$ with $L = a/[2\sin(\theta/2)]$. We can thus write

$$e^{-i\mathbf{q}_1\cdot\mathbf{r}} e^{i(\mathbf{b}+\mathbf{K})\cdot\boldsymbol{\phi}(\mathbf{r})} = e^{i\mathbf{g}\cdot\mathbf{r}} e^{i(\mathbf{b}+\mathbf{K})\cdot\mathbf{u}(\mathbf{r})}, \tag{74}$$

with $\mathbf{q}_1 = (a/L)\mathbf{K}\times\hat{z}$ the momentum separation between the two layers, and $\mathbf{g} = (a/L)\mathbf{b}\times\hat{z}$ a moiré reciprocal lattice vector. The right-hand side of eq. 74 explicitly has the periodicity of the moiré. In the following, we first approximate $h(\mathbf{r})$ with the average interlayer distance. We will incorporate the effect of corrugation below using the local-stacking approximation to interpolate $h(\mathbf{r})$.

i. Absence of corrugation: only in-plane relaxation

For constant $h = \langle h(\mathbf{r})\rangle$ we obtain

$$
\begin{aligned}
T_{\sigma\sigma'}(\mathbf{r}) &= e^{i\mathbf{q}_1\cdot\mathbf{r}} \sum_{\mathbf{b}} e^{i\mathbf{b}\cdot\boldsymbol{\delta}_{\sigma\sigma'}}\, \breve{t}_\perp(\mathbf{b}+\mathbf{K},h)\left[e^{i\mathbf{g}\cdot\mathbf{r}} e^{i(\mathbf{b}+\mathbf{K})\cdot\mathbf{u}(\mathbf{r})}\right] \\
&= e^{i\mathbf{q}_1\cdot\mathbf{r}} \sum_{\mathbf{g}} e^{i\mathbf{g}\cdot\mathbf{r}}\, \breve{T}_{\sigma\sigma'}(\mathbf{g}),
\end{aligned}
\tag{75}
$$

where

$$
\begin{aligned}
\breve{T}_{\sigma\sigma'}(\mathbf{g}) &= \sum_{\mathbf{b}} e^{i\mathbf{b}\cdot\boldsymbol{\delta}_{\sigma\sigma'}}\, \breve{t}_\perp(\mathbf{b}+\mathbf{K},h)\left[\frac{1}{A_{\mathrm{m}}}\int_{\text{moiré cell}} d^2\,\mathbf{r}\, e^{-i\mathbf{g}\cdot\mathbf{r}} e^{-i\mathbf{q}_1\cdot\mathbf{r}} e^{i(\mathbf{b}+\mathbf{K})\cdot\boldsymbol{\phi}(\mathbf{r})}\right] \\
&= \sum_{\mathbf{b}} e^{i\mathbf{b}\cdot\boldsymbol{\delta}_{\sigma\sigma'}}\, \breve{t}_\perp(\mathbf{b}+\mathbf{K},h)\left[\frac{1}{A_{\mathrm{m}}}\int_{\text{moiré cell}} d^2\,\mathbf{r}\, e^{-i\mathbf{q}\cdot\mathbf{r}} e^{i(\mathbf{b}+\mathbf{K})\cdot\boldsymbol{\phi}(\mathbf{r})}\right],
\end{aligned}
\tag{76}
$$

with $\mathbf{q} = \mathbf{q}_1 + \mathbf{g}$. We can thus instead write

$$
T_{\sigma\sigma'}(\mathbf{r}) = \sum_{\mathbf{q}} e^{i\mathbf{q}\cdot\mathbf{r}}\, \breve{T}_{\sigma\sigma'}(\mathbf{q}). \tag{77}
$$

The Fourier components $\breve{t}_\perp(\mathbf{b}+\mathbf{K},h_0)$ only depend on the magnitude $|\mathbf{b}+\mathbf{K}|$ [see eq. 72 and decay rapidly with $|\mathbf{b}+\mathbf{K}|$ such that we only retain a few shells. The values are listed in Table S2 using the hopping function of Ref. [14].

**Table S2|** Largest three Fourier components of the hopping amplitude $t_\perp(\mathbf{b}+\mathbf{K},h)$ in units meV for $h = 3.35$ Å. We use the parameters of Ref. (in principle h should be the average interlayer distance $h_0 \approx 3.43$ Å).

| $\lvert\mathbf{b}+\mathbf{K}\rvert/\lvert\mathbf{K}\rvert$ | 1 | 2 | $\sqrt{7}$ |
|---|---|---|---|
| $\breve{t}_\perp(\mathbf{b}+\mathbf{K},h)$ | 110.889 | 1.55865 | 0.0617242 |

ii. Corrugation only: no in-plane relaxation

In the absence of in-plane relaxation, we obtain

$$
\begin{aligned}
\breve{T}_{\sigma\sigma'}(\mathbf{q}) &= \frac{1}{A_{\mathrm{m}}A_{\mathrm{g}}}\sum_{\mathbf{b}'} e^{i\mathbf{b}'\cdot\boldsymbol{\delta}_{\sigma\sigma'}} \int_{\text{moiré cell}} d^2\,\mathbf{r}\, e^{ia/L(\mathbf{b}'-\mathbf{b})\cdot(\hat{z}\times\mathbf{r})} \int d^2\mathbf{r}'\, e^{-i(\mathbf{b}'+\mathbf{K})\cdot\mathbf{r}'}\, t_\perp[\mathbf{r}'+h(\mathbf{r})\hat{z}] \\
&= \frac{e^{i\mathbf{b}\cdot\boldsymbol{\delta}_{\sigma\sigma'}}}{A_{\mathrm{g}}^2}\sum_{\mathbf{b}'} \int_{\text{graphene cell}} d^2\,\mathbf{r}\, e^{-i\mathbf{b}'\cdot(\mathbf{r}-\mathbf{r}'+\boldsymbol{\delta}_{\sigma\sigma'})} \int d^2\mathbf{r}'\, e^{-i(\mathbf{b}+\mathbf{K})\cdot\mathbf{r}'}\, t_\perp(\mathbf{r}'+h[\mathbf{r}]\hat{z}) \\
&= \frac{e^{i\mathbf{b}\cdot\boldsymbol{\delta}_{\sigma\sigma'}}}{A_{\mathrm{g}}} \int d^2\mathbf{r}\, e^{-i(\mathbf{b}+\mathbf{K})\cdot\mathbf{r}}\, t_\perp(\mathbf{r}+h[\mathbf{r}-\boldsymbol{\delta}_{\sigma\sigma'}]\hat{z}),
\end{aligned}
\tag{78}
$$

where in the second line, we let $\mathbf{b}' \to \mathbf{b}-\mathbf{b}'$ and we used the Dirac comb formula

$$
\sum_{\mathbf{a}} \delta\,(\mathbf{r}-\mathbf{a}) = \frac{1}{A_{\mathrm{g}}}\sum_{\mathbf{b}} e^{i\mathbf{b}\cdot\mathbf{r}}, \tag{79}
$$

with $\mathbf{a}$ monolayer lattice vectors. This recovers the result from Ref. [20]. For the first shell $\breve{T}_{\mathrm{AA}} \approx 79.7$ meV and $\breve{T}_{\mathrm{AB}} \approx 97.5$ using the hopping function of Ref. [14].

iii. Both in-plane relaxation and out-of-plane corrugation

Fully accounting for both the in-plane and out-of-plane relaxation, we have

$$
\begin{aligned}
\breve{T}_{\sigma\sigma'}(\mathbf{q}) &= \frac{1}{A_{\mathrm{m}}}\sum_{\mathbf{b}} e^{i\mathbf{b}\cdot\boldsymbol{\delta}_{\sigma\sigma'}} \int_{\text{moiré cell}} d^2\,\mathbf{r}\, e^{-i\mathbf{q}\cdot\mathbf{r}} e^{i(\mathbf{b}+\mathbf{K})\cdot\boldsymbol{\phi}(\mathbf{r})} \breve{t}_{\perp}[\mathbf{b}+\mathbf{K}, h(\mathbf{r})] \\
&= \frac{2\pi}{A_{\mathrm{m}}A_{\mathrm{g}}}\sum_{\mathbf{b}} e^{i\mathbf{b}\cdot\boldsymbol{\delta}_{\sigma\sigma'}} \int_{\text{moiré cell}} d^2\,\mathbf{r}\, e^{-i\mathbf{q}\cdot\mathbf{r}} e^{i(\mathbf{b}+\mathbf{K})\cdot\boldsymbol{\phi}(\mathbf{r})} \int_0^{\infty} d\,r'\, r' J_0(|\mathbf{b}+\mathbf{K}|r')\, t_{\perp}[\mathbf{r}'+h(\mathbf{r})\hat{z}],
\end{aligned}
\tag{80}
$$

where we used the two-center approximation. A simple model for the corrugation is given by [20]

$$
h(\mathbf{r}) = h[\boldsymbol{\phi}(\mathbf{r})] = h_0 + 2h_1 \sum_{n=1}^{2} \cos\,[\mathbf{b}_n \cdot \boldsymbol{\phi}(\mathbf{r})], \tag{81}
$$

with

$$
h_0 = \frac{h_{\mathrm{AA}} + 2h_{\mathrm{AB}}}{3}, \qquad h_1 = \frac{h_{\mathrm{AA}} - h_{\mathrm{AB}}}{9}, \tag{82}
$$

where $h_{\mathrm{AA}} \approx 3.6$ Å and $h_{\mathrm{AB}} \approx 3.35$ Å is the interlayer distance for AA [$\boldsymbol{\phi} = (0,0)$] and AB/BA [$\boldsymbol{\phi} = \left(0, \pm a/\sqrt{3}\,\right)$] stacking, respectively [42].

iv. Symmetry constraints on the moiré tunnelling

In this section, show that each interlayer moiré star, which contains either three or six $\mathbf{q}$ vectors (see eq. 77), is generally characterized by one real and one complex tunnelling amplitude. Here we will make use of the symmetries that conserve the valley index and

$$
\begin{aligned}
3\boldsymbol{\phi}(\mathcal{C}_{3z}\mathbf{r}) &= \mathcal{C}_{3z}\boldsymbol{\phi}(\mathbf{r}), & h(\mathcal{C}_{3z}\mathbf{r}) &= h(\mathbf{r}), \\
\boldsymbol{\phi}(\mathcal{C}_{2y}\mathbf{r}) &= -\mathcal{C}_{2y}\boldsymbol{\phi}(\mathbf{r}), & h(\mathcal{C}_{2y}\mathbf{r}) &= -h(\mathbf{r}) \\
\boldsymbol{\phi}(-\mathbf{r}) &= -\boldsymbol{\phi}(\mathbf{r}) & h(-\mathbf{r}) &= h(\mathbf{r}).
\end{aligned}
\tag{83}
$$

For concreteness, we take a coordinate system in which the zigzag direction of the untwisted monolayers is along the $x$ axis with $\mathbf{K} = 4\pi\hat{x}/(3a)$ the position of the Dirac point.

First, we consider the in-plane rotation symmetry by $120°$ degrees:

$$
\begin{aligned}
\breve{T}_{\sigma\sigma'}(\mathcal{C}_{3z}\mathbf{q}) &= \frac{1}{A_{\mathrm{m}}}\int_{\text{moiré cell}} d^2\,\mathbf{r}\, e^{-i\mathbf{q}\cdot\mathbf{r}} \sum_{\mathbf{b}} e^{i\mathbf{b}\cdot\boldsymbol{\delta}_{\sigma\sigma'}}\, e^{i\mathcal{C}_{3z}^{-1}(\mathbf{b}+\mathbf{K})\cdot\boldsymbol{\phi}(\mathbf{r})} \breve{t}_{\perp}(\mathbf{b}+\mathbf{K}, h) \\
&= \frac{e^{i(\mathcal{C}_{3z}\mathbf{K}-\mathbf{K})\cdot\boldsymbol{\delta}_{\sigma\sigma'}}}{A_{\mathrm{m}}}\int_{\text{moiré cell}} d^2\,\mathbf{r}\, e^{-i\mathbf{q}\cdot\mathbf{r}} \sum_{\mathbf{b}} e^{i\mathbf{b}\cdot\boldsymbol{\delta}_{\sigma\sigma'}}\, e^{i(\mathcal{C}_{3z}\mathbf{b}-\mathbf{b})\cdot\boldsymbol{\delta}_{\sigma\sigma'}} e^{i(\mathbf{b}+\mathbf{K})\cdot\boldsymbol{\phi}(\mathbf{r})} \breve{t}_{\perp}[\mathcal{C}_{3z}(\mathbf{b}+\mathbf{K}), h]
\end{aligned}
\tag{84}
$$

where we changed the integration variable and let $\mathbf{b} \to \mathcal{C}_{3z}(\mathbf{b}+\mathbf{K}) - \mathbf{K}$ which is still a reciprocal vector. Also note that $e^{i(\mathcal{C}_{3z}\mathbf{b}-\mathbf{b})\cdot\boldsymbol{\delta}_{\sigma\sigma'}} = e^{i\mathbf{b}\cdot(\mathcal{C}_{3z}^{-1}\boldsymbol{\delta}_{\sigma\sigma'}-\boldsymbol{\delta}_{\sigma\sigma'})} = 1$. Hence,

$$
\breve{T}_{\sigma\sigma'}(\mathcal{C}_{3z}\mathbf{q}) = e^{i(\mathcal{C}_{3z}\mathbf{K}-\mathbf{K})\cdot\boldsymbol{\delta}_{\sigma\sigma'}} \breve{T}_{\sigma\sigma'}(\mathbf{q}). \tag{85}
$$

Next, we consider the out-of-plane $\mathcal{C}_{2y}$ symmetry which exchanges the layers:

$$
\begin{aligned}
\breve{T}_{\sigma\sigma'}(\mathcal{C}_{2y}\mathbf{q}) &= \frac{1}{A_{\mathrm{m}}}\int_{\text{moiré cell}} d^2\,\mathbf{r}\, e^{-i\mathbf{q}\cdot\mathbf{r}} \sum_{\mathbf{b}} e^{i\mathbf{b}\cdot\boldsymbol{\delta}_{\sigma\sigma'}}\, e^{-i\mathcal{C}_{2y}(\mathbf{b}+\mathbf{K})\cdot\boldsymbol{\phi}(\mathbf{r})}\breve{t}_{\perp}(\mathbf{b}+\mathbf{K},h) \\
&= \frac{e^{-i(\mathcal{C}_{2y}\mathbf{K}+\mathbf{K})\cdot\boldsymbol{\delta}_{\sigma\sigma'}}}{A_{\mathrm{m}}}\int_{\text{moiré cell}} d^2\,\mathbf{r}\, e^{-i\mathbf{q}\cdot\mathbf{r}} \sum_{\mathbf{b}} e^{-i\mathbf{b}\cdot\mathcal{C}_{2y}\boldsymbol{\delta}_{\sigma\sigma'}}\, e^{i(\mathbf{b}+\mathbf{K})\cdot\boldsymbol{\phi}(\mathbf{r})}\breve{t}_{\perp}\left[-\mathcal{C}_{2y}(\mathbf{b}+\mathbf{K}),h\right]
\end{aligned}
\tag{86}
$$

where we changed the integration variable and let $\mathbf{b} \rightarrow -\mathcal{C}_{2y}(\mathbf{b}+\mathbf{K}) - \mathbf{K}$. Also note that $\left(\mathcal{C}_{2y}\mathbf{K}+\mathbf{K}\right)\cdot \boldsymbol{\delta}_{\sigma\sigma'} = 0$ and $-\mathcal{C}_{2y}\boldsymbol{\delta}_{\sigma\sigma'} = \boldsymbol{\delta}_{\sigma'\sigma} + \mathbf{a}$. Hence, we obtain

$$
\breve{T}_{\sigma\sigma'}\left(-q_x, q_y\right) = \breve{T}_{\sigma'\sigma}(\mathbf{q}) = \left[\breve{T}_{\sigma\sigma'}(\mathbf{q})\right]^*, \tag{87}
$$

where the last equality follows from $\mathcal{C}_{2z}$ through $\boldsymbol{\phi}(-\mathbf{r}) = -\boldsymbol{\phi}(\mathbf{r})$ and the reality of $\breve{t}_{\perp}(\mathbf{b}+\mathbf{K},h)$. We find numerically that all these conditions hold. We also see that the realness of $\breve{t}_{\perp}(\mathbf{k},h)$, which is a consequence of the two-centre approximation, implies that $\breve{T}_{\sigma\sigma}(\mathbf{q})$ is real. Therefore, in general each interlayer moiré star is characterized by a real $\breve{T}_{\mathrm{AA}}$ and one complex $\breve{T}_{\mathrm{AB}}$. The latter can be chosen real if there exists a representative $\mathbf{q}$ parallel to the $x$ axis.

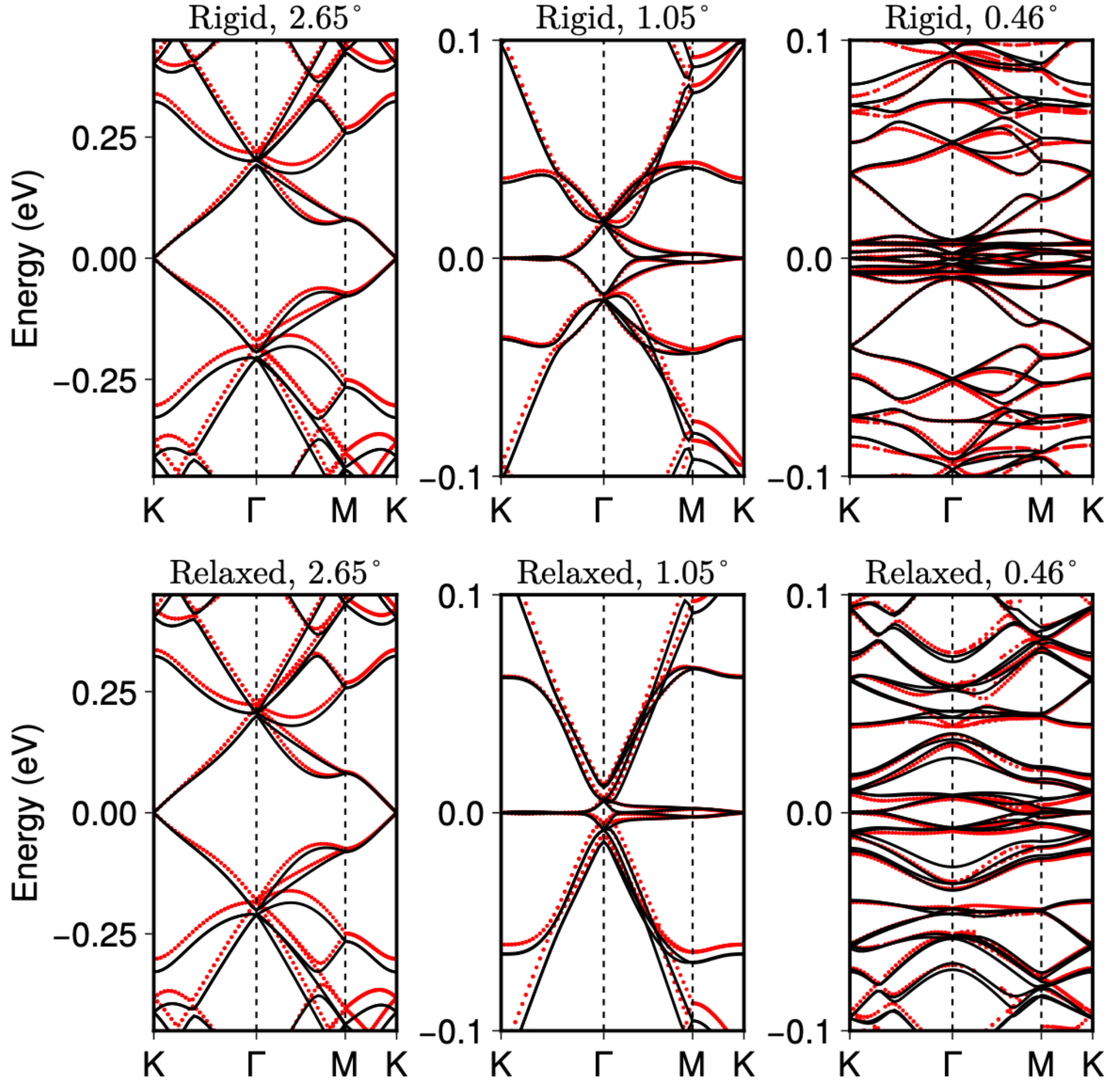

**Figure S15|** Band structures of twisted bilayer graphene at twist angles $\theta = 2.65°$, 1.05°, and 0.46°. Red dots show results from the tight-binding model, while black lines correspond to the continuum model. The top row shows the rigid (unrelaxed) case, and the bottom row shows the relaxed case using atomic displacement fields from continuum

elasticity. The comparison highlights the effect of lattice relaxation and validates the continuum model against the tight-binding benchmark.

**c. Comparison with the tight-binding model**

To benchmark the continuum model, we compare it with the full tight-binding (TB) calculation discussed in Sec. 3. In the TB approach, the relaxed atomic positions obtained from continuum elasticity (see Sec. 2) are directly used to construct the lattice, and the TB Hamiltonian is built by computing the hopping amplitudes between the relaxed atomic sites within the Slater–Koster approximation. Figure 5 shows the resulting spectra at twist angles $\theta = 2.65°$, $1.05°$, and $0.46°$, for both rigid and relaxed configurations. The inclusion of lattice relaxation in the TB calculation produces characteristic modifications of the flat bands near charge neutrality, and become significant for marginal twist angles, illustrating the importance of structural reconstruction in the moiré lattice. The TB method, however, is restricted to commensurate twist angles and becomes computationally prohibitive at the very small angles relevant to experiments. For this reason, we turn to the continuum formulation, which can be applied at arbitrary small twist angles. In Figure S15, continuum model results (black lines) are compared with TB spectra (red dots). The continuum model reproduces the TB bands with high fidelity in the low-energy window around charge neutrality. Deviations are visible only at higher energies, where higher-order corrections become significant. These include gradient corrections to the pseudo gauge fields and to the moiré tunnelling, as well as trigonal warping of the Dirac cone. Overall, the close agreement with the TB benchmark justifies the use of the continuum model in the small-angle regime explored in the main text.

**11.4 Hartree corrections**

Based on the single-particle Hamiltonian we constructed above; we then include the interaction by considering Hartree corrections. The Hartree contribution to the Hamiltonian is generally expressed as [47,48]:

$$\begin{aligned} V_H(\mathbf{r}) &= \int d^2\mathbf{r}\, \Psi^\dagger(\mathbf{r})\, v_H(\mathbf{r})\, \Psi(\mathbf{r}), \\ v_H(\mathbf{r}) &= \int d^2\mathbf{r}'\, V(\mathbf{r}-\mathbf{r}')\, \rho(\mathbf{r}'), \end{aligned} \tag{88}$$

where $V(\mathbf{r}) = e^2/\epsilon r$ represents the Coulomb interaction, and $\rho(\mathbf{r})$ denotes the charge density fluctuations arising from contributions of all occupied wave functions. Taking the Fourier transform, the Hartree potential becomes

$$v_H(\mathbf{k}) = \sum_{\mathbf{g}\neq\mathbf{0}} V\,(\mathbf{g})\, \rho(\mathbf{g})\, \Psi^\dagger(\mathbf{k})\, \Psi(\mathbf{k}-\mathbf{g}), \tag{89}$$

where $V(\mathbf{g})$ and $\rho(\mathbf{g})$ are the Fourier components of the Coulomb interaction and charge density, respectively. Specifically, the Fourier components of the charge density are given by

$$\rho(\mathbf{g}) = 4 \sum_{n,\mathbf{k}\in\mathrm{occ}} \sum_{\xi,\mathbf{g}'} \phi_{\mathbf{g}'}^{\xi,n,*}\,(\mathbf{k})\, \phi_{\mathbf{g}+\mathbf{g}'}^{\xi,n}(\mathbf{k}), \tag{90}$$

where the factor of 4 comes from spin and valley, $n$ labels the band indices, $\mathbf{k}$ is the momentum in the moiré Brillouin zone, and the summation is carried over all occupied states. Here $\phi_{\mathbf{g}}^{\xi,n}(\mathbf{k})$ are the Fourier components of the Bloch wave function of band $n$ for sublattice $\xi$ obtained by numerically diagonalizing the continuum model. The equations outlined above need to be solved self-consistently. Starting with the non-interacting $\rho(\mathbf{g})$ values, we proceed by diagonalizing the full Hamiltonian to obtain new eigenvalues and eigenvectors. The new density matrix is then computed using eq. 90. These steps are iterated until a self-consistent solution is found. In Figure S16, we compare the band structure for relaxed marginally twisted bilayer graphene including self-consistent Hartree interactions for different filling $n/n_\mathrm{B}$. Here $n_\mathrm{B} = 4/A_\mathrm{m}$ is the density per moiré band with $A_\mathrm{m} = \sqrt{3}L^2/2$ the area of the moiré unit cell. For small twist angles using $a = 2.46$ Å, we have $n_\mathrm{B} \approx 2.325(\theta/1°)^2 \times 10^{12}$ cm$^{-2}$. We find the Hartree potential modifies the bands near the Fermi level, and these changes become increasingly pronounced at larger fillings due to stronger electrostatic repulsion as charge accumulates in the moiré unit cell. Such band reshaping also has an important impact on screening: since the polarizability is governed by particle–hole excitations near the Fermi level, Hartree corrections directly influence the effective interaction strength and thereby affect superconducting behavior in the active bands.

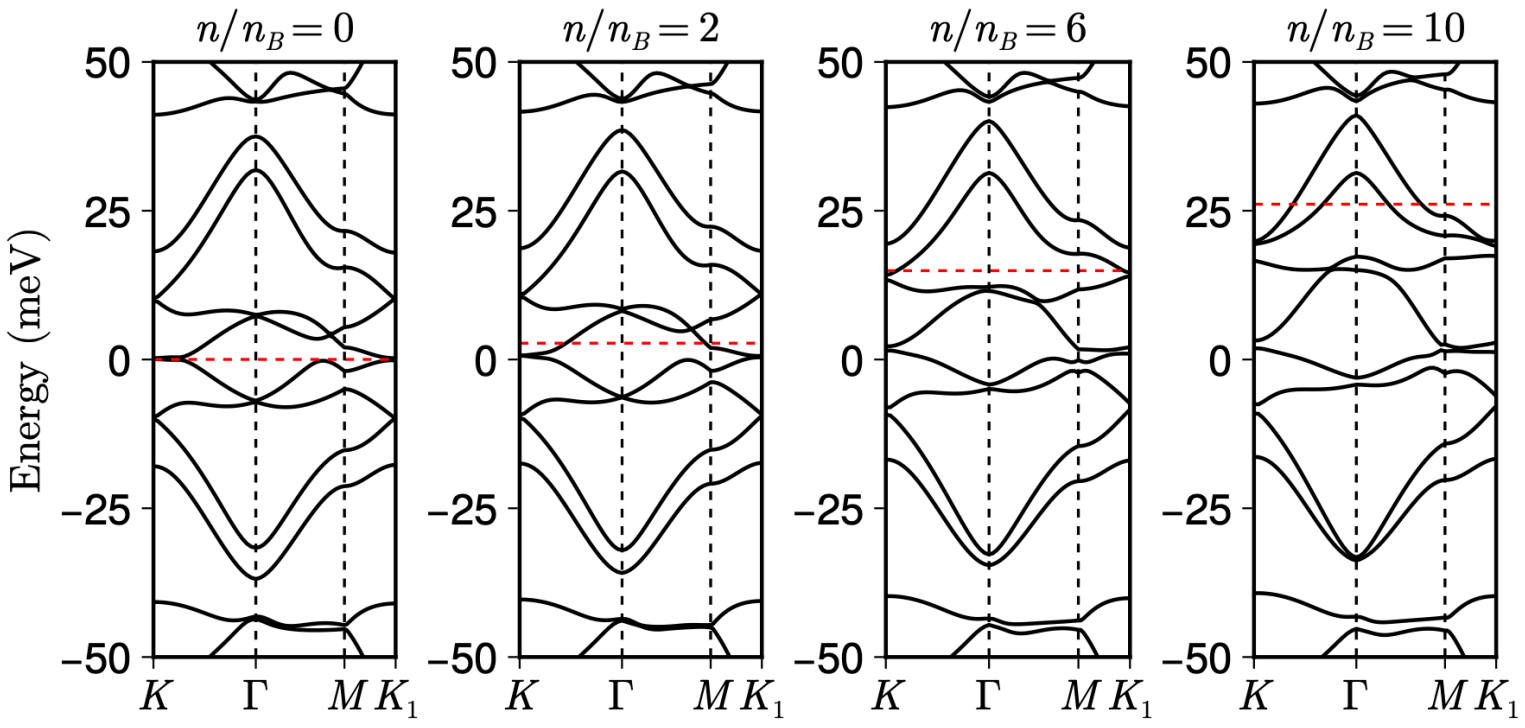

**Figure S16|** Band structure for relaxed marginally twisted bilayer graphene including self-consistent Hartree interactions for different filling $n/n_\mathrm{B}$ (where $n_\mathrm{B}$ is the density corresponding to 4 electrons per moiré unit cell). Horizontal red dashed lines indicate the Fermi energy.

**11.5 Comparison of results from different models**

We now summarize the comparison across different models, as shown in Figure S16. For a narrow averaging window, the rigid model exhibits sharp peaks in the DOS due to Van Hove singularities, which are strongly enhanced near $\theta = 0.5°$. Introducing strain and, more importantly, lattice relaxation and Hartree corrections, broadens these features and reduces the singular behavior. When a larger averaging window is considered, the DOS from the interacting relaxed model remains significantly enhanced compared to monolayer or bilayer graphene, while being less sharply peaked than in the rigid model. This demonstrates that relaxation and Hartree effects smooth out the large DOS fluctuations yet still produce a large enhancement that is crucial for the screening properties discussed in the main text. It is important to note that most previous studies neglect two key effects: (i) band renormalization due to electron–electron interactions and (ii) strain, which is almost always present in real devices and can substantially modify the band structure. In Figure S17, we compute the band structure and density of states for relaxed marginally twisted bilayer graphene in the presence of 0.3% uniaxial heterostrain (along the zigzag direction). The presence of strain will gap out the Dirac cone and lead to the strong renormlization to the low energy bands. In the right panel of Figure S17, we compare the averaged density of states over an energy window of 10 meV corresponding to the plasmon frequency. The red and blue dashed line correspond to $2\times$ and $10\times$ the monolayer case, respectively. The heterostrain strongly modifies the DOS at low filling factor, and the effect might vary sample by sample.

By explicitly incorporating both effects, our calculations of the top TBG band structure go beyond the state-of-the-art. The resulting DOS is shown in Figure S16, highlighting the strong sensitivity of the screening-layer DOS to modelling assumptions. This sensitivity implies that an exact quantitative prediction is unlikely. Nevertheless, to reproduce the qualitative experimental trends, a model must satisfy two essential criteria: (i) the DOS should increase with carrier density, and (ii) its order of magnitude should be consistent with experimental observations.

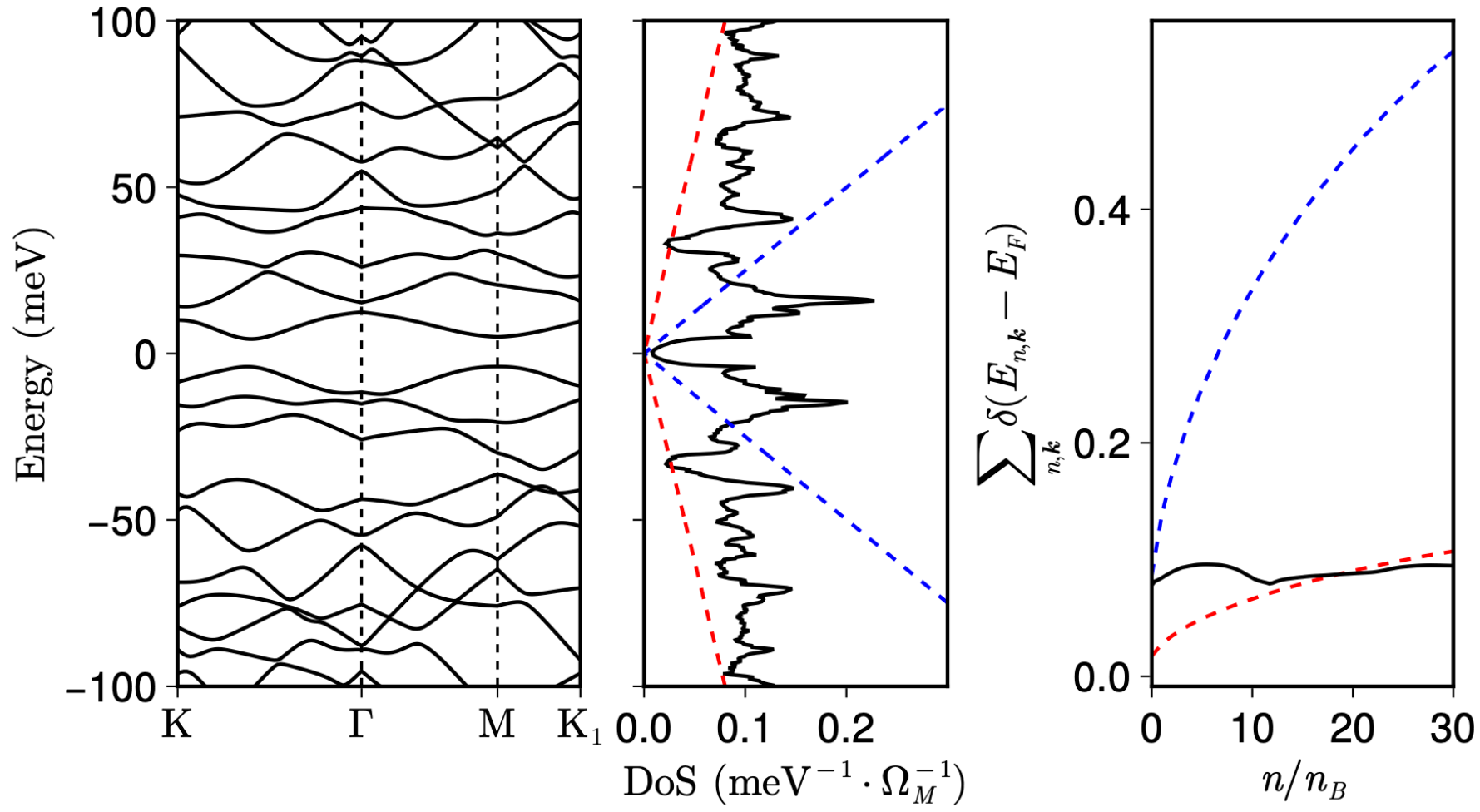

**Figure S17|** (left panel) The band structure and density of states for relaxed marginally twisted bilayer graphene in the presence of 0.3% uniaxial heterostrain (along the zigzag direction) not yet including self-consistent Hartree. (right panel) Averaged density of states over an energy window of 10 meV corresponding to the plasmon frequency. The red and blue dashed line correspond to 2 × and 10 × the monolayer case, respectively [the black line is noninteracting, the interacting case is in the cluster].

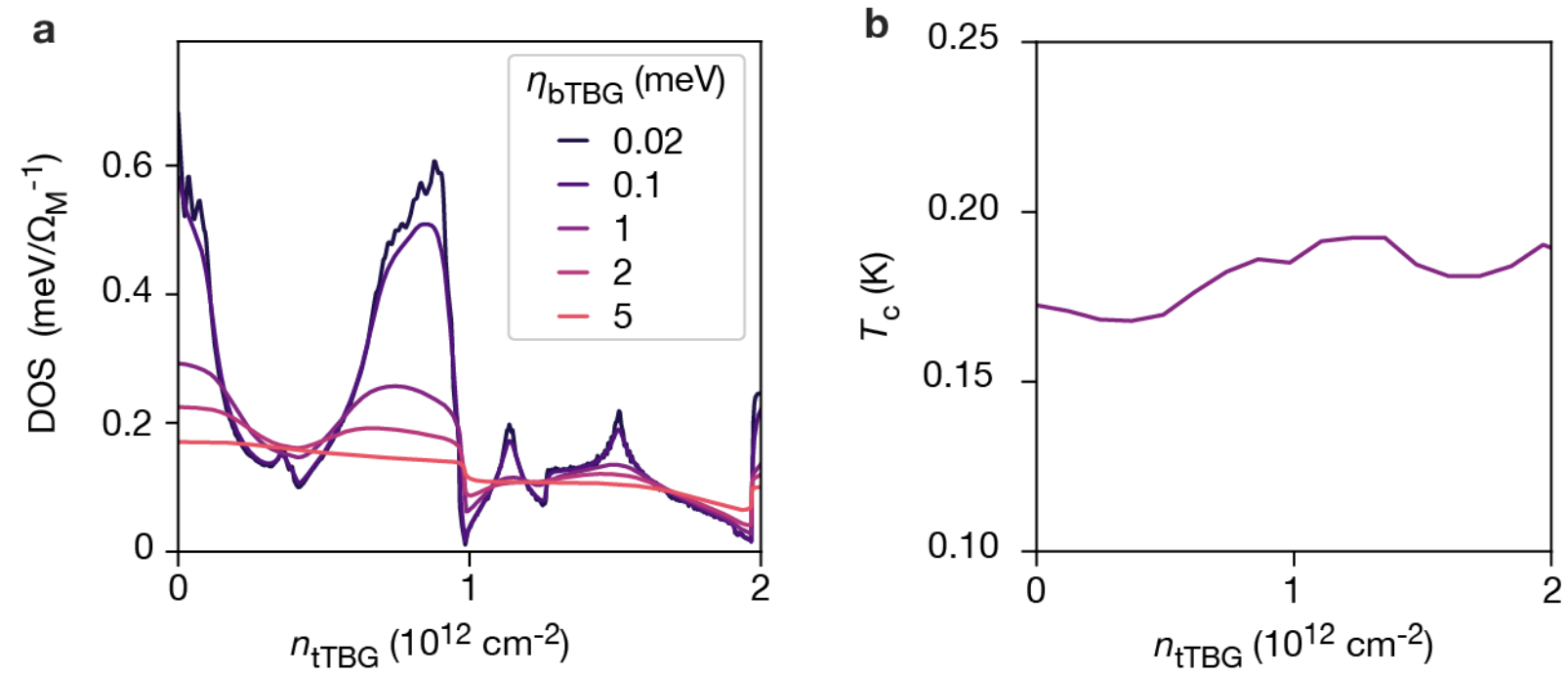

**Figure S18|** (a) Calculated density of states as a function of carrier density $n_{tBG}$ for different pairing strengths $\eta_{bTBG}$. The DOS exhibits sharp peaks, which can be seen to broaden and flatten as the pairing strength increases. (b) Superconducting critical temperature $T_c$ as a function of carrier density $n_{tBG}$. The critical temperature shows a non-monotonic behavior, with a dome-like shape, which is a key characteristic of superconductivity in these systems. This figure illustrates the importance of averaging the DOS for screening the effects of superconductivity, as discussed in the main text.

To balance accuracy with tractability, we approximate the top TBG layer by a scaled monolayer graphene band structure, characterized by a scaling factor $\alpha$. From our experimental data – particularly the suppression of correlated insulating states – we estimate $\alpha \sim 10$, which differs only by a factor of two from values $\alpha = 4 - 5$ obtained in more sophisticated top TBG models (see Figure S18). Given the complexity of the problem, we regard this level of agreement as reasonable and supportive of our simplified approach.

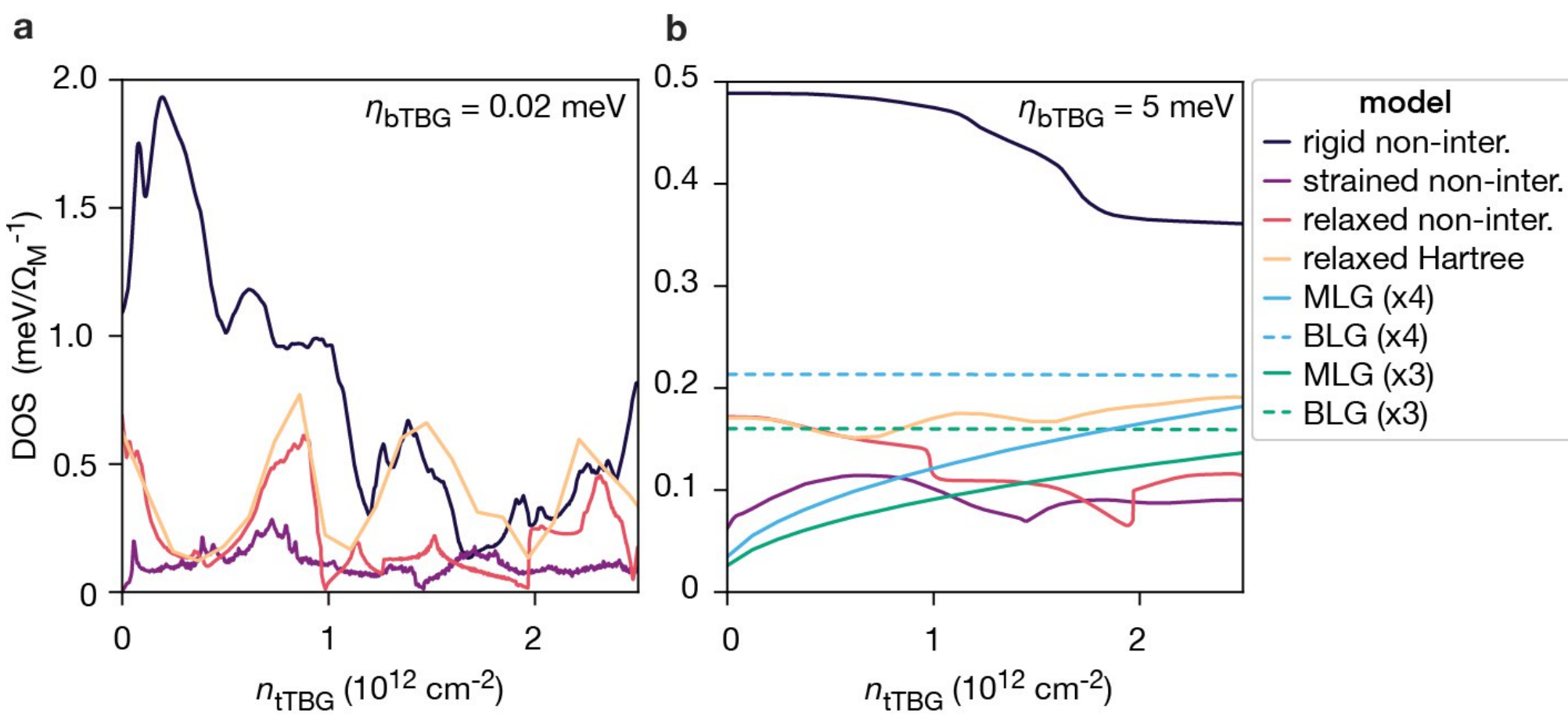

**Figure S19**| Theoretical calculations of the averaged DoS in TBG with $\theta = 0.46°$ obtained using different assumptions (a) DOS averaged over a window η=0.02meV for three different models (rigid non-interacting, non-interacting with 0.3% uniaxial heterostrain along the x axisand non-interacting including lattice relaxation and Hartree interactions including lattice relaxation) (b) Same for an averaging window η=5meV, with comparison to the DOS equivalent to 3 times that of monolayer graphene (MLG) and 3 times that of bilayer graphene (BLG) . Colour coding is the same for all panels. The enhanced DoS in the rigid model arises due to proximity of $\theta = 0.46°$ to the second magic angle of TBG 0.5° predicted in previous work [18], which is not stable against reconstruction.

**11.6 Importance of averaging the density of states for screening of superconductivity**

The DOS of the screening subsystem enters theory only after being averaged over the characteristic pairing frequency, which is set by properties of the superconducting subsystem. The DOS of the superconducting subsystem enters directly into the gap equation but is fixed at filling ν=-2. The experimentally tunable control parameter is therefore the carrier density and the corresponding DOS of the screening subsystem.

Depending on the superconducting pairing mechanism, the DOS has to be averaged over a specific energy. In the case of a conventional phonon-mediated pairing, e.g. Bardeen–Cooper–Schrieffer (BCS), the critical temperature is given by $k_B T_c = 1.134 E_D e^{-1/N(0)}$, where $E_D$ is the Debye cutoff energy and $N(0)$ is the DOS at the Fermi level. For a fixed Debye cutoff energy, the critical temperature is fully determined by the density of states at the Fermi level. However, a more complete theory, e.g. Eliashberg theory, shows that the pairing interaction has both momentum and frequency dependence [49]. Therefore, the effective screening DOS that enters the pairing problem is not simply $N(0)$ but rather an average over an energy window set by the bosonic cutoff and weighted by the interaction kernel.

In moiré materials the DOS varies strongly with energy. In our experiment, the neutrality points of the screening layer correspond to local DOS minima, with an energy scale varying of about 5-10meV. This energy scale is comparable to the pairing energy in the proposed plasmon mechanism. For this reason, if the characteristic bosonic energy is higher than the characteristic bandwidth of the screening layer, the local DOS cannot enhance or suppress superconductivity, and the critical temperature is set by the average DOS behavior. To illustrate this point, we have calculated the superconducting critical temperature $T_c$ of magic-angle TBG assuming a plasmon-mediated pairing mechanism and using a simplified miniband structure for the top TBG layer. The results are shown in Figure S18. In our case, the large DOS from the bottom TBG layer provides strong self-screening, which suppresses the contribution of the DOS at the Fermi level in the long-wavelength limit. On the other hand, the particle–hole channel nature of the RPA contribution gives weight to the DOS at finite frequencies. As a result, the superconducting transition temperature correlates more closely with the DOS averaged over the plasmon frequency range, rather than with the DOS strictly at the Fermi level.

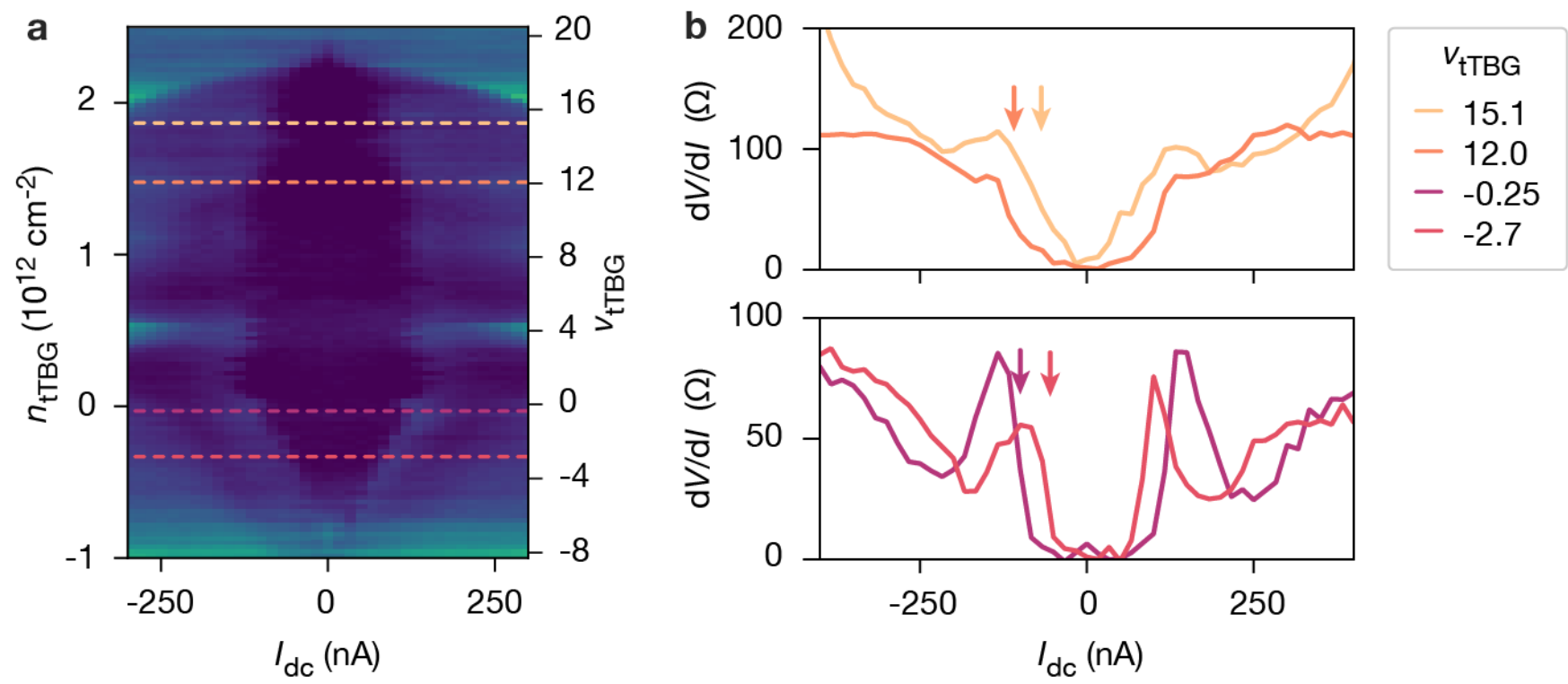

**Figure S20| Reduction of the critical current at high $n^{\mathrm{tTBG}}$ (a)** $dV/dI(I_{\mathrm{dc}}, n^{\mathrm{tTBG}})$ measured at T = 50 mK. The critical current is shown to reduce as $|n^{\mathrm{tTBG}}|$ is decreased. The dashed lines indicate the densities at which the line cuts of panel b where extracted. Same color scale as in figure 3 of the main text. (**b**) Line traces extracted from map (a) with the same color-coding. Critical current is defined at the point at which the resistance is half of the normal state resistance, highlighted with color-coded vertical arrows. The critical current is reduced as $|n^{\mathrm{tTBG}}|$ increases. Note: zero-resistance peaks attributed to contact artefacts

### 11.7 Estimation of the screening length scale

To analyze our results, first we estimated the characteristic length scale for the screening in our experiment. For a metallic screening gate on one side, placed at a distance $l$ from TBG embedded into hBN, electron-electron interactions are described by the Fourier transform of the screened potential and can be written as:

$$V(\mathbf{q}) = \frac{2\pi e^2}{q\sqrt{\epsilon_z \epsilon_{\parallel}}}\left[1 - e^{-2ql\sqrt{\epsilon_{\parallel}/\epsilon_z}}\right] \tag{91}$$

The first term in this expression stands for direct Coulomb potential, whereas the second accounts for image charges in the gate; those are mirrored at a distance $2d\sqrt{\frac{\epsilon_{\parallel}}{\epsilon_z}}$, dependent on the anisotropy of dielectric parameters of hBN [50]: c-axis $\epsilon_{zz}$~3 (ref [51]) and in-plane $\epsilon_{\parallel}$~6.5 (ref. [52]). The latter sensitivity to dielectric parameters can be traced to the scaling up out-of-plane ($z \to z' = z\sqrt{\epsilon_{||}/\epsilon_{zz}}$) coordinate while solving the Poisson equation inside hBN, $\left[\epsilon_{zz}\partial_z^2 - \epsilon_{||}q^2\right]V = 0$. Note that this formula neglects screening by TBG itself. As a result, this approach gave an upper estimation for the relevant thickness $l_*$ of the dielectric spacer.

To estimate the relevant distance between TBG and the metallic layer at which the screening of Coulomb interaction would affect the superconducting critical temperature we noted that the superconducting 'domes' in this system appeared at ½ fillings of the mini Brillouin zone of moiré superlattice. As a result, the Cooper pairs are formed of electrons near a Fermi level with a typical size of $\sim\frac{1}{2}g$, where $g = \frac{4\pi}{3\lambda} = \frac{4\pi}{3a/\theta}$ is the moiré Bragg vector corresponding to the period, $\lambda = a/\theta$, where $a$ is graphene's lattice constant, and $\theta$~0.02 for magic angle TBG (with a 1° twist). The relevant interaction of electrons has a typical momentum transfer $q \sim \frac{1}{2} g$. To reduce interaction twice, one should require that $\frac{4\pi}{3a/\theta} l \sqrt{\epsilon_{||}/\epsilon_z} \lesssim \ln 2$, therefore we obtained:

$$l_* \lesssim 0.7 \frac{3}{4\pi} \sqrt{\frac{\epsilon_{zz}}{\epsilon_{\|}}} \lambda \sim 0.7 \frac{3}{4\pi\theta} \sqrt{\frac{\epsilon_{zz}}{\epsilon_{\|}}} a \sim 1.5\text{nm}, \tag{92}$$

which corresponds to the thickness of 5 hBN layers.

**12. Estimations of the interface transparency for the proximity effect**

We also considered the possibility that the proximity effect between a thin superconductor and a normal metal [53,54] caused the suppression of the superconducting gap observed in our device. This effect can be thought as the detour of Cooper pairs into the normal metal where they eventually disappear/collapse. This is the case for a highly transparent superconducting-normal interface. However, in the case of a significant interfacial/tunnel resistivity $\rho$ to enter the normal metal, Cooper pairs may not be allowed to take such a detour.

To estimate the possibility for such a scenario, we started by considering the expression for the microscopic supercurrent $I$ associated with the jump of a Cooper pair into the normal metal [55], given by $I \approx \Delta/eR$, where $R$ is the tunnel resistance over an area commensurate with the typical size of a Cooper pair, and $\Delta$ is the superconducting gap. $R$ could be estimated as $R \approx \rho/\xi^2$ where $\xi$ is the superconducting coherence length and we used $\rho$ as the resistivity, in units of $\Omega \cdot \text{cm}^{-2}$. The characteristic time scale of such problem is the decay rate of tunnelling Cooper pairs, given by $1/\tau \approx I/e$. To enable such process, Cooper pairs must tunnel sufficiently fast so that the associated uncertainty $h/\tau$ is larger than the superconducting gap $\Delta$. Otherwise (at long $\tau > h/\Delta$), Cooper pairs are reflected by the resistance barrier. The above equations result into the following inequality required to prevent Cooper pairs from efficient tunnelling into the normal metal:

$$\Delta > \frac{h}{\tau} \approx \frac{hI}{e} \approx \frac{\left(\frac{h}{e^2}\right)\Delta}{R} \; or \; \frac{\rho}{\xi^2} > \frac{h}{e^2}. \tag{93}$$

We note that eq. 93 also follows from a dimensional analysis and must be very robust with respect to microscopic derivations.

To exhibit interlayer jumps, electrons in high angle twisted system should change their momentum by a value comparable to the intervalley distance in single layer graphene, which is at the order of the size of the graphene Brillouin zone. Such dramatic change would be impossible without short range impurities or atomic defects. According to earlier work [56], the interface resistivity $\rho$ between two graphene layers twisted by a large angle is well above $10^{-4}$ $\Omega$ $\text{cm}^{-2}$ and is expected to be even higher for our high-quality device encapsulated with hBN. This yields $R \approx \rho/\xi^2 > 10^6$ $\Omega >> h/e^2$ (considering $\xi \sim 50$ nm as in ref. [57,58]), hence, our barriers were insufficiently transparent to cause any suppression of superconductivity due to the superconducting proximity effect.

**13. Commensurate construction for twisted tetralayer systems**

In this section we discuss how to construct commensurate structures for tetralayer systems. We first review the standard $(m, r)$ construction for twisted bilayers of hexagonal lattices. We then explain how this construction is extended to a twisted tetralayer, where the twist angles $\theta_{12}$ and $\theta_{34}$ are fixed and the remaining angle $\theta_{23}$ is chosen to make the full structure commensurate.

For a single layer we choose the triangular Bravais lattice vectors

$$\boldsymbol{a}_1 = a(1,0), \qquad \boldsymbol{a}_2 = a\left(1/2, \sqrt{3}/2\right), \tag{94}$$

a twisted bilayer with relative twist angle $\theta$ is commensurate if there exist two linearly independent vectors that are Bravais translations of both layers. For hexagonal lattices this condition can be parameterized by two integers $(m, r)$ with $m \geq 0$ and $r \geq 1$. For a given $(m, r)$, the supercell vectors (written in the basis of layer 1) are

$$\boldsymbol{T}_1(m, r) = m\boldsymbol{a}_1 + (m + r)\boldsymbol{a}_2, \tag{95}$$

$$\boldsymbol{T}_2(m, r) = -(m + r)\boldsymbol{a}_1 + (2m + r)\boldsymbol{a}_2. \tag{96}$$

The corresponding commensurate twist angle is

$$\cos\theta(m, r) = \frac{3m^2 + 3mr + \frac{1}{2}r^2}{3m^2 + 3mr + r^2}. \tag{97}$$

Equations 95-97 generate all commensurate twisted bilayer structures.

We now consider a twisted tetralayer in which the twist angles between layers (1,2) and (3,4) are fixed to $\theta_{12}$ and $\theta_{34}$, respectively. Each bilayer defines a triangular moiré lattice in real space, with primitive vectors $\boldsymbol{T}_{11}, \boldsymbol{T}_{12}$ for the (1,2) moiré lattice and $\boldsymbol{T}_{21}, \boldsymbol{T}_{22}$ for the (3,4) moiré lattice. The corresponding moiré lattice constants are $a_M^{12} = |\boldsymbol{T}_{11}| = |\boldsymbol{T}_{12}|$ and $a_M^{34} = |\boldsymbol{T}_{21}| = |\boldsymbol{T}_{22}|$. The vectors $\boldsymbol{T}_{11}, \boldsymbol{T}_{12}$ are obtained from a chosen bilayer pair $(m_{12}, r_{12})$ using Eqs. 95-96, and similarly for $\boldsymbol{T}_{21}, \boldsymbol{T}_{22}$ from $(m_{34}, r_{34})$.

The remaining task is to choose the intermediate twist angle $\theta_{23}$ such that the two moiré lattices admit a common supercell. We first define the ratio of the two moiré lattice constants,

$$\rho \equiv a_M^{34}/a_M^{12}. \tag{98}$$

When $\rho \neq 1$, commensurability requires matching both the relative orientation and the length scale, which cannot be achieved with a single $(m, r)$ pair.

To proceed, we represent two-dimensional vectors as complex numbers. We introduce

$$\omega = e^{i\pi/3} = \frac{1}{2} + i\frac{\sqrt{3}}{2}, \tag{99}$$

and define

$$\beta(m, r) = m + (m + r)\omega, \qquad m \geq 1, r \geq 1. \tag{100}$$

Up to an overall real scale factor, $\beta(m, r)$ represents a superlattice vector of a triangular lattice. A commensurate (or near-commensurate) condition between the two moiré lattices is then

$$a_M^{12}\beta(m_1, r_1) = \rho e^{i\theta_{23}} a_M^{12}\beta(m_2, r_2), \tag{101}$$

or equivalently

$$\rho e^{i\theta_{23}} = \frac{\beta(m_1, r_1)}{\beta(m_2, r_2)}. \tag{102}$$

Taking the modulus and argument yields

$$\rho_{\text{cand}} = \left|\frac{\beta(m_1, r_1)}{\beta(m_2, r_2)}\right|, \qquad \theta_{23,\text{cand}} = \arg\left(\frac{\beta(m_1, r_1)}{\beta(m_2, r_2)}\right) \text{ (mod } 60°\text{)}. \tag{103}$$

Using $|u + u\omega|^2 = u^2 + uv + v^2$, one finds

$$|\beta(m, r)|^2 = 3m^2 + 3mr + r^2, \tag{104}$$

and therefore

$$\rho_{\text{cand}}^2 = \frac{3m_1^2 + 3m_1 r_1 + r_1^2}{3m_2^2 + 3m_2 r_2 + r_2^2}. \tag{105}$$

For a target $\rho$, we search integer pairs $(m_1, r_1)$ and $(m_2, r_2)$ such that $\rho_{\text{cand}}$ matches $\rho$ within a chosen tolerance; each match yields a candidate $\theta_{23}$.

The derivation above implicitly assumes that $\boldsymbol{T}_{11}$ and $\boldsymbol{T}_{21}$ lie along the $x$ direction, while $\boldsymbol{T}_{12}$ and $\boldsymbol{T}_{22}$ are oriented at 30°. However, the primitive vectors obtained from Eqs. 95-96 do not in general satisfy this orientation. To convert the relative rotation angle obtained above into a global twist angle, this offset must be taken into account.

Let $\theta_{12}^0$ and $\theta_{34}^0$ denote the angles between $\boldsymbol{T}_{11}$ and $\boldsymbol{T}_{21}$ and the $x$ axis, respectively. The global rotation angle should then be replaced by

$$\theta_{23} \to \theta_{12}^0 - \theta_{34}^0 + \theta_{23} - \theta_{12} \text{ (mod } 60°\text{)}. \tag{106}$$

With this convention, the global rotation angles of the four layers are

$$\phi_1 = 0, \qquad \phi_2 = \theta_{12}, \qquad \phi_3 = \theta_{12} + \theta_{23}, \qquad \phi_4 \to \theta_{12} + \theta_{23} + \theta_{34}^0. \tag{107}$$

In practice, we enumerate integer pairs $(m_1, r_1, m_2, r_2)$ up to a chosen cutoff to obtain all commensurate angles below a given supercell size. Once a candidate commensurate structure is identified, we explicitly verify commensurability in real space. We therefore have

$$\boldsymbol{T}_1 = m_1 \boldsymbol{T}_{11} + (m_1 + r_1)\boldsymbol{T}_{12} = m_2 \boldsymbol{T}_{21} + (m_2 + r_2)\boldsymbol{T}_{22}, \tag{108}$$

$$\boldsymbol{T}_2 = -(m_1 + r_1)\boldsymbol{T}_{11} + (2m_1 + r_1)\boldsymbol{T}_{12} = -(m_2 + r_2)\boldsymbol{T}_{21} + (2m_2 + r_2)\boldsymbol{T}_{22}. \tag{109}$$

As an explicit example, we choose $(m_1, r_1) = (5, 1)$ for the $(1, 2)$ bilayer, which gives a twist angle $\theta_{12} = 6.01°$, and $(m_2, r_2) = (25, 1)$ for the $(3, 4)$ bilayer, which gives $\theta_{34} = 1.25°$. The corresponding moiré lattice basis vectors, obtained from Eqs. 95-96, are

$$\begin{aligned} \boldsymbol{T}_{11} &= (19.68, 12.78)\ \text{Å}, \qquad \boldsymbol{T}_{12} = (-1.23, 23.43)\ \text{Å}, \\ \boldsymbol{T}_{21} &= (97.17, 57.52)\ \text{Å}, \qquad \boldsymbol{T}_{22} = (-1.23, 112.91)\ \text{Å}. \end{aligned} \tag{110}$$

These vectors correspond to moiré lattice constants $a_M^{12} = 23.47$ Å and $a_M^{34} = 112.92$ Å, giving a ratio $\rho = 4.812$. From the orientation of $\boldsymbol{T}_{11}$ and $\boldsymbol{T}_{21}$, we find the offset angles $\theta_{12}^0 = 33°$ and $\theta_{34}^0 = 30.62°$.

Choosing $(m_1, r_1, m_2, r_2) = (9, 2, 2, 1)$, the candidate ratio $\rho_{\text{cand}}$ matches the target value $\rho = a_M^{34}/a_M^{12}$ within numerical tolerance, and the relative rotation angle obtained from the $\beta$ construction is $\theta_{23} = 47.2°$. After accounting for the orientation offsets, the resulting global rotation angle is $\theta_{23} = 43.57°$. The corresponding tetralayer unit cell is shown in Fig. S21.

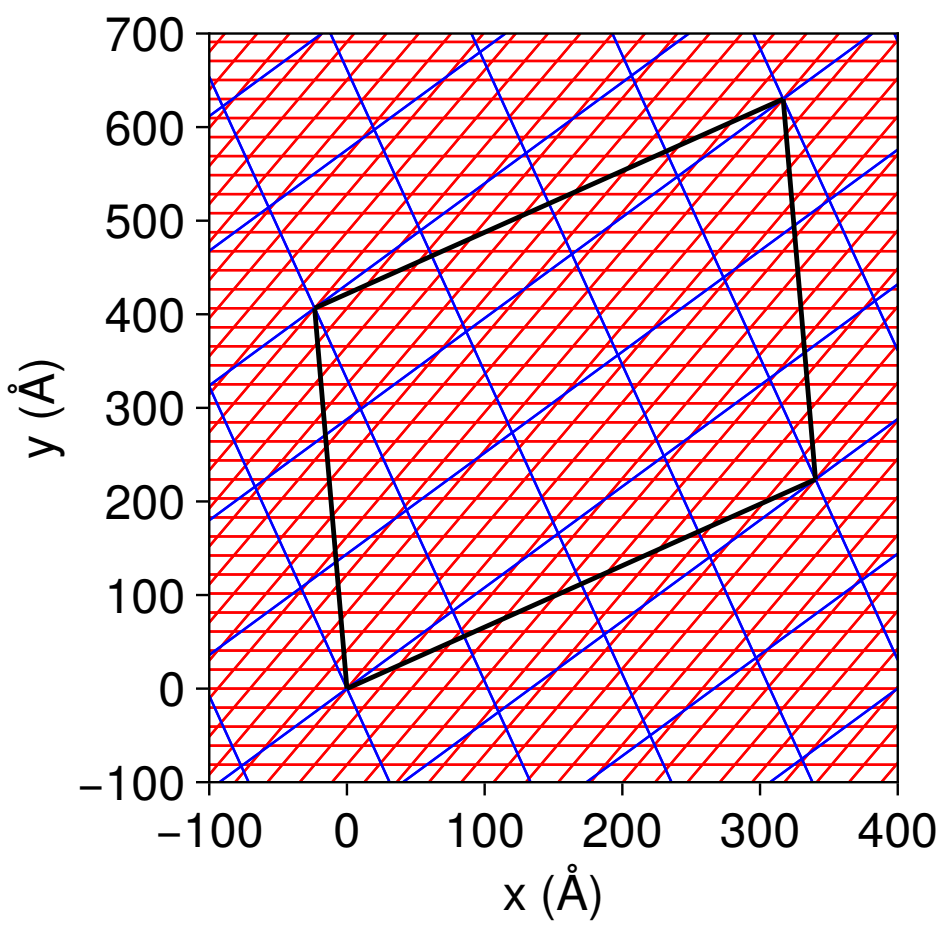

**Figure S21|** Illustration of the commensurate tetralayer construction in real space. The red and blue lattices represent the moiré patterns of the (1,2) and (3,4) bilayers, respectively. Here we choose $(m, r) = (5, 1)$ for the (1,2) bilayer, which gives a twist angle $\theta_{12} = 6.01°$, and $(m, r) = (26, 1)$ for the $(3, 4)$ bilayer, which gives $\theta_{34}$. The black polygon denotes the common supercell corresponding to $(m_1, r_1, m_2, r_2) = (9, 2, 2, 1)$, which yields an intermediate twist angle $\theta_{23} = 43.57°$. This construction ensures that the full four-layer structure is commensurate and periodic.

## Supplementary References